\documentclass[iop,apj,twocolumn,numberedappendix,twocolappendix]{emulateapj}
\usepackage{apjfonts}
\usepackage{graphicx}
\usepackage{natbib}
\usepackage{color}

\newcommand{\GILDAS}{\texttt{GILDAS}}
\newcommand{\CLIC}{\texttt{CLIC}}

\newcommand{\MAPPING}{\texttt{MAPPING}}

\newcommand{\ie} {{\em i.e.}}
\newcommand{\eg} {{\em e.g.}}
\newcommand{\etal} {et al.}

\newcommand{\nds}[1]{\emm{\displaystyle#1}} 
\newcommand{\paren}[1]  {\nds{\left(  #1 \right) }} 
\newcommand{\bracket}[1]{\nds{\left[  #1 \right] }} 
\newcommand{\cbrace}[1]{\nds{\left\{  #1 \right\} }} 
\newcommand{\aver}[1] {\nds{\left< #1 \right>}}  %
\newcommand{\abs}[1] {\nds{\left| #1 \right|}}  %

\newcommand{\about}{\emm{\sim}}

\renewcommand{\ion}[2]{\mbox{#1{\sc #2}}}
\newcommand{\HI}{\ion{H}{i}}   

\renewcommand{\H} {\mbox{H}}         
\newcommand{\HH}  {\mbox{H$_2$}}     
\newcommand{\twCO}{\mbox{$^{12}$CO}} 
\newcommand{\thCO}{\mbox{$^{13}$CO}} 
\newcommand{\twC}{\mbox{$^{12}$C}} 
\newcommand{\thC}{\mbox{$^{13}$C}} 
\newcommand{\twCp}{\mbox{$^{12}$C$^+$}} 
\newcommand{\thCp}{\mbox{$^{13}$C$^+$}} 
\newcommand{\Cp}{\mbox{C$^+$}} 
\newcommand{\HCOp}{\emr{HCO^+}}

\newcommand{\Jone}{\mbox{(1--0)}}

\newcommand{\Xco}{\emm{X_\emr{CO}}}
\newcommand{\N}[1]{\emm{N_{#1}}}
\newcommand{\NCO}{\N{\emr{CO}}}
\newcommand{\NHH}{\N{\HH{}}}
\newcommand{\W}[1]{\emm{W_\emr{#1}}}
\newcommand{\WCO}{\W{CO}}
\newcommand{\R}{\emm{R}}
\newcommand{\Rsun}{\emm{R_\sun}}
\newcommand{\Tpeak}{\emm{T_\emr{peak}}}

\newcommand{\Av}{\emr{A_v}} 
\newcommand{\Ebv}{\emm{E_\emr{B-V}}}

\newcommand{\emm}[1]{\ensuremath{#1}}   
\newcommand{\emr}[1]{\emm{\mathrm{#1}}} 
\newcommand{\unit}[1]{\emm{\, \emr{#1}}}
\newcommand{\sci}[2]{\emm{#1\times10^{#2}}}

\newcommand{\K}{\unit{K}}
\newcommand{\mK}{\unit{mK}}
\newcommand{\mum}{\unit{\mu{}m}}
\newcommand{\mm}{\unit{mm}}

\newcommand{\m}{\unit{m}}
\newcommand{\kms}{\unit{km\,s^{-1}}}
\newcommand{\Kkms}{\unit{K\,km\,s^{-1}}}
\newcommand{\Kkmspcpc}{\unit{K\,km\,s^{-1}\,pc^{2}}}
\newcommand{\pscmpKkms}{\unit{cm^{-2}/(K\,km\,s^{-1})}}
\newcommand{\Kpccm}{\unit{K\,cm^{-3}}}
\newcommand{\MHz}{\unit{MHz}}
\newcommand{\GHz}{\unit{GHz}}
\newcommand{\pc}{\unit{pc}}
\newcommand{\kpc}{\unit{kpc}}
\newcommand{\Mpc}{\unit{Mpc}}
\newcommand{\Msol}{\unit{M_\odot}}
\newcommand{\pccm}{\unit{cm^{-3}}}
\newcommand{\pscm}{\unit{cm^{-2}}}

\newcommand{\Tas}{\emm{T_\emr{A}^*}}
\newcommand{\Tmb}{\emm{T_\emr{mb}}}
\newcommand{\Tsys}{\emm{T_\emr{sys}}}
\newcommand{\Beff}{\emm{B_\emr{eff}}}
\newcommand{\Feff}{\emm{F_\emr{eff}}}

\newcommand{\Wco}{\emm{W_\emr{CO}}}
\newcommand{\Lco}{\emm{L_\emr{CO}}}
\newcommand{\Mhh}{\emm{M_{\HH}}}
\newcommand{\D}[1]{\emm{\Delta #1}}
\newcommand{\uncert}[1]{\emm{\delta #1}}
\newcommand{\independent}{\emm{\eta}}

\newcommand{\rmin}{\emm{r_\emr{min}}}

\newcommand{\dprim}{\emm{d_\emr{prim}}}

\newcommand{\Afwhm}{\emm{\theta_\emr{fwhm}}}
\newcommand{\wavelength}{\emm{\lambda}}

\renewcommand{\vec}[1]{\emm{\mathbf{#1}}}
\newcommand{\vgal}{\emm{\vec{v_\emr{gal}}}}
\newcommand{\vobs}{\emm{v_\emr{obs}}}
\newcommand{\vcent}{\emm{v_\emr{cent}}}
\newcommand{\vturb}{\emm{\sigma_\emr{turb}}}
\newcommand{\usys}{\emm{u_\emr{sys}}}
\newcommand{\up}{\emm{u_p}}
\newcommand{\vp}{\emm{v_p}}
\newcommand{\vz}{\emm{v_z}}
\renewcommand{\i}{\emm{i}}
\newcommand{\beamwidth}{\emm{\theta}}
\newcommand{\cvi}{\emm{\abs{\vec{\emr{grad}(\vcent{})}}}}

\newcommand{\A}{\emm{A}} \newcommand{\Grav}{\emm{G}}
\newcommand{\Height}[1]{\emm{H_\emr{#1}}}
\newcommand{\sigvz}{\emm{\sigma_z}}
\newcommand{\sigvstar}{\emm{\sigma_{\star}}}
\newcommand{\siggas}{\emm{\Sigma_\emr{gas}}}
\newcommand{\sigstar}{\emm{\Sigma_{\star}}}
\newcommand{\rhostar}{\emm{\rho_{\star}}} \newcommand{\rhogas}{\emm{\rho}}
\newcommand{\Pgas}{\emm{P}}
\newcommand{\Qgas}{\emm{Q_\emr{gas}}}
\newcommand{\Qstar}{\emm{Q_\star}}

\newcommand{\TabWhirlpool}{%
  \begin{table*}
    \caption{Parameters for NGC\,5194.}
    \centering{} %
    \begin{tabular}{lccl}
      \hline \hline
      Parameter & NGC 5194 & Notes & Ref. \\
      \hline
      Morphological type              & SA(s)bc pec                                                &                                & \citet{devaucouleurs91}\\
      Activity type                   & Seyfert 2                                                  &                                & \citet{veron06}\\
      Kinematic center                & $13^\emr{h}29^\emr{m}52.7087^\emr{s}; +47\degr11'42.789''$ & $\alpha,\delta$ J2000          & \citet{hagiwara07}\\
                                      & $1.80'',0.81''$                                            & Offset from phase center       & \\
      Distance                        & $7.6\pm1\Mpc$                                              & $1''=37\pm5\pc$                & \citet{ciardullo02}\\
      Systemic velocity               & $471.7\pm0.3\kms$                                          & LSR, radio convention          & \citet{shetty07}\\
      Mean CO inclination             & $21 \pm 3\degr$                                            &                                & \citet{colombo12b}\\
      Mean position angle             & $173 \pm 3\degr$                                           &                                & \citet{colombo12b}\\
      Emitting surface                & $1.9\times10^{8}\pc^2$                                     & $\Wco \ge 3\sigma$             & This work, Appendix~\ref{sec:co:luminosity}\\
      Total CO luminosity$^{a}$       & $1.4\times10^{9}\Kkmspcpc$                                 & in [LSR$-120$,LSR$+120$]\kms{} & This work, Appendix~\ref{sec:co:luminosity}\\ 
      Total molecular mass$^{a}$      & $6.2\times10^{9}\Msol$                                     & Helium included                & This work, Appendix~\ref{sec:co:luminosity}\\
      Mean brightness$^{b}$           & $7.6\Kkms$                                                 &                                & This work, Appendix~\ref{sec:co:luminosity}\\
      Mean mass surface density$^{b}$ & $33\Msol\pc^2$                                             &                                & This work, Appendix~\ref{sec:co:luminosity}\\
      \hline
    \end{tabular}\\
    $^{a}$ Using the IRAM-30m data. $^{b}$ The mean brightness and mass
    surface density are computed using the area with significant emission,
    \ie{}, $\Wco \ge 3\sigma$.
    \label{tab:m51}
  \end{table*}}

\newcommand{\TabObsSDmain}{%
  \begin{table*}
    \caption{Parameters of the 30m observations.}
    \centering{} %
    \begin{tabular}{rccccccccc}
      \hline \hline
      Molecule      & Frequency  & \Feff{} & \Beff{} & Resol.$^{a}$ & Resol.$^{b}$ & Map Size   & Time$^{c}$ & \Tsys{}  & $\sigma$ \\
      \& Transition &       GHz  &         &         & \kms{}       & arcsec       & arcmin$^2$ & hours      & K [\Tas] & mK [\Tmb]      \\
      \hline
      \twCO{} \Jone{} & 115.271202 & 0.95 & 0.75 & 5.20/5.00 & 21.3/22.5 & $53$ $(\sim 6\times10)$ & 17.3/41 & 285 & 16\\
      \thCO{} \Jone{} & 110.201354 & 0.95 & 0.76 & 5.44/5.00 & 22.3/23.6 & $53$ $(\sim 6\times10)$ & 17.3/41 & 140 & 7.5\\
      \hline
    \end{tabular}\\
    $^{a}$ The two values correspond to the backend natural channel 
    spacing and to the channel spacing used to match the PdBI channel spacing.\\
    $^{b}$ The two values correspond to the natural FWHM of the beam
    and to the map resolution after gridding through convolution with a Gaussian.\\
    $^{c}$ Two values are given for the integration time: the on-source
    time and the telescope time.
    \label{tab:obs:30m:main}
  \end{table*}}

\newcommand{\TabObsPdBImain}{%
  \begin{table*}
    \caption{Parameters of the PdBI observations.}
    \centering{} %
    \begin{tabular}{rcccccc}
      \hline \hline
      Molecule & Transition & \multicolumn{2}{c}{Frequencies [\GHz{}]} & \multicolumn{3}{c}{Velocity [\kms{}]} \\
               &            &  Rest         & Tuned       & LSR & Tuned & Resolution \\
      \hline
      \twCO{}  & \Jone{}    & 115.271202    & 115.090     & 471.7       & 0     & 5 \\
      \hline
    \end{tabular}
    \begin{tabular}{cccccc}
      \hline
      \hline
      Mosaic \# & \multicolumn{2}{c}{Projection center (J2000)} & $N_\emr{fields}$ & Beam         & PA     \\
                & R.A                & Dec.                     &                  & $''\times''$ & $\deg$ \\ 
      \hline
      Top    & $13^h 29^m 54.921^s$ & $47^\circ 12' 21.589''$ & 30 & $ 1.12 \times 0.92$ & 61.5 \\
      Bottom & $13^h 29^m 50.504^s$ & $47^\circ 11' 03.642''$ & 30 & $ 1.20 \times 1.01$ & 84.0 \\
      Full   & $13^h 29^m 52.532^s$ & $47^\circ 11' 41.982''$ & 60 & $ 1.16 \times 0.97$ & 73.0 \\
      \hline
    \end{tabular}
    \label{tab:obs:pdbi:main}
  \end{table*}}

\newcommand{\TabNoise}{%
  \begin{table*}
    \caption{Median noise levels of the different \twCO{} \Jone{} cubes.}
    \centering{} %
    \begin{tabular}{ccccccccccc}
      \hline
      \hline
      Cube type & Resolution & \multicolumn{9}{c}{$1\sigma$ Noise levels} \\
                &   arcsec   & (\mK{}[\Tmb{}]) & \multicolumn{4}{c}{(\Kkms{})} & \multicolumn{4}{c}{$(\Msol\pc^{-2})$} \\
                &            &                 & $5\kms$ & $25\kms$ & $50\kms$ & $100\kms$ & $5\kms$ & $25\kms$ & $50\kms$ & $100\kms$ \\
      \hline
      IRAM-30m                     &   22.5 &    16 &   0.08 &   0.18 &   0.25 &   0.36 &   0.35 &   0.78 &   1.11 &   1.57 \\
      \hline
      Hybrid synthesis             &    6.0 &    35 &   0.17 &   0.39 &   0.55 &   0.78 &   0.76 &   1.70 &   2.40 &   3.40 \\
      Hybrid synthesis             &    3.0 &   106 &   0.53 &   1.18 &   1.67 &   2.37 &   2.30 &   5.15 &   7.28 &  10.30 \\
      Hybrid synthesis             &    1.1 &   394 &   1.97 &   4.40 &   6.23 &   8.81 &   8.56 &  19.15 &  27.08 &  38.30 \\
      \hline
      PdBI-only                    &    6.0 &    38 &   0.19 &   0.43 &   0.60 &   0.85 &   0.83 &   1.85 &   2.62 &   3.71 \\
      PdBI-only                    &    3.0 &    95 &   0.47 &   1.06 &   1.50 &   2.11 &   2.06 &   4.60 &   6.50 &   9.19 \\
      PdBI-only                    &    1.1 &   396 &   1.98 &   4.43 &   6.27 &   8.86 &   8.62 &  19.26 &  27.24 &  38.53 \\
      \hline
      Hybrid synthesis - PdBI-only &    6.0 &    14 &   0.07 &   0.15 &   0.22 &   0.31 &   0.30 &   0.66 &   0.94 &   1.33 \\
      Hybrid synthesis - PdBI-only &    3.0 &    25 &   0.12 &   0.27 &   0.39 &   0.55 &   0.53 &   1.19 &   1.69 &   2.39 \\
      Hybrid synthesis - PdBI-only &    1.1 &    33 &   0.17 &   0.37 &   0.52 &   0.74 &   0.72 &   1.61 &   2.27 &   3.22 \\
      \hline
    \end{tabular}
    \label{tab:noise}
  \end{table*}}

\newcommand{\TabObsPdBIdetail}{%
  \begin{table}
    \caption{Detailed parameters for the PdBI observations.}
    \centering{} %
      \begin{tabular}{ccccccr}
        \hline \hline
        Config. & $N_\emr{ant}$ & Mosaic & Int. Time$^{a}$ & \Tsys{} & Seeing & \multicolumn{1}{c}{Obs. date} \\
        &               & \#     & hours           & K       & arcsec & \\
        \hline
        D & 5 & Top & 2.0/09.5 & 150-400 & 1.70 & 29 Aug. 2009 \\
          & 5 & Bot & 2.2/09.5 & 200-400 & 1.70 & 02 Oct. 2009 \\
        C & 6 & Top & 4.3/09.5 & 180-250 & 0.68 & 27 Oct. 2009 \\
          & 6 & Bot & 3.3/08.5 & 180-300 & 0.85 & 28 Oct. 2009 \\
        B & 6 & Top & 1.8/08.0 & 200-250 & 0.37 & 20 Feb. 2010 \\
          & 6 & Top & 1.2/05.0 & 200-250 & 0.60 & 02 Mar. 2010 \\
          & 6 & Top & 1.5/05.0 & 160-230 & 0.50 & 04 Mar. 2010 \\
          & 6 & Top & 3.3/07.5 & 180-280 & 0.29 & 11 Mar. 2010 \\
          & 6 & Bot & 5.1/11.0 & 150-250 & 0.34 & 08 Mar. 2010 \\
          & 6 & Bot & 4.2/11.5 & 200-300 & 0.50 & 09 Mar. 2010 \\
          & 6 & Bot & 4.6/09.0 & 160-220 & 0.30 & 12 Mar. 2010 \\
        A & 6 & Top & 1.3/03.0 & 180-200 & 0.32 & 14 Dec. 2009 \\
          & 6 & Top & 3.0/08.5 & 180-220 & 0.28 & 17 Dec. 2009 \\
          & 6 & Top & 5.2/11.0 & 160-230 & 0.15 & 18 Jan. 2010 \\
          & 6 & Top & 2.2/05.0 & 200-600 & 0.27 & 22 Jan. 2010 \\
          & 6 & Top & 3.6/07.5 & 180-230 & 0.27 & 12 Feb. 2010 \\
          & 6 & Top & 1.5/04.5 & 160-230 & 0.16 & 13 Feb. 2010 \\
          & 6 & Bot & 3.2/05.5 & 170-220 & 0.31 & 15 Dec. 2009 \\
          & 6 & Bot & 1.5/07.5 & 180-400 & 0.35 & 03 Jan. 2010 \\
          & 6 & Bot & 3.0/08.5 & 180-400 & 0.44 & 05 Jan. 2010 \\
          & 6 & Bot & 3.2/05.5 & 180-230 & 0.18 & 23 Jan. 2010 \\
          & 6 & Bot & 2.3/06.5 & 250-300 & 0.41 & 29 Jan. 2010 \\
        \hline
      \end{tabular}
    $^{a}$ Two values are given for the integration time: the good on-source
    time (as if observed with 6 antennas) and the telescope time.
    \label{tab:obs:pdbi:detail}
  \end{table}}

\newcommand{\TabObsSDdetail}{%
  \begin{table}
    \caption{Detailed parameters for the IRAM-30m observations}
    \centering{} %
      \begin{tabular}{cccc}
        \hline \hline
        Observing date & Time$^{a}$ & \Tsys{}$^{b}$ & Water vapor \\
        & hours      & K [\Tas]      & mm          \\
        \hline
        18/05/2010 & 3.8/8.0 & 279 & 4.3-8.8 \\
        19/05/2010 & 2.9/8.0 & 275 & 3.6-6.1 \\
        20/05/2010 & 3.4/8.0 & 273 & 2.6-8.5 \\
        21/05/2010 & 3.6/9.0 & 287 & 5.8-9.0 \\
        22/05/2010 & 3.6/8.0 & 297 & 6.5-9.4 \\
        \hline
      \end{tabular}\\
    $^{a}$ Two values are given for the integration time: the on-source
    time and the telescope time.\\
    $^{b}$ The \Tsys{} values are given for the \twCO{} \Jone{} frequency.
    \label{tab:obs:30m:detail}
  \end{table}}

\newcommand{\TabFlux}{%
  \begin{table}
    \caption{Flux (in Jy) of the amplitude and phase calibrators used
      during the PdBI calibration.}
    \centering{} %
    \begin{tabular}{cccc}
      \hline
      \hline
      Date & 1418$+$546 &  1308$+$326 & J1332$+$473 \\
      \hline
      29-aug-2009 & 1.05 &  --- & 0.46 \\
      02-oct-2009 & 1.10 & 2.58 & ---  \\
      27-oct-2009 & 1.06 &  --- & 0.22 \\
      28-oct-2009 & 1.02 & 2.58 & ---  \\
      14-dec-2009 & 1.11 &  --- & 0.22 \\
      15-dec-2009 & 1.10 & 2.02 & ---  \\
      17-dec-2009 & 1.16 &  --- & 0.21 \\
      03-jan-2010 & 1.23 & 2.26 & ---  \\
      05-jan-2010 & 1.23 & 2.16 & ---  \\
      18-jan-2010 & 1.22 &  --- & 0.23 \\
      22-jan-2010 & 1.16 &  --- & 0.24 \\
      23-jan-2010 & 1.16 & 1.93 & ---  \\
      25-jan-2010 & 1.22 & 1.83 & ---  \\
      29-jan-2010 & 1.06 & 1.83 & ---  \\
      12-feb-2010 & 0.98 & 1.91 & 0.22 \\
      13-feb-2010 & 1.00 &  --- & 0.22 \\
      20-feb-2010 & 0.92 &  --- & 0.18 \\
      02-mar-2010 & 0.92 &  --- & 0.19 \\
      04-mar-2010 & 0.94 &  --- & 0.20 \\
      08-mar-2010 & 0.94 & 1.92 & ---  \\
      09-mar-2010 & 0.96 & 1.95 & ---  \\
      11-mar-2010 & 0.96 &  --- & ---  \\    
      12-mar-2010 & 0.96 &  --- & ---  \\  
      \hline
    \end{tabular}
    \label{tab:fluxes}
  \end{table}}

\newcommand{\TabHIstackedFit}{%
  \begin{table}
    \caption{Results of the dual Gaussian fit (see Fig.~\ref{fig:stacked:hi}).}
    \centering{} %
    \begin{tabular}{ccccc}
      \hline
      \hline
      Gaussian &  \Tpeak{}   & Velocity &  Width   & Area \\
         \#    & (K[\Tmb{}]) & (\kms{}) & (\kms{}) & (\Kkms{})\\
      \hline
      1 & 5.7 & $-1.4 \pm 0.4$ & $80 \pm 2$ & $490 \pm 13$\\
      2 & 7.2 & $+2.5 \pm 0.2$ & $34 \pm 1$ & $260 \pm 14$\\
      \hline
    \end{tabular}
    \label{tab:stacked:fit:hi}
  \end{table}}

\newcommand{\TabComp}{%
  \begin{table}
    \caption{Comparison of registration and total flux within different \twCO{} \Jone{} datasets.}
    \centering{} %
    \begin{tabular}{cccc}
      \hline
      \hline
      Data set  & Offset & Integrated flux   & Flux difference \\
                & arcsec & $10^{8}\Kkmspcpc$ & \% of PdBI+30m flux \\
      \hline
      IRAM-30m  &  $0$ & 7.82 & $-0.5$ \\  
      PdBI+30m  & $<1$ & 7.86 & $+0.0$ \\  
      CARMA+45m & $<1$ & 7.41 & $-5.7$\\  
      BIMA+12m  & $<1$ & 7.03 & $-10.6$\\  
      \hline
    \end{tabular}
    \label{tab:comp}
  \end{table}}

\newcommand{\TabTotalFlux}{%
    \begin{table}
      \caption{Total CO flux measured within the PdBI+30m data cube,
        as measured from integrated intensity images constructed using
        different techniques to isolate significant emission (see
        text).}
      \centering %
      \begin{tabular}{llc}
        \hline 
        \hline 
        Method & Parameters & Total CO luminosity \\
               &            & [$\times 10^{8} \Kkmspcpc$] \\
        \hline
        Direct sum          & $\mathbf{[-120,+120]\kms}$     & {\bf 9.1} \\ 
        \hline
        Sigma-clipping      & $\Tmb > 1\sigma$                &  417 \\ 
                            & $\Tmb > 2\sigma$                &   16 \\ 
                            & $\mathbf{\Tmb > 3\sigma}$       & {\bf 7.0} \\ 
                            & $\Tmb > 4\sigma$                &  4.6 \\ 
                            & $\Tmb > 5\sigma$                &  3.5 \\ 
        \hline
        Dilated mask        & $(t,p) = (3,1.2)$               &   10 \\ 
                            & $(t,p) = (3,2)$                 &  6.9 \\ 
                            & $(t,p) = (4,1.2)$               &  7.6 \\ 
                            & $(t,p) = (4,2)$                 &  5.8 \\ 
                            & $\mathbf{(t,p) = (5,1.2)}$      & {\bf 7.2} \\ 
                            & $(t,p) = (5,2)$                 &  5.6 \\ 
        \hline
        Smooth \& mask      & $(\theta,m) = (2.0,3)$          &  8.7 \\ 
                            & $(\theta,m) = (2.0,5)$          &  5.6 \\ 
                            & $(\theta,m) = (2.4,3)$          &  8.7 \\ 
                            & $(\theta,m) = (2.4,5)$          &  6.0 \\ 
                            & $(\theta,m) = (3.0,3)$          &  8.7 \\ 
                            & $(\theta,m) = (3.0,5)$          &  6.4 \\ 
                            & $(\theta,m) = (3.6,3)$          &  8.6 \\ 
                            & $\mathbf{(\theta,m) = (3.6,5)}$ & {\bf 6.5} \\ 
        \hline
        \HI{} prior         & $10\kms$ window                 &  3.2 \\ 
                            & $20\kms$ window                 &  5.0 \\ 
                            & $\mathbf{50\kms}$ {\bf window}  & {\bf 8.0} \\ 
        \hline
        Combined            &                                 & {\bf 8.5} \\ 
        \hline
      \end{tabular}
      \label{tab:total:flux}
    \end{table}}

\newcommand{\TabErrorBeam}{%
  \begin{table*}
    \caption{Parameters used to model the IRAM-30m beam at 115.271202\GHz{}. 
      Adapted from~\citet{greve98}.}
    \centering{} %
    \begin{tabular}{lcccc}
      \hline \hline
                         & Main beam     & 1st error beam & 2nd error beam & 3rd error beam \\
      \hline
      Origin             & Diffraction & Large scale   & Panel Frame   & Panel        \\
                         &   pattern   & deformations  & misalignments & deformations \\
      Correlation length &    ---      &  $2.5-3.5\m$  &  $1.5-2.0\m$  & $0.3-0.5\m$  \\ 
      Surface rms        &    ---      & $\la 0.06\mm$ &   $0.055\mm$  &  $0.055\mm$  \\
      Beamwidth (FWHM)   &  $21.3''$   &    $230''$    &   $330''$     &   $1840''$   \\
      Power amplitude    &    1.00     & \sci{9.0}{-4} & \sci{2.7}{-4} & \sci{3.1}{-5}\\
      Relative power     &   71.8\%    &     7.4\%     &     4.7\%     &    16.1\%   \\
      \hline
    \end{tabular}\\
    \label{tab:error:beam}
  \end{table*}}

\newcommand{\FigPdBIvsTHM}{%
  \begin{figure*}
    \centering
    \includegraphics[width=\textwidth]{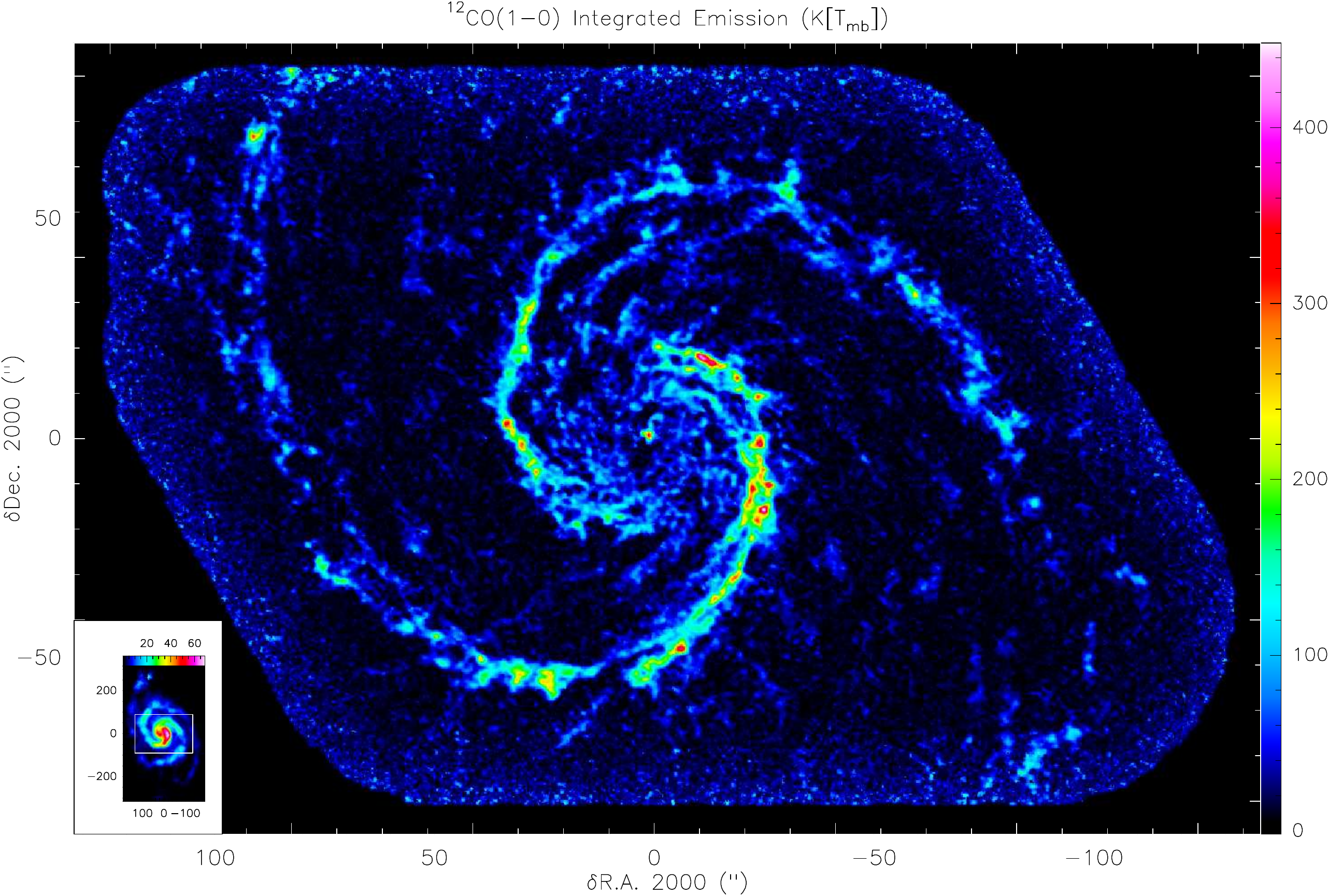}
    \caption{\twCO{} \Jone{} integrated emission of the inner
      $\sim10\kpc\times6\kpc$ of the galaxy NGC\,5194 (M51a). The
      coordinate offsets are relative to the nucleus of NGC\,5194 (see
      Table~\ref{tab:m51}). This image results from the joint deconvolution
      of the IRAM-30m single-dish and Plateau de Bure interferometer data
      sets. The image inserted at the bottom left is the \twCO{} \Jone{}
      integrated emission of the full M51 system (\ie{}, NGC\,5194 +
      NGC\,5195) as observed by the IRAM-30m telescope. The white
      horizontal rectangle shows the PAWS field of view. The images were
      scaled such that the angular resolution of both data sets occupies
      the same size on paper. In other words, the PAWS image shows the
      center of the small image zoomed by a factor of 21.}
    \label{fig:pdbi-vs-30m}
  \end{figure*}}

\newcommand{\FigSkyCover}{%
  \begin{figure}[t]
    \centering \includegraphics[width=\hsize{}]{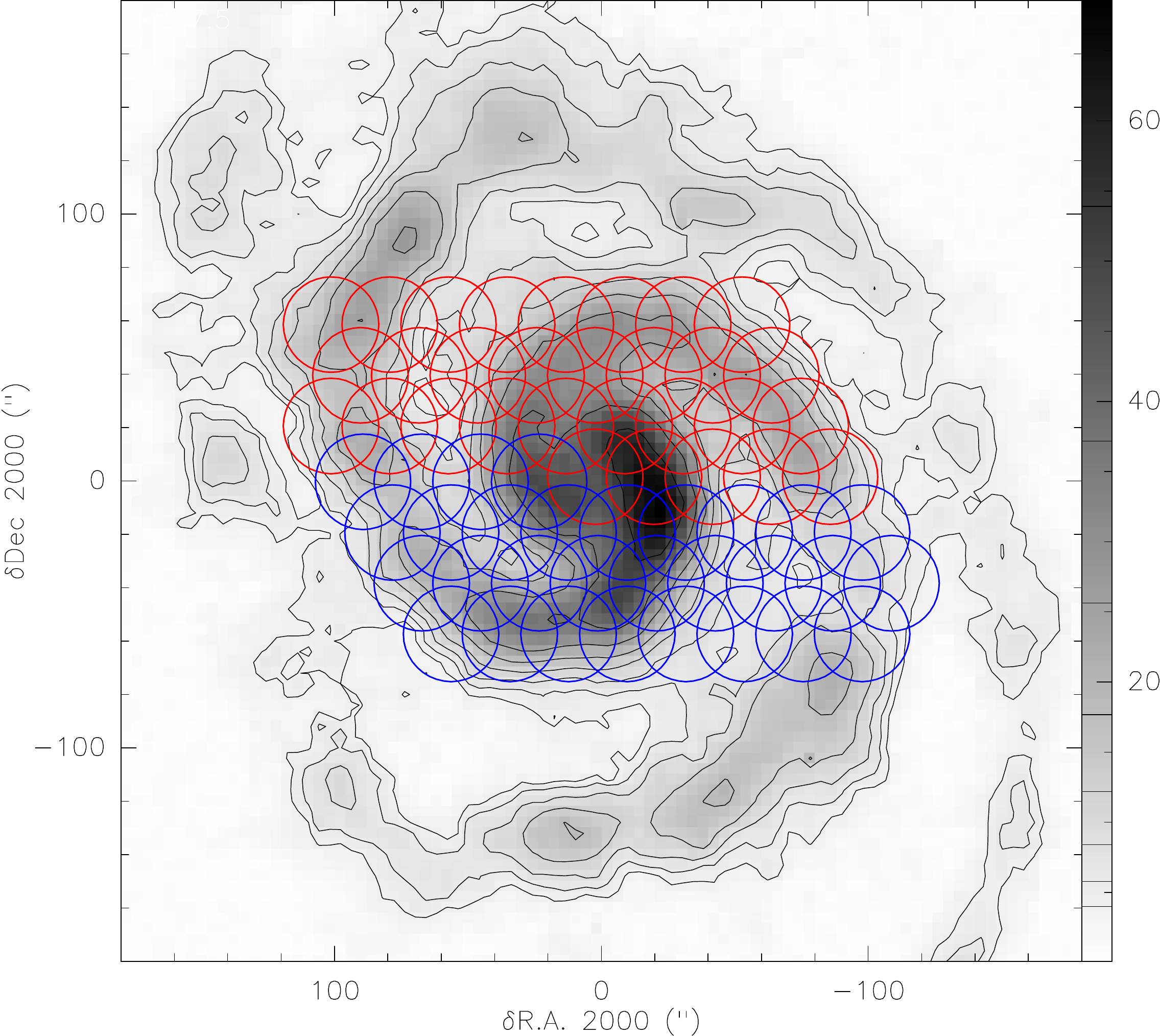}
    \caption{Pointing pattern of the two observed PdBI mosaics overlaid on the 
      integrated emission of the \twCO{} \Jone{} line observed with the
      IRAM-30m. The 30 fields of each mosaic are displayed as red and blue
      circles, respectively. The diameter of each circle is
      $\wavelength/\dprim$, where \wavelength{} is the observation
      wavelength and \dprim{} is the diameter of an interferometer antenna,
      \ie{}, 15\m{} for the PdBI. In our case, we have $\wavelength/\dprim
      \sim 36''$.}
    \label{fig:skycover}
  \end{figure}}

\newcommand{\FigSDmoments}{%
  \begin{figure*}[t]
    \centering 
    \includegraphics[width=\hsize{}]{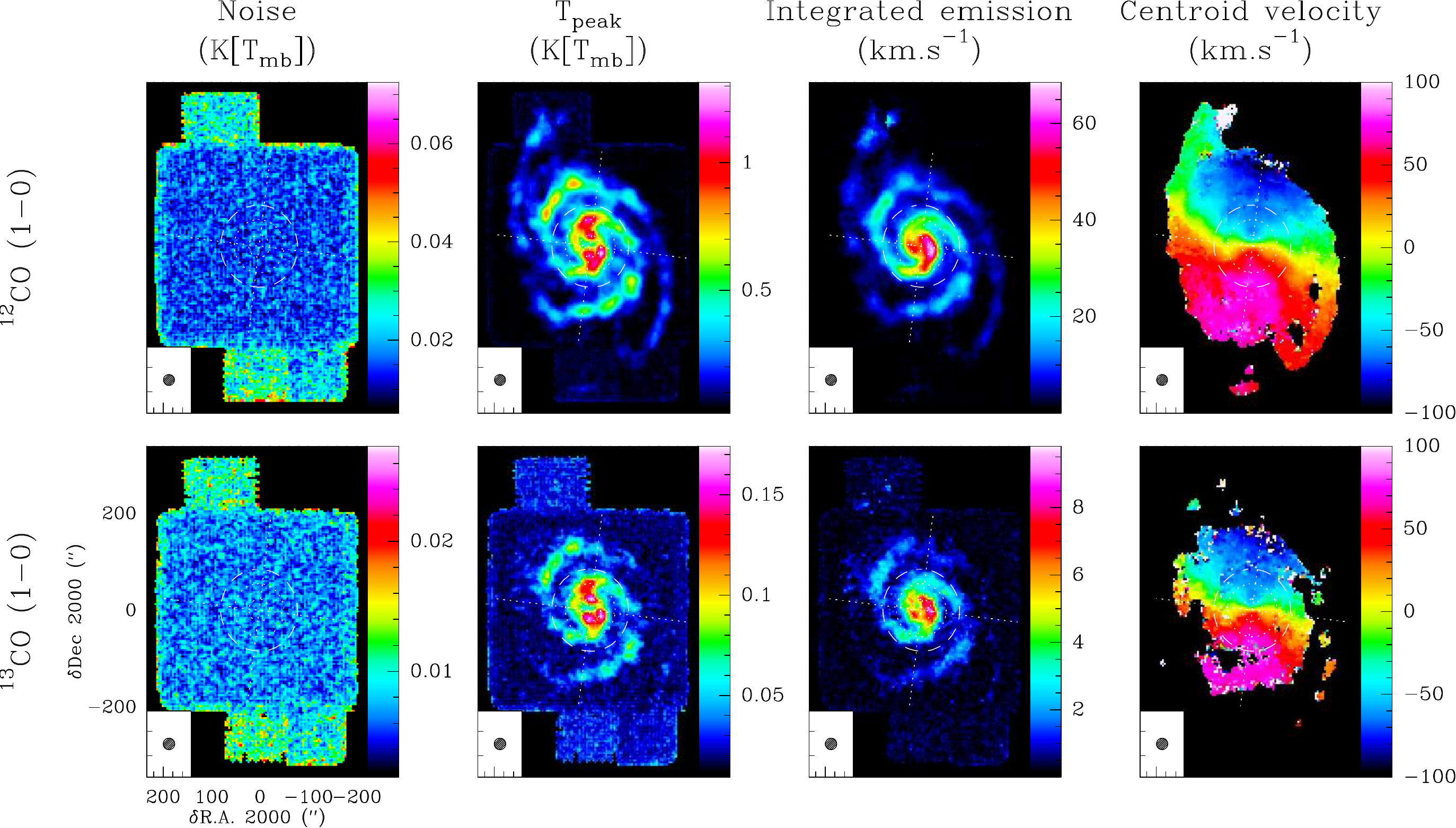}
    \caption{Spatial distributions of the rms noise, peak intensity,
      integrated intensity and centroid velocity of the \twCO{} (upper
      panels) and \thCO{} (lower panels) \Jone{} emission as observed with
      the IRAM-30m telescope.  The angular resolution is displayed as a
      circle in the bottom left corner of each panel. The intensity scale
      is shown on the right side of each panel with units indicated in the
      caption title. The major and minor kinematic axes are displayed as
      perpendicular dotted lines. The dotted circles show the two inner
      corotation resonances at radii equal to $23''$ and $55''$, while the
      dashed circle marks the start of the material arms at a radius equal
      to $85''$~\citep{meidt12b}.}
    \label{fig:moments:30m}
  \end{figure*}}

\newcommand{\FigDeconvolution}{%
  \begin{figure}[t]
    \centering
    \includegraphics[width=\hsize{}]{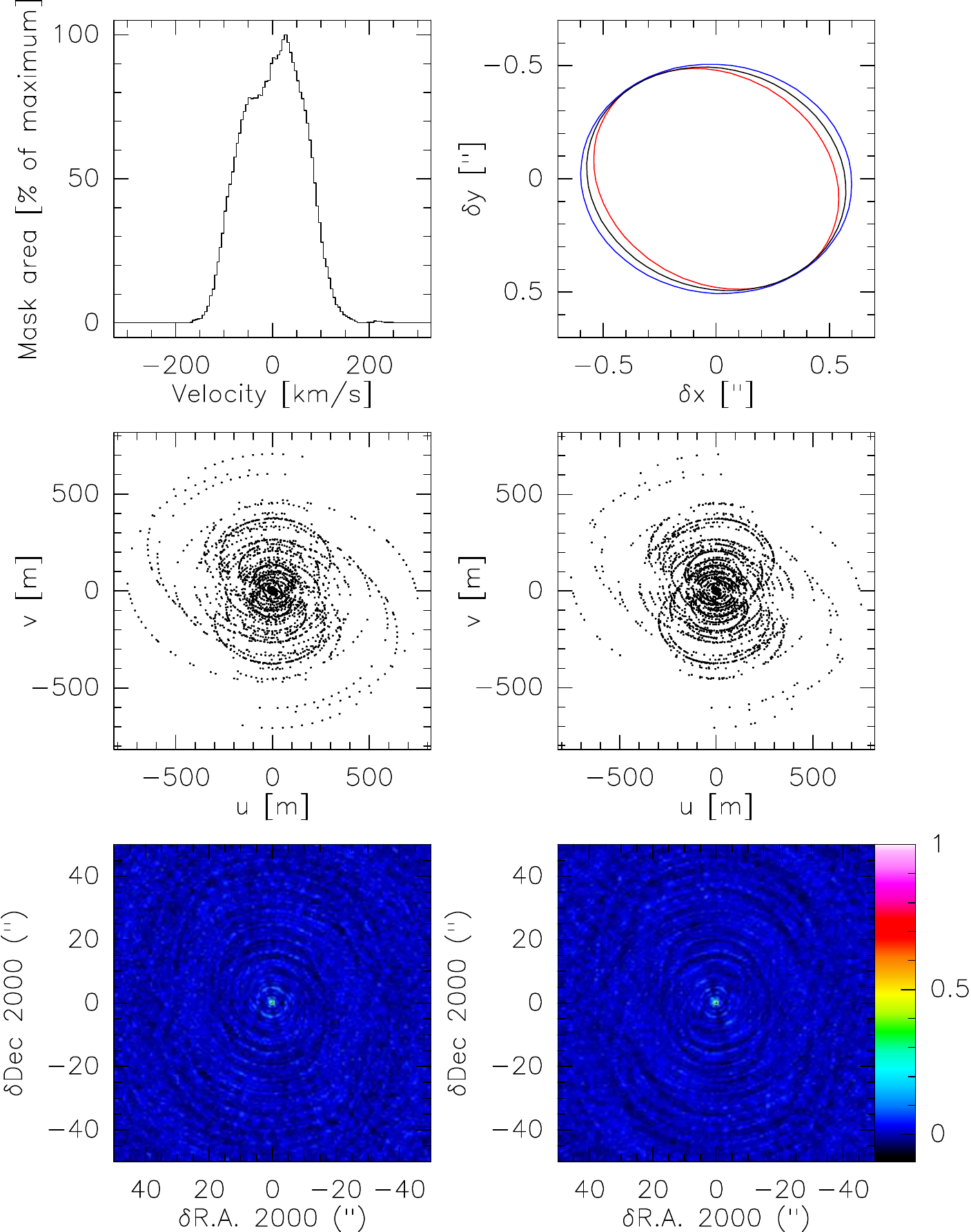}
    \caption{{\bf Top, left:} Area of the \texttt{CLEAN} mask, normalized
      to its maximum value, as a function of the channel velocity.  {\bf
        Top, right:} Ellipses representing the mean Gaussian beams for the
      top (red) and bottom (blue) mosaics. The black ellipse represents the
      Gaussian beam used for the restoration of the \texttt{CLEAN}
      component list. {\bf Middle:} $uv$ coverages for the fields of the
      top (left panel) and bottom (right panel) mosaics. {\bf Bottom:} Zoom
      of the dirty beams for the top (left panel) and bottom (right panel)
      mosaics.}
    \label{fig:deconv}
  \end{figure}}

\newcommand{\FigDeconvolvedFlux}{%
  \begin{figure}[t]
    \centering
    \includegraphics[width=\hsize{}]{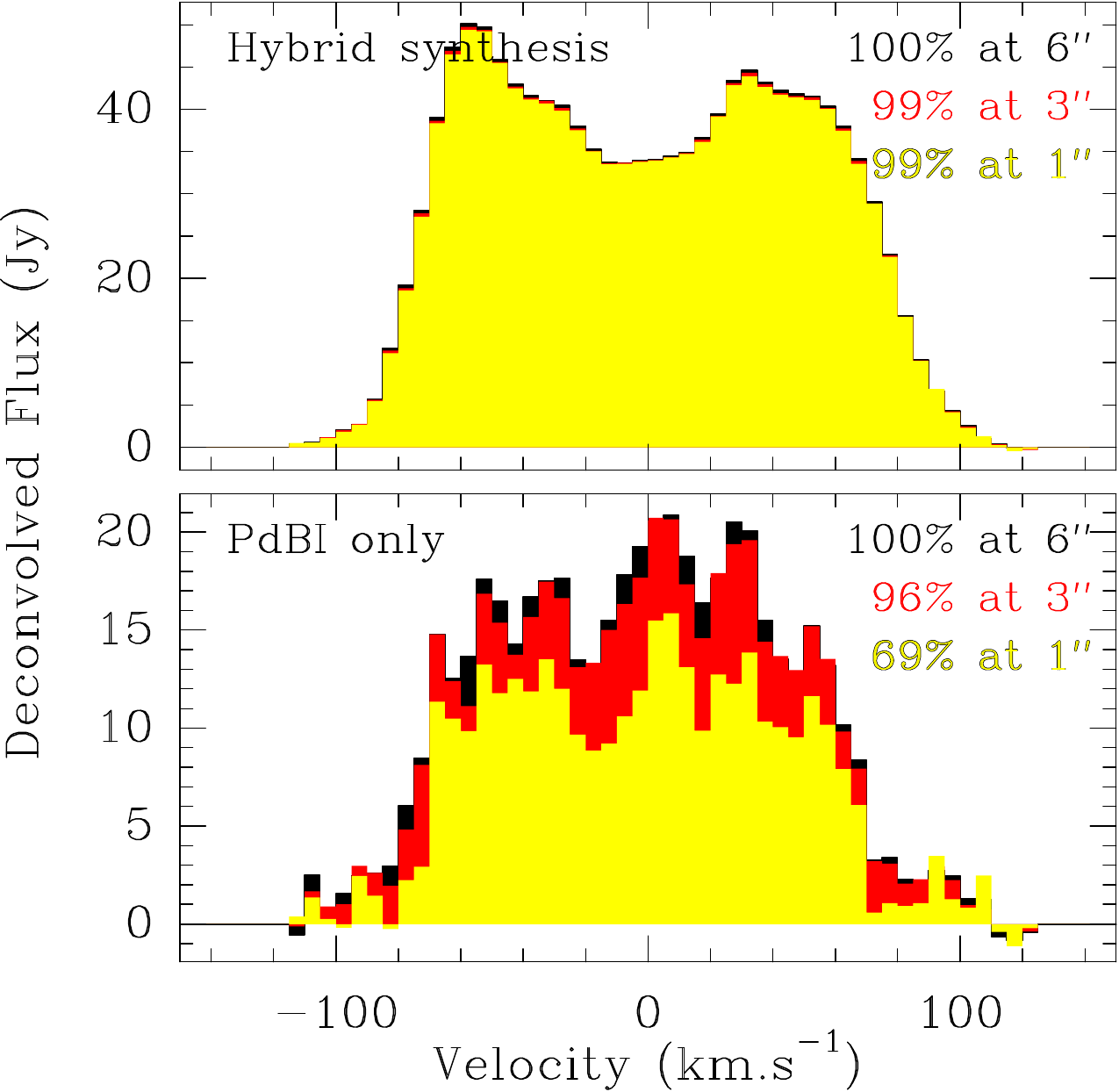}
    \caption{Deconvolved flux as a function of velocity for the hybrid
      synthesis (top) and the PdBI only (bottom) data sets. Yellow, red,
      and black curves correspond to cubes imaged at an angular resolution
      of $1''$, $3''$, and $6''$, respectively. The percentage of total
      flux recovered compared to the $6''$ cube is indicated in the top
      right corner using the same color coding.}
    \label{fig:flux:deconv}
  \end{figure}}

\newcommand{\FigMeanSpectra}{%
  \begin{figure}[t]
    \centering
    \includegraphics[width=\hsize{}]{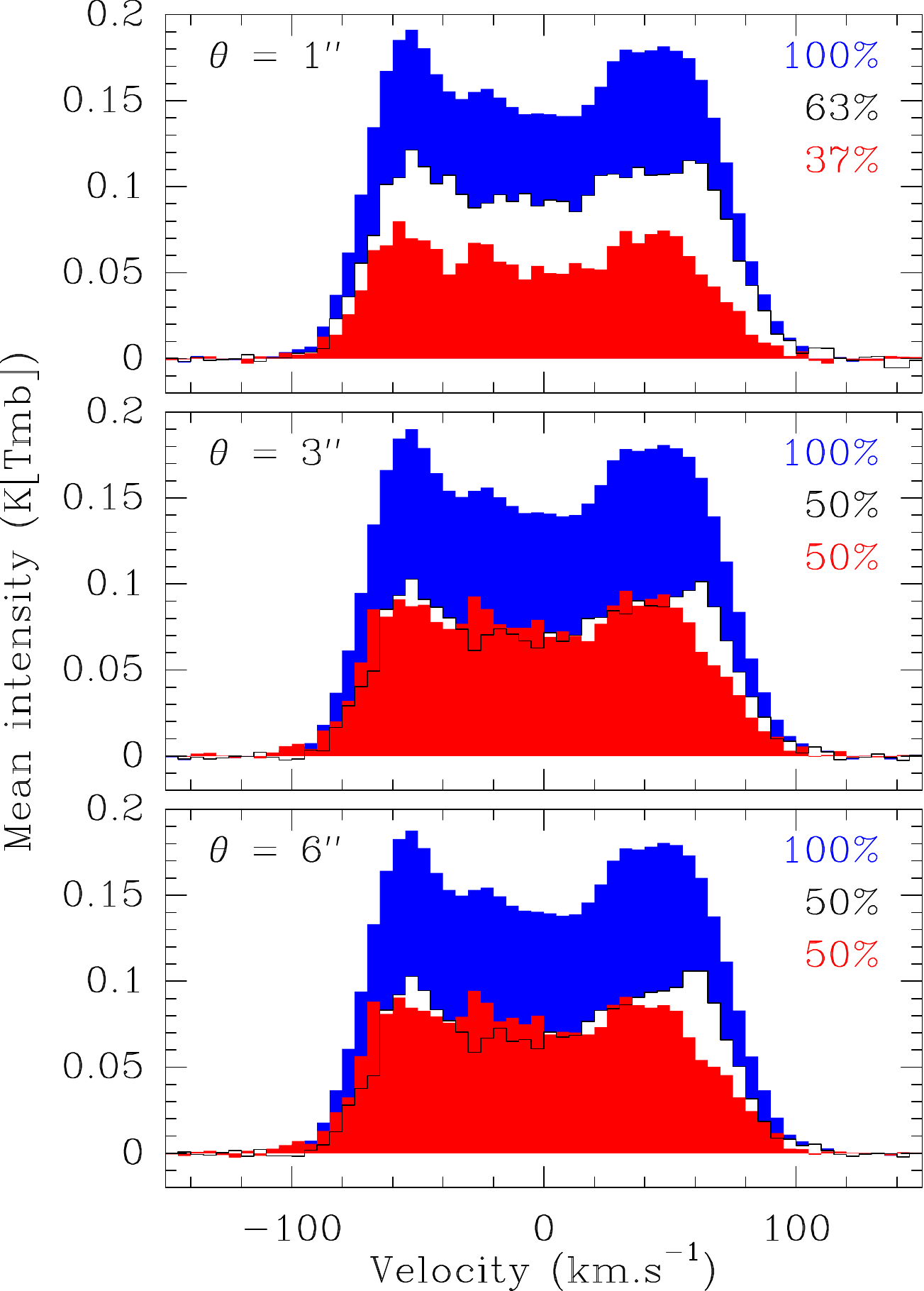}
    \caption{Spectra, averaged over the field of view, of the hybrid 
      synthesis (blue), PdBI-only (red) and filtered (= hybrid-PdBI, white)
      emission. The percentage of flux recovered in each spectrum is
      written in the top right corners with the same color code. The
      computations were done at the angular resolution displayed at the top
      left corner of each panel.}
    \label{fig:mean-spectra}
  \end{figure}}

\newcommand{\FigFluxComp}{%
  \begin{figure}[t]
    \centering
    \includegraphics[width=\hsize{}]{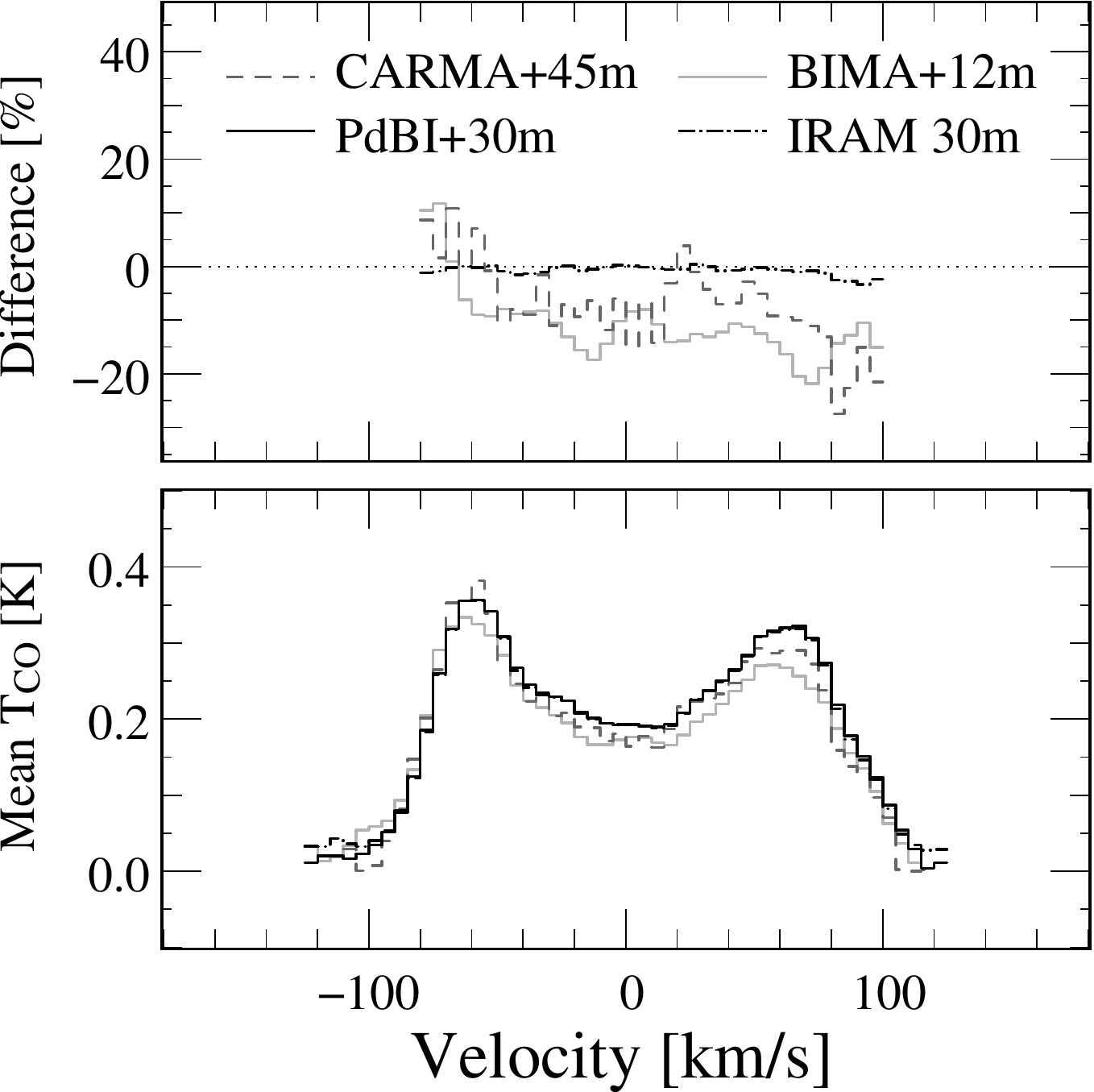}
    \caption{Channel by channel flux comparison of the PdBI+30m,
      IRAM-30m, CARMA and BIMA data cubes over the PAWS field of view. The
      bottom panel shows the mean CO temperature as a function of the
      velocity and the top panel shows the relative difference with respect
      to the PdBI+30m data set. The dotted horizontal line indicates
      perfect agreement.}
    \label{fig:flux:comp}
  \end{figure}}

\newcommand{\FigAmpVsRadiusWide}{%
  \begin{figure}[t]
    \centering
    \includegraphics[width=\hsize{}]{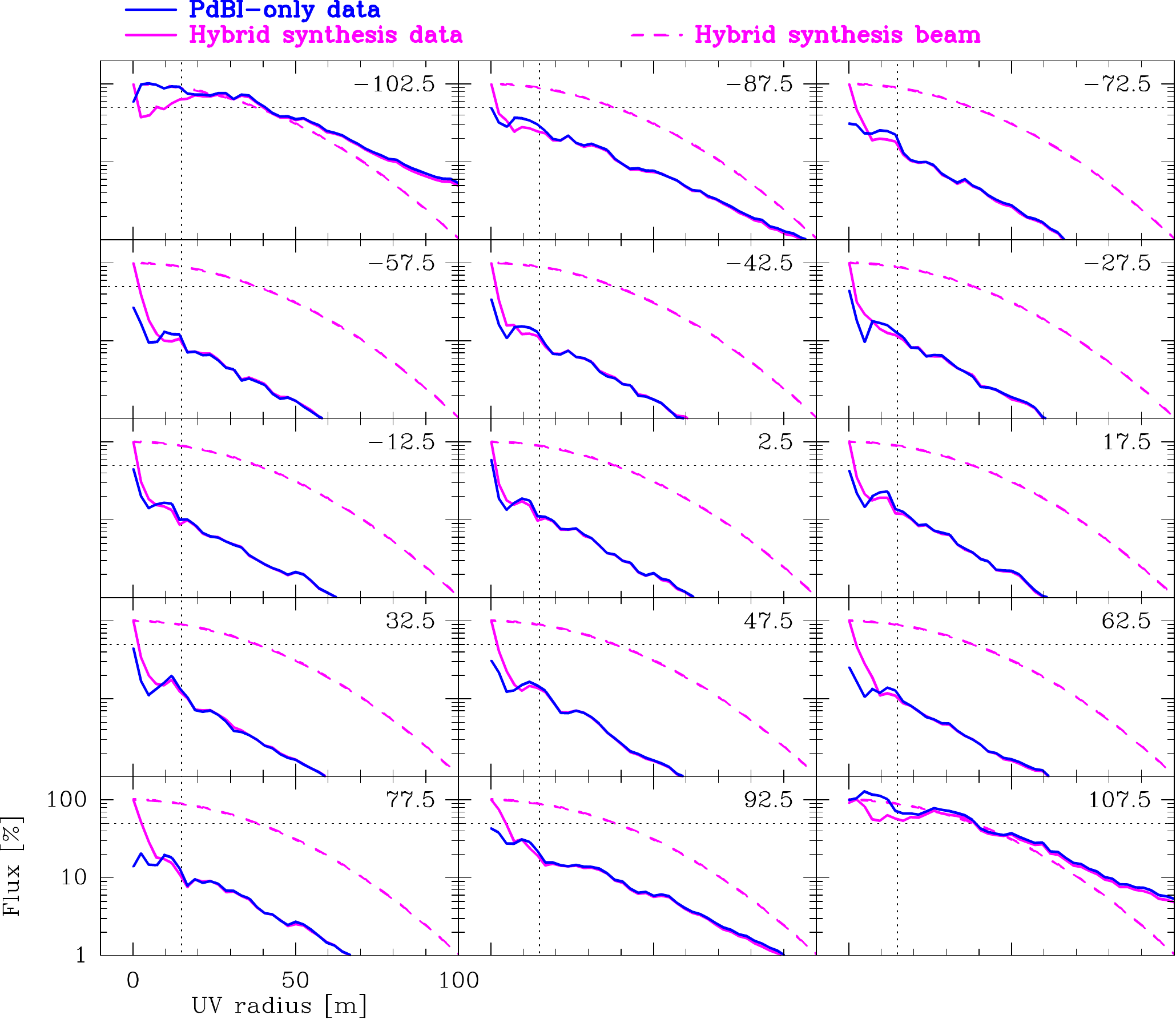}
    \caption{Variations of the azimuthal average of the Fourier transform 
      amplitude as a function of $uv$ radius for every third velocity
      channel between $-102.5$ and $+107.5\kms$. On the y-axis, we plot the
      percentage of the maximum flux at the zero $uv$ radius.  The pink
      dashed curve displays the theoretical shape for a signal-free
      channel, \ie{}, the Fourier Transform of the hybrid synthesis beam.
      The pink and blue solid curves show, respectively, the hybrid
      synthesis and PdBI-only data imaged at $6''$ resolution.  The dotted
      vertical line indicates the minimum $uv$ radius measured by the
      interferometer (\ie{} $\sim15\m$). The dotted horizontal line
      represents a 50\% reduction in flux.}
    \label{fig:amp:rad:100m}
  \end{figure}}

\newcommand{\FigAmpVsRadiusNarrow}{%
  \begin{figure}[t]
    \centering
    \includegraphics[width=\hsize{}]{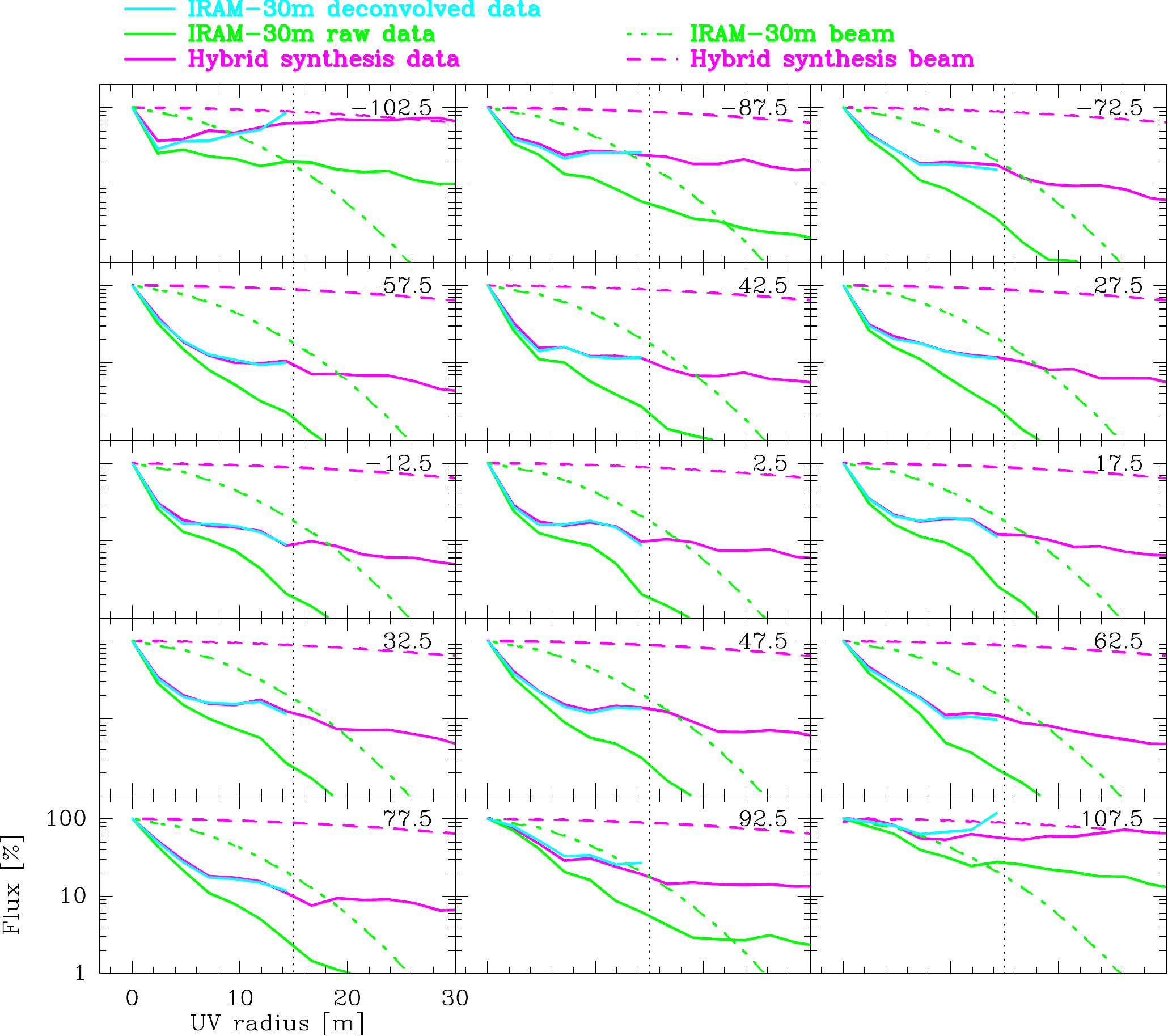}
    \caption{Zoom of the variations of the azimuthal average of the Fourier transform 
      amplitude as a function of $uv$ radius for every third velocity
      channel between $-102.5$ and $+107.5\kms$. The y-axis shows the
      percentage of the maximum flux at zero $uv$ radius.  The dashed pink
      and green curves display the Fourier Transforms of the hybrid
      synthesis and the IRAM-30m beams, respectively.  The solid pink curve
      shows the hybrid synthesis data imaged at $6''$ resolution.  The
      solid green and cyan curves represent, respectively, the IRAM-30m
      data before and after deconvolution from the $22.5''$ 30m beam and
      convolution with a $6''$ beam. The dashed vertical line indicates the
      minimum $uv$ radius measured by the interferometer (\ie{},
      $\sim15\m$). }
    \label{fig:amp:rad:30m}
  \end{figure}}

\newcommand{\FigAmpVsRadiusDiff}{%
  \begin{figure}[t]
    \centering
    \includegraphics[width=\hsize{}]{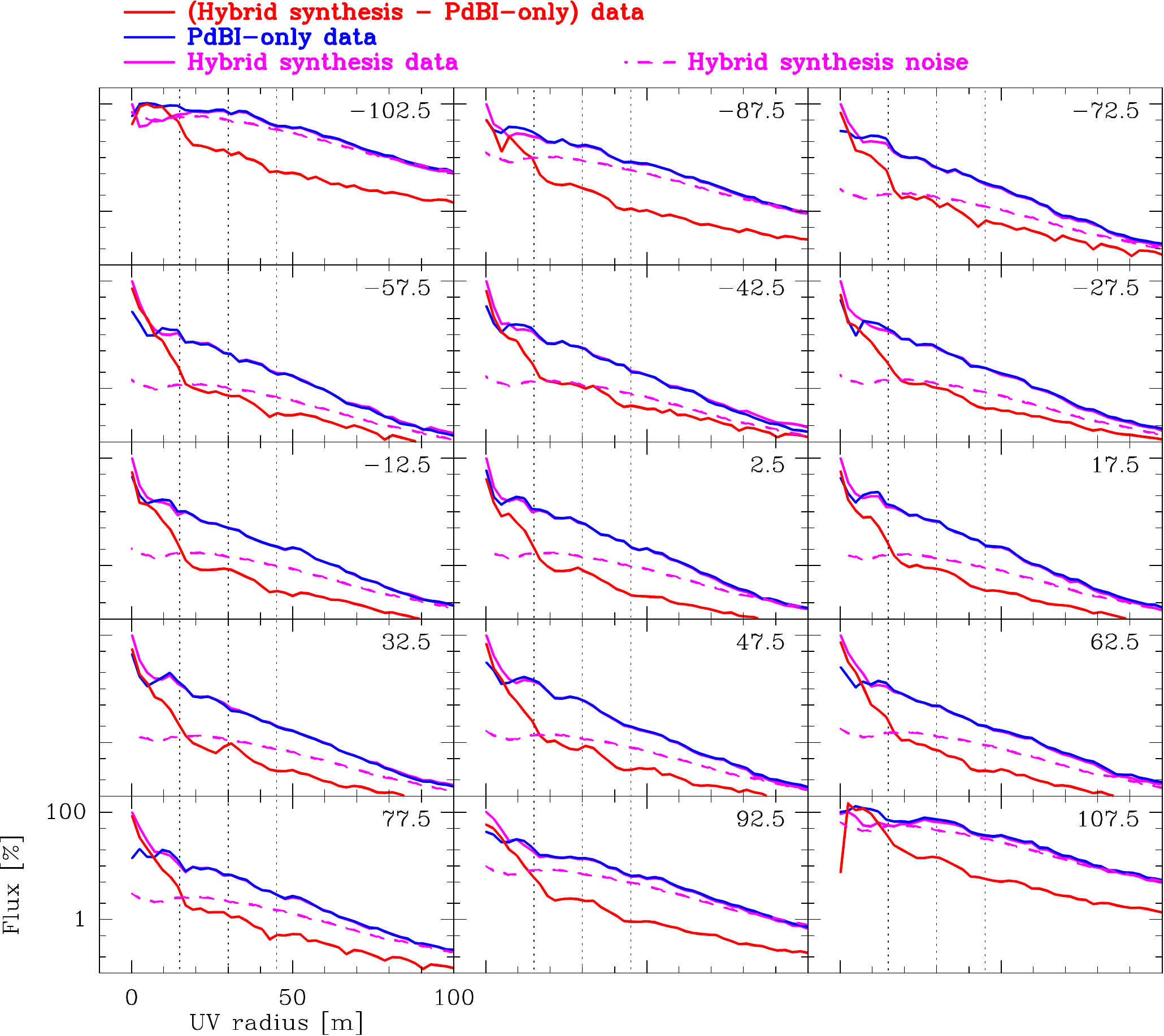}
    \caption{Same as Fig.~\ref{fig:amp:rad:100m} but the pink dot-dashed
      curve displays the average noise level as a function of the $uv$
      radius. The red curve shows the difference between the hybrid
      synthesis and the PdBI-only at $6''$ resolution. The 3 dotted
      vertical lines indicate \dprim{}, 2\dprim{}, and 3\dprim{}, where
      \dprim{} is the primary beam diameter of the interferometer.}
    \label{fig:amp:rad:diff}
  \end{figure}}

\newcommand{\FigNoiFilling}{%
  \begin{figure}[t]
    \centering %
    \includegraphics[height=0.55\textheight{}]{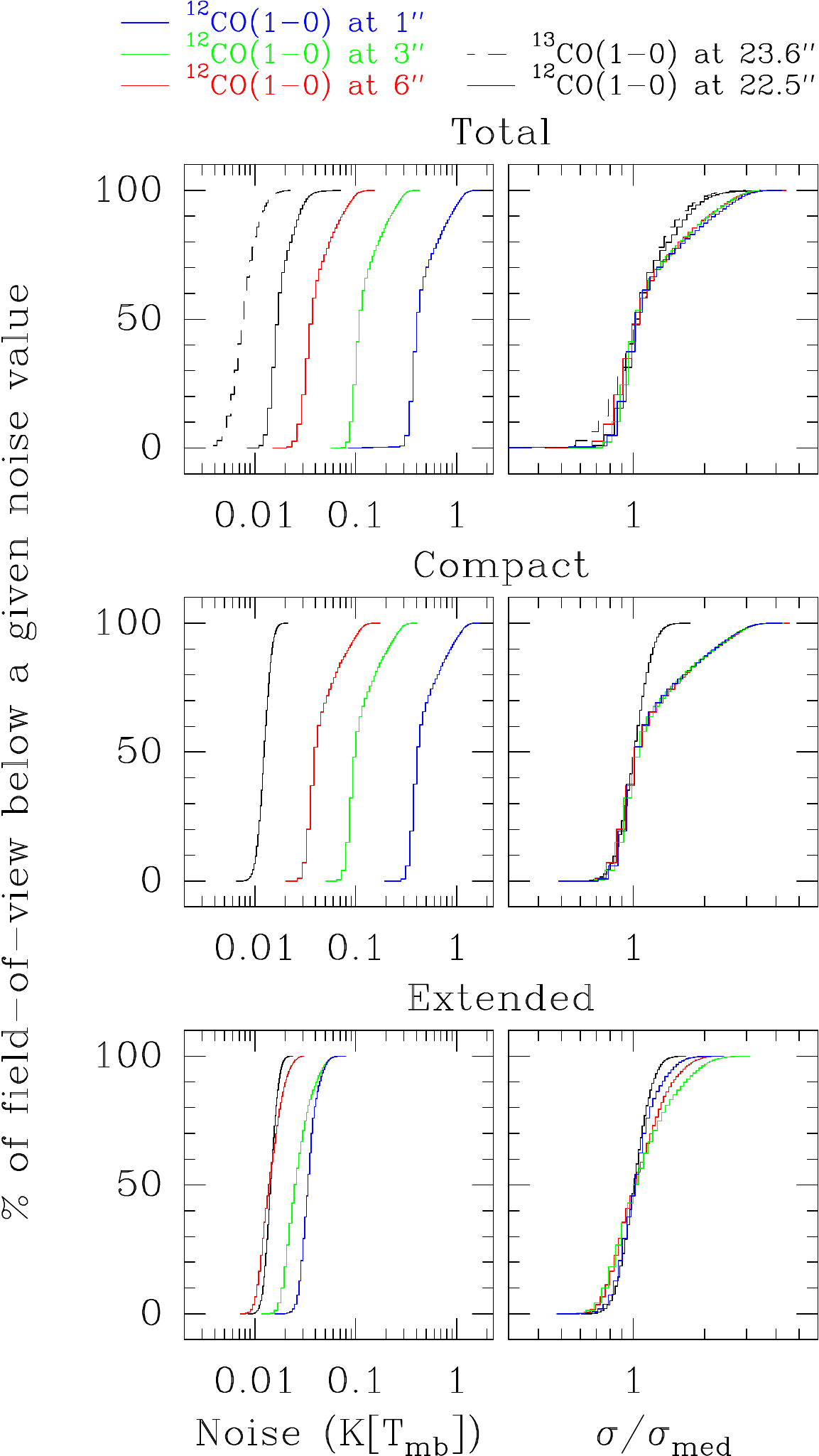}
    \caption{Cumulative histograms of the rms noise for the \thCO{}
      \Jone{} (dashed line) and \twCO{} \Jone{} (solid lines) cubes. The
      IRAM-30m cubes are displayed in black, the interferometric cubes at
      1, 3, and $6''$ are displayed in blue, green, and red, respectively.
      \textbf{Top:} Hybrid synthesis or IRAM-30m data cubes.
      \textbf{Middle:} PdBI-only data cubes or the $6''$ extended data cube
      smoothed to $22.5''$ and subtracted from the IRAM-30m cube.
      \textbf{Bottom:} PdBI-only cubes subtracted from the hybrid synthesis
      cubes or the $6''$ PdBI-only data cube smoothed to $22.5''$ and
      subtracted from the IRAM-30m cube. \textbf{Left:} Histograms of
      absolute noise values. \textbf{Right:} Histograms of noise normalized
      by the median noise value.}
    \label{fig:filling:noi}
  \end{figure}}

\newcommand{\FigSigFilling}{%
  \begin{figure}[t]
    \centering %
    \includegraphics[height=0.55\textheight{}]{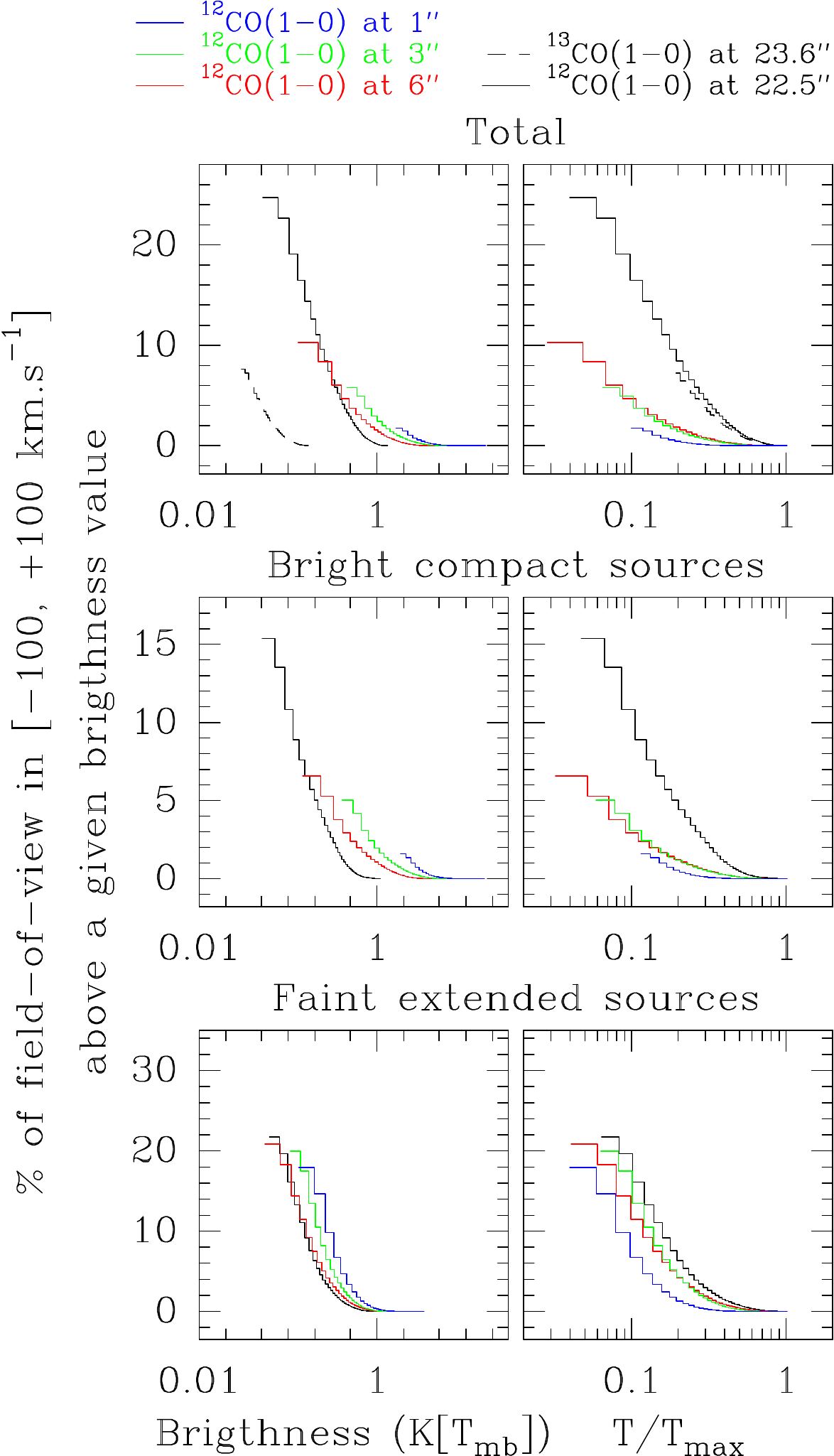}
    \caption{Cumulative histograms of the signal above the $5\sigma$-level 
      for the \thCO{} \Jone{} (dashed line) and \twCO{} \Jone{} (solid
      lines) cubes. The IRAM-30m cubes are displayed in black, the
      interferometric cubes at 1, 3, and $6''$ are displayed in blue,
      green, and red, respectively.  \textbf{Top:} Hybrid synthesis or
      IRAM-30m data cubes.  \textbf{Middle:} PdBI-only data cubes or the
      $6''$ extended data cube smoothed to $22.5''$ and subtracted from the
      IRAM-30m cube.  \textbf{Bottom:} PdBI-only cubes subtracted from the
      hybrid synthesis cubes or the $6''$ PdBI-only data cube smoothed to
      $22.5''$ and subtracted from the IRAM-30m cube. \textbf{Left:}
      Histograms of raw brightnesses. \textbf{Right:} Histograms of
      brightnesses normalized to the maximum brightness.}
    \label{fig:filling:sig}
  \end{figure}}

\newcommand{\FigPAWSmomentsMixed}{%
  \begin{figure*}[t]
    \centering
    \includegraphics[width=\hsize{}]{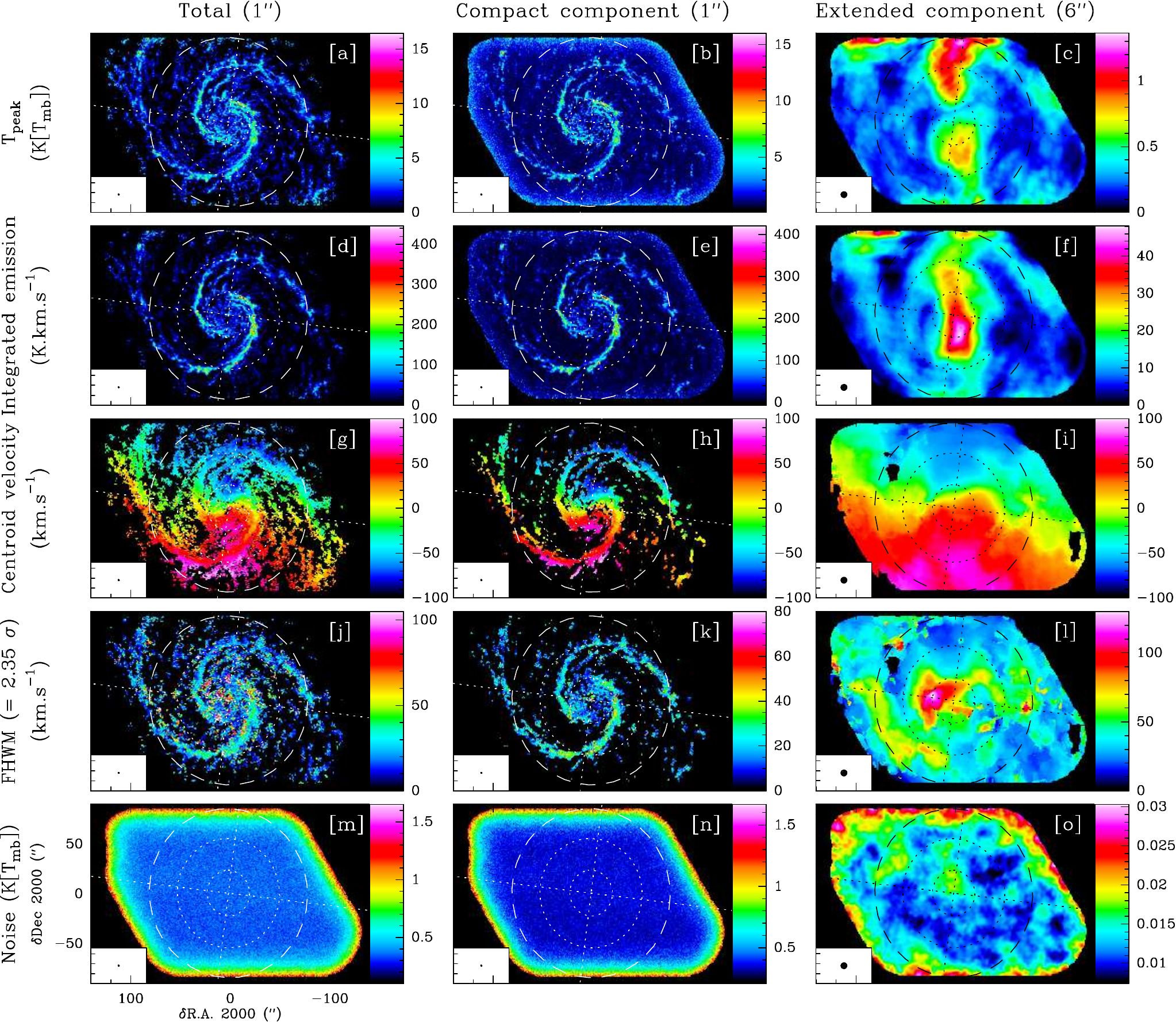}
    \caption{Comparison of the spatial distribution (from top to bottom) 
      of the peak intensity, integrated intensity, centroid velocity, the
      line full width at half maximum (\ie{}, 2.35 times the standard
      deviation in velocity), and rms noise of the \twCO{} \Jone{} emission
      for the hybrid synthesis (PdBI + 30m, left column) and the PdBI-only
      (middle column) cubes at $\sim1''$ resolution, and the
      $6''$-resolution subtracted cube (right column). The angular
      resolution is indicated by a circle in the bottom left corner of each
      panel. The intensity scale is shown on the right-hand side of each
      panel.  The major and minor axes are displayed as perpendicular
      dotted lines.  The dotted circles show the two inner corotation
      resonances at radii equal to $23''$ and $55''$, while the dashed
      circle indicates the start of the material arms at a radius equal to
      $85''$~\citep{meidt12b}.}
    \label{fig:moments:paws:mixed}
  \end{figure*}}

\newcommand{\FigPAWSaverages}{%
  \begin{figure}[t]
    \centering
    \includegraphics[width=0.98\hsize{}]{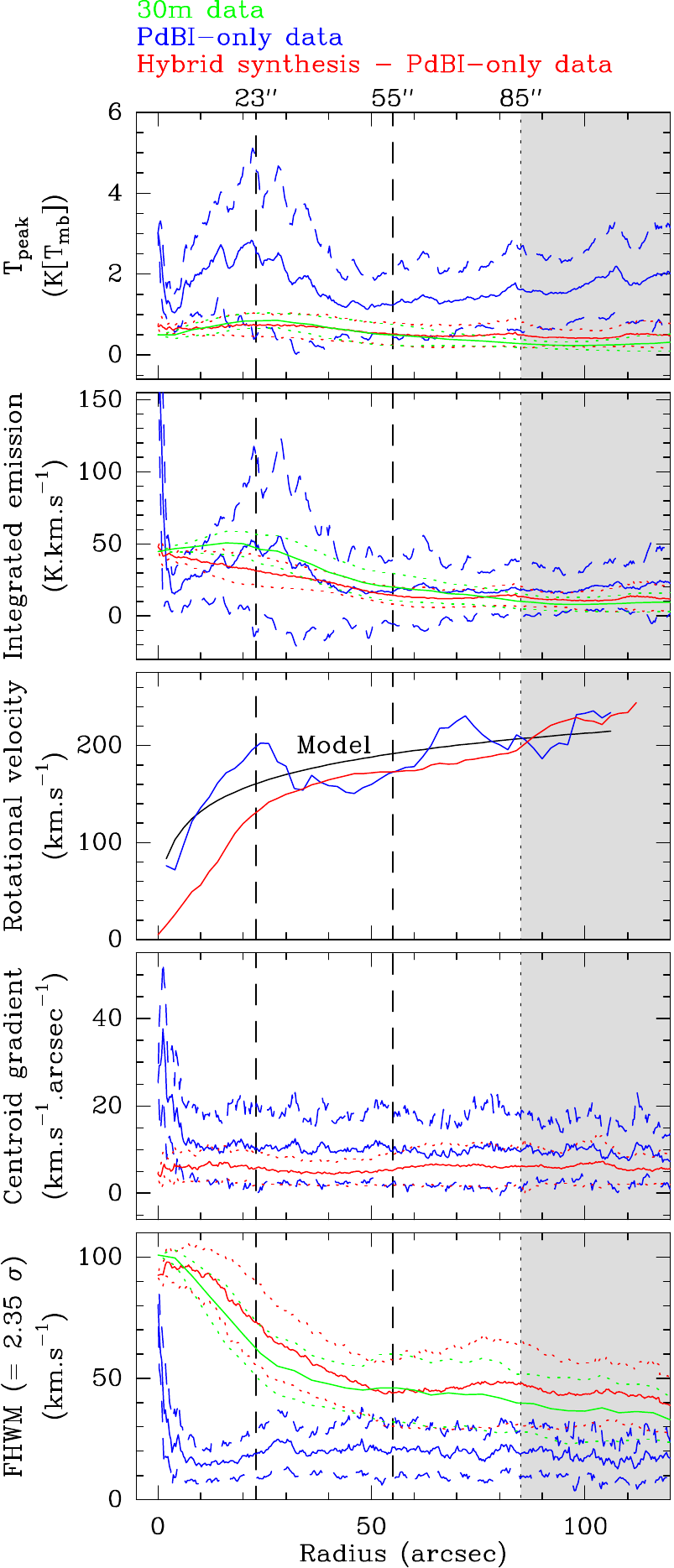}
    \caption{Deprojected azimuthal averages around the kinematic center for 
      the IRAM-30m (green curves), the PdBI-only (blue curves) and the
      subtraction of the PdBI-only from the hybrid synthesis (red curves)
      cubes. The solid lines display the averages while the dashed lines
      give the azimuthal averages plus/minus the azimuthal standard
      deviations.  From top to bottom, the panels present the peak
      temperature, line integrated emission, the rotational velocity, the
      modulus of the centroid velocity gradient, and 2.35 times the line
      2nd order moment as a function of radius. The vertical dashed lines
      indicate the two inner corotation resonances at radii equal to $23''$
      and $55''$, while the vertical dotted line show the start of the
      material arms at a radius equal to $85''$~\citep{meidt12b}.  The
      radial zone where the averages are affected by edge effects (see
      Sect.~\ref{sec:azimuthal-averages}) is highlighted in grey.}
    \label{fig:aver:paws}
  \end{figure}}

\newcommand{\FigPVmajor}{%
  \begin{figure*}[t]
    \centering
    \includegraphics[width=\hsize{}]{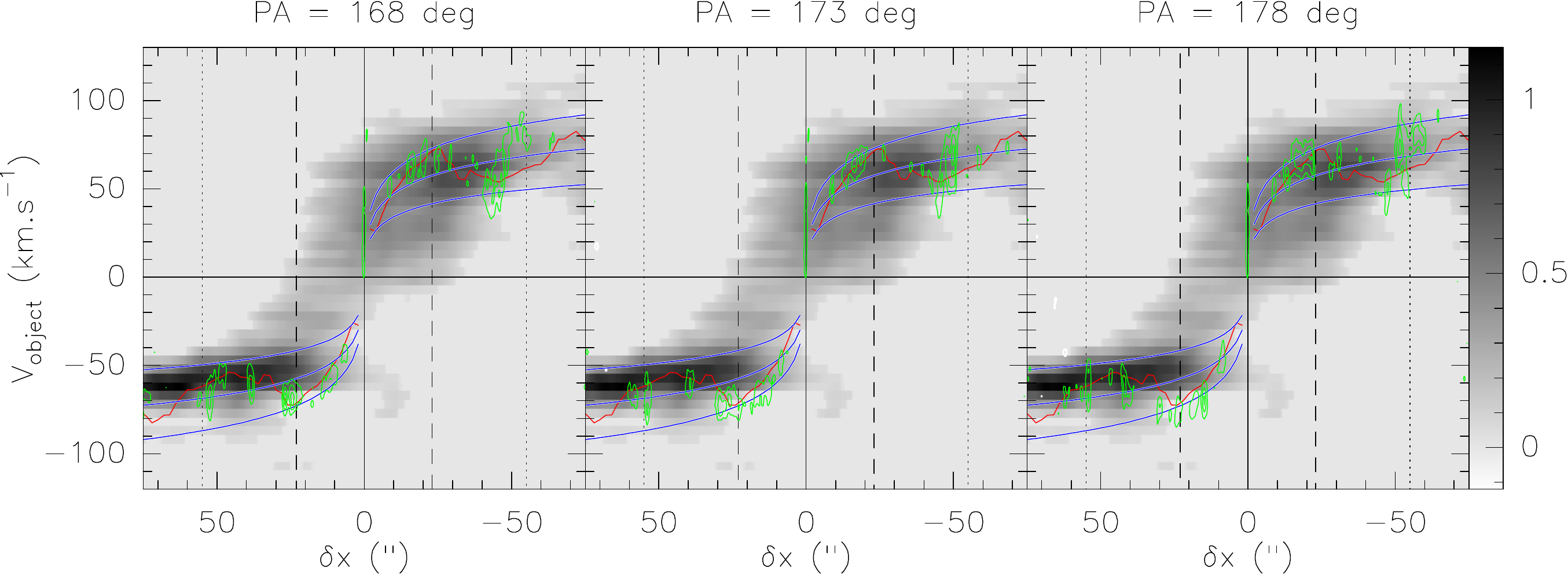}
    \caption{Position-velocity diagrams of the extended emission 
      (hybrid synthesis minus PdBI-only) imaged at $6''$ along 3 axes at PA
      $= 173\degr$ (major axis, central panel), and $173\pm6\degr$ (left
      and right panels) shown in grey-scale. The scale is shown in the
      right-hand side (\K{} [\Tmb{}]).  The green contours (at levels 2, 4,
      6, 8, 10 \K{} [\Tmb{}]) display the hybrid synthesis emission at
      $1''$ resolution.  The red curve is the azimuthally averaged
      rotational velocity profile and the central blue curve is a smooth
      3-parameters fit of the red curve with a fixed inclination of
      $21\degr$. The two other blue curves diplay the same model but for an
      inclination of $15\degr$ and $27\degr$. The vertical lines show the
      two inner corotation resonances at a radius of $23''$ and
      $55''$~\citep{meidt12b}.}
    \label{fig:pv:major}
  \end{figure*}}

\newcommand{\FigHIstackedSpectra}{%
  \begin{figure}[t]
    \centering
    \includegraphics[width=\hsize{}]{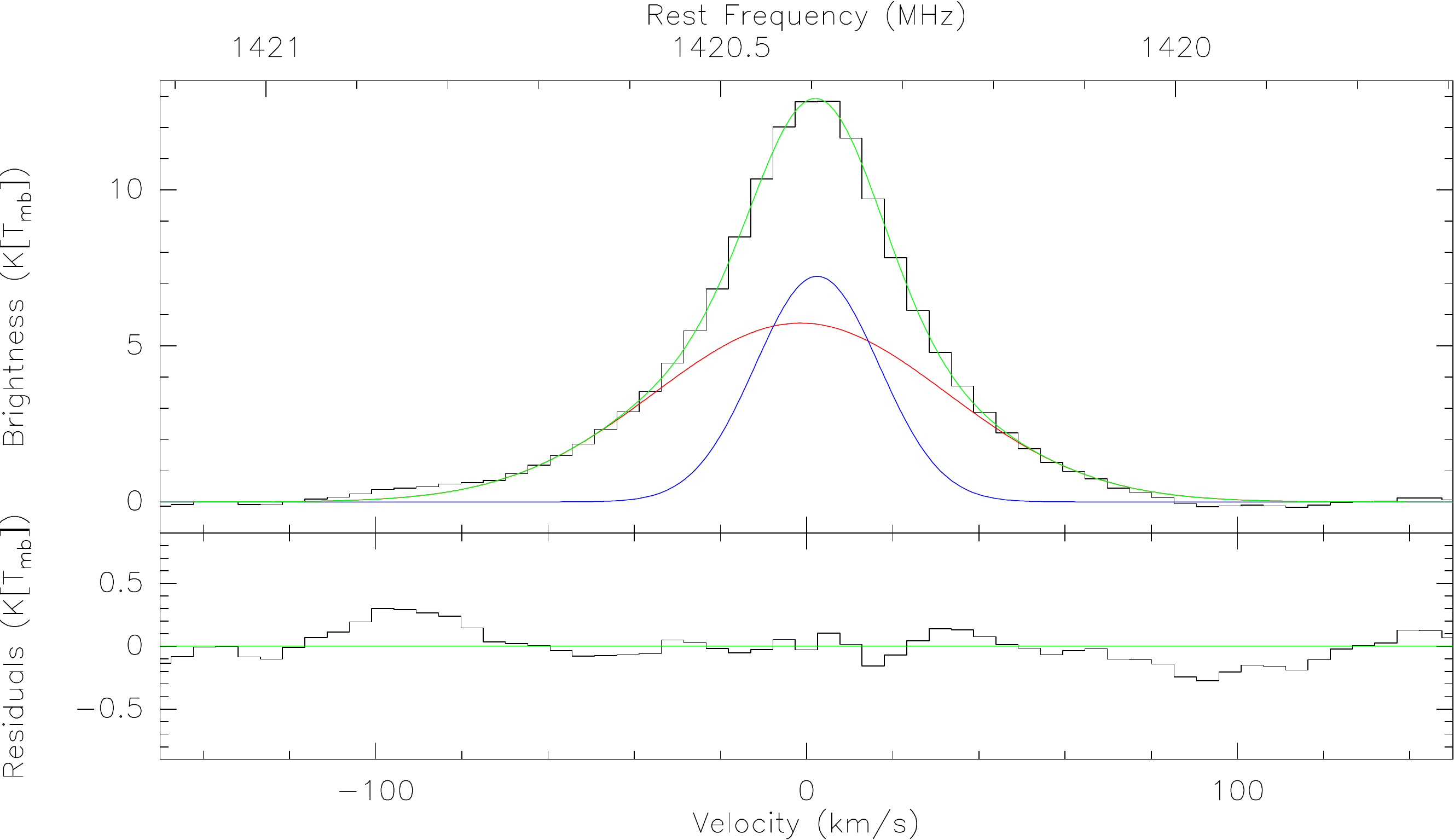}
    \caption{Dual Gaussian decomposition of the THINGS HI
      spectrum averaged over the PAWS field of view \emph{after} having
      aligned all the individual spectra along the velocity axis according
      to their centroid velocity value. \textbf{Top:} Brightness of the
      average spectra as a black histogram. The dual Gaussian fit is
      displayed as the green line and the two individual Gaussians are
      shown in red and blue.  \textbf{Bottom:} Residual brightness after
      subtraction of the dual Gaussian fit.}
    \label{fig:stacked:hi}
  \end{figure}}

\newcommand{\FigScaleHeight}{%
  \begin{figure*}[t]
    \centering
    \includegraphics[width=0.75\hsize{}]{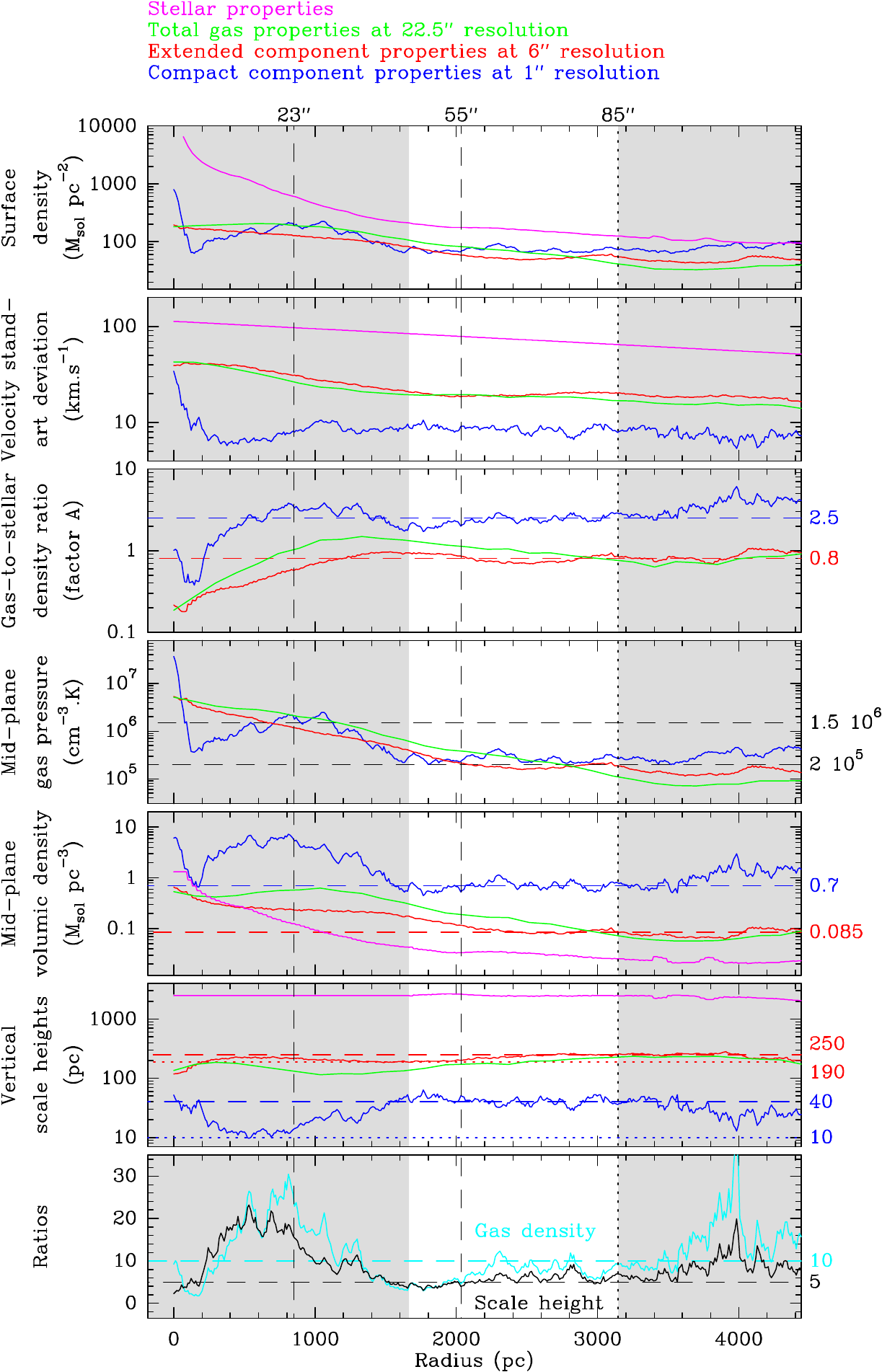}
    \caption{Azimuthal averages around the kinematic center, 
      \textbf{from top to bottom:} Mass surface densities, vertical
      velocity standard deviations (\ie{}, line full width at half maximum
      divided by 2.35), gas-to-stellar density ratios (A factor), gas
      thermal + turbulent mid-plane pressures, gas mid-plane densities,
      vertical scale heights for the stars (pink curves), the IRAM-30m
      (green curves), the PdBI-only (blue curves) and the subtracted (red
      curves) cubes. The bottom panel displays the ratios of the scale
      heights (black) and mid-plane densities (cyan) computed from the
      PdBI-only and subtracted data sets. The horizontal lines indicates
      typical values of the different parameters for the compact and
      extended components. The vertical dashed lines indicate the two inner
      corotation resonances at radii equal to $23''$ and $55''$, while the
      vertical dotted line show the start of the material arms at a radius
      equal to $85''$~\citep{meidt12b}. The radial zones where the results
      should be interpreted with caution (see
      Sect.~\ref{sec:scale-height:application}) are highlighted in grey.}
    \label{fig:scale-height}
  \end{figure*}}

\newcommand{\FigFWHMvsCVI}{%
  \begin{figure}[t]
    \centering
    \includegraphics[width=\hsize{}]{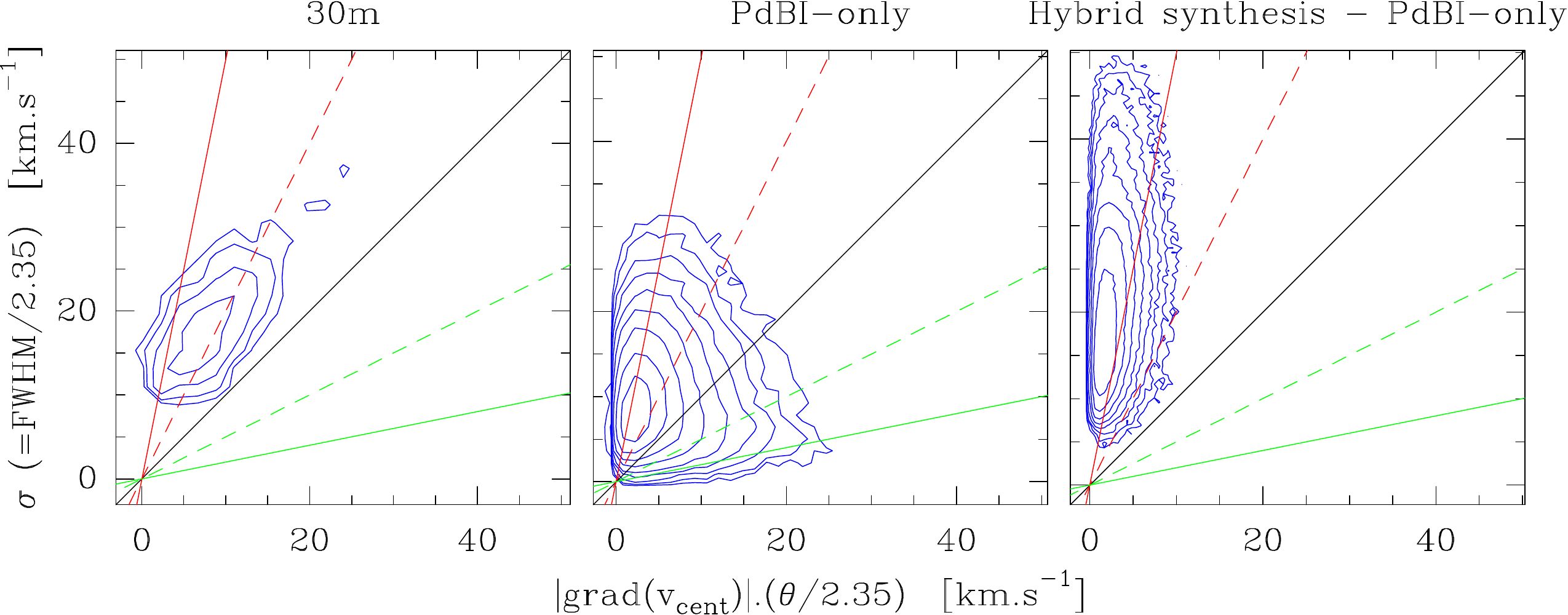}
    \caption{Joint distributions of the line 2nd order moment as a function
      of the centroid velocity gradient modulus times the half primary beam
      width for the 30m (left), PdBI-only (middle) and subtracted (right)
      cubes. The lines have a slope of 0.2 (green, solid), 0.5 (green,
      dashed), 1 (black, solid), 2 (red, dashed), and 5 (red, solid).}
    \label{fig:fwhm:cvi}
  \end{figure}}

\newcommand{\FigThCOvsTwCO}{%
  \begin{figure*}[t]
    \centering
    \includegraphics[width=\hsize{}]{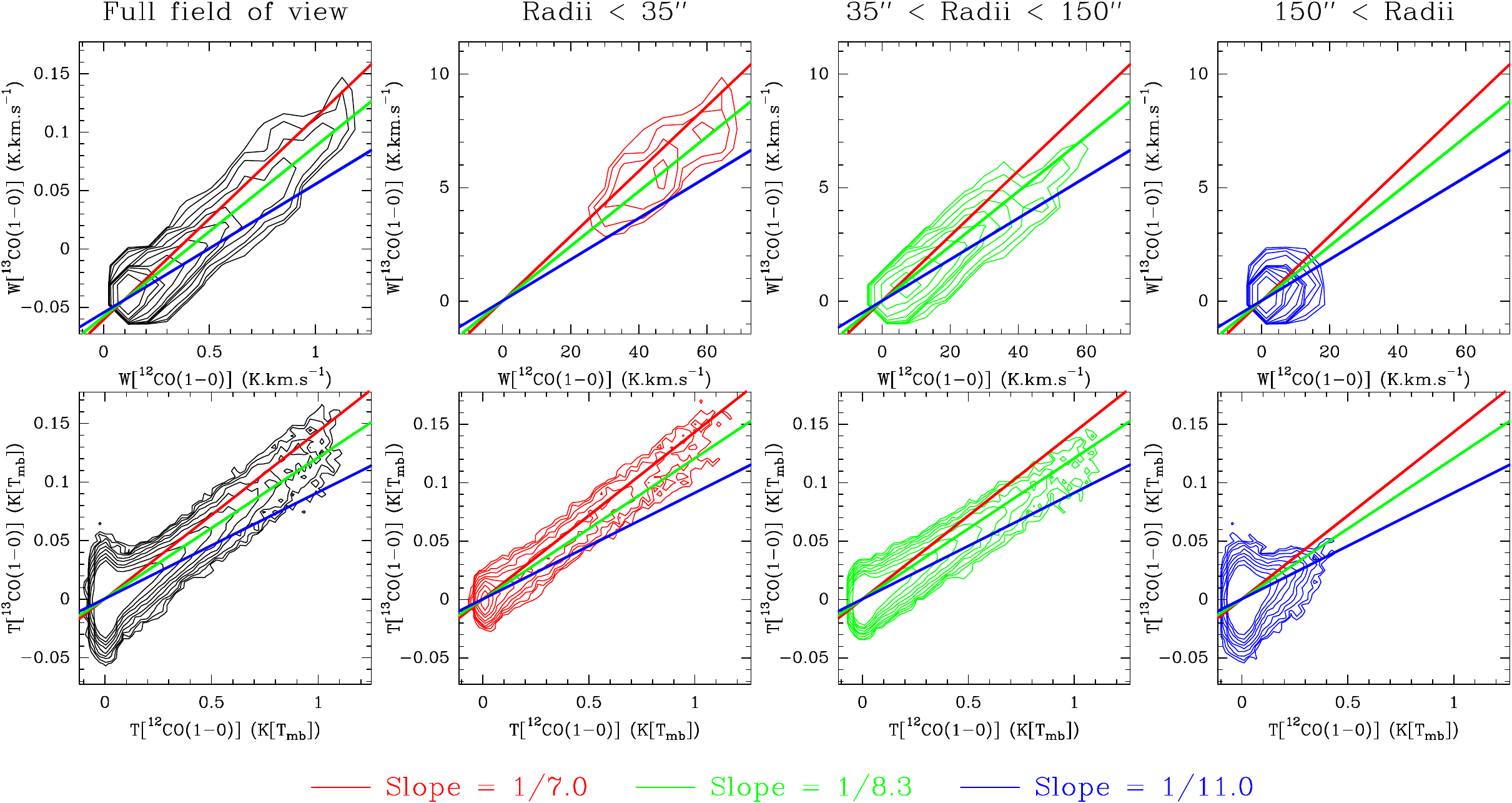}
    \caption{Joint distributions of the \thCO{} \Jone{} emission as a
      function of the \twCO{} \Jone{} emission. {\bf Top row:} Brightness
      integrated over the line profile. {\bf Bottom row:} Brightness in
      5\kms{} channels. Contour levels are set to 2, 4, 8, ... 2048 and 8,
      16, 32, ... 2048 points per pixel respectively for the top and bottom
      rows. {\bf First column:} Full field-of-view (black contours).  {\bf
        Second column:} Radii below $35''$ (red contours).  {\bf Third
        column:} Radii between $35''$ and $150''$ (green contours).  {\bf
        Fourth column:} Radii larger than $150''$ (blue contours). The 3
      same straight lines display 3 different \twCO{}/\thCO{} emission
      ratios on each panel: 7 (red), 8.3 (green) and, 11 (blue).}
    \label{fig:13co-vs-12co}
  \end{figure*}}

\newcommand{\FigMassDistribution}{%
  \begin{figure*}[t]
    \centering
    \includegraphics[width=\hsize{}]{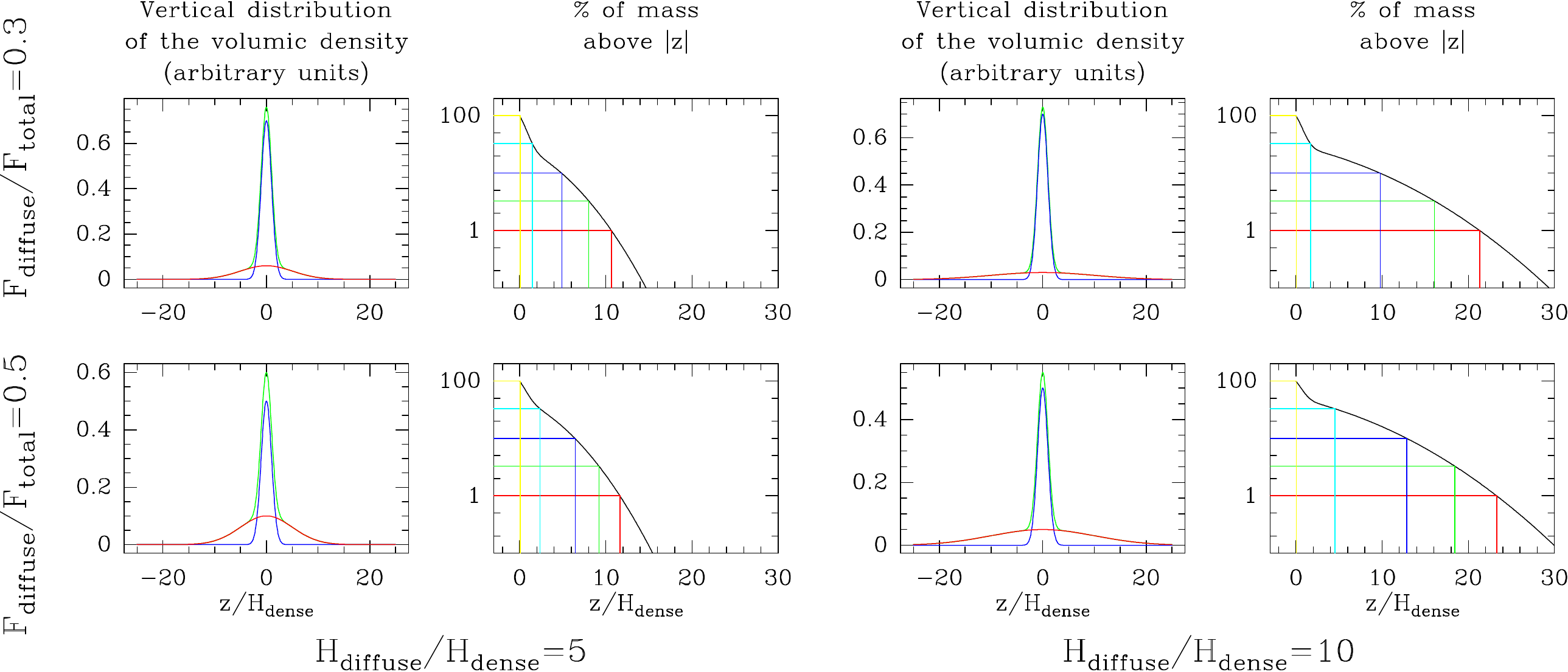}
    \caption{\textbf{First and third columns:} Vertical distributions of 
      the total volume density (green), decomposed into the sum of two
      Gaussian of different scale height. The broad and narrow Gaussian are
      respectively called diffuse (red) and dense (blue). \textbf{Second
        and fourth columns:} Percentage of the vertical mass above a given
      altitude in scale height unit of the dense Gaussian for the total
      distributions plotted in the first and third columns. The vertical
      lines indicate the altitudes above which 1\% (red), 3\% (green), 10\%
      (blue), 33\% (cyan), and 100\% (yellow) of the mass is located.  The
      ratio of the integrated masses of the diffuse over the total
      (diffuse+dense) are 0.3 for the top row and 0.5 for the bottom row.
      The two right (respectively left) columns are for a diffuse scale
      height 5 (respectively 10) times higher than the dense scale height.
      This allows us to quantify the mass of molecular gas which is
      extra-planar in different scenarii.}
    \label{fig:mass:dist}
  \end{figure*}}

\newcommand{\FigTotalFlux}{%
  \begin{figure*}
    \begin{center}
      \includegraphics[width=\textwidth{}]{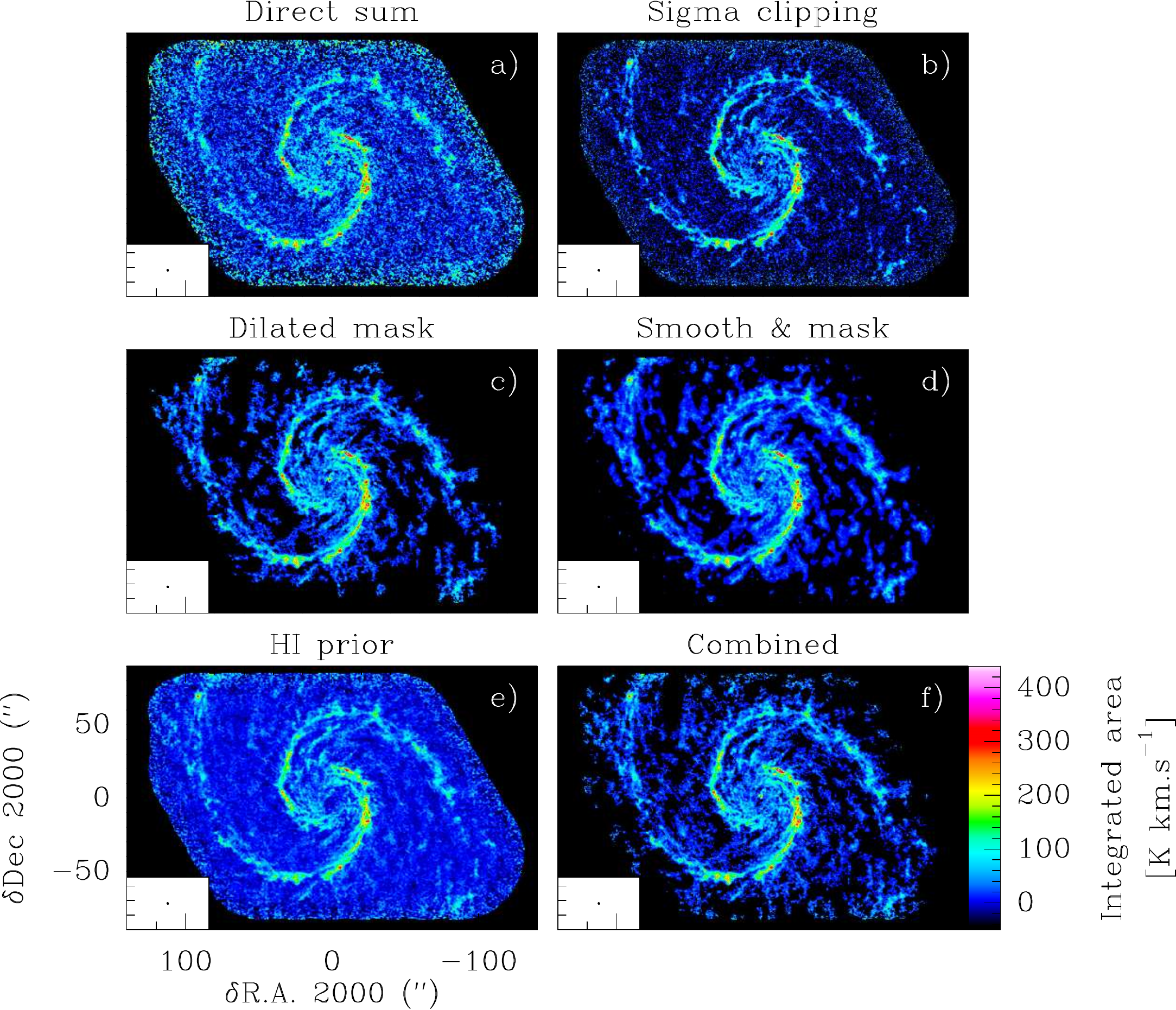}
      \caption{Integrated intensity images constructed from the
        hybrid synthesis cube using different techniques: [a] direct sum of
        all pixels; [b] sigma-clipping method with a threshold of
        $5\sigma$; [c] dilated mask method with $(t,p) = (5,1.2)$; [d]
        smooth-and-mask method with $(\theta,m) = (3.6,5)$; [e] \HI{} prior
        method, with a 50\kms{} integration window; [f] the combined method
        (see text). All images are presented on the same intensity scale,
        which is displayed near the bottom right image.}
      \label{fig:total:flux}
    \end{center}
  \end{figure*}}

\newcommand{\FigErrorBeam}{%
  \begin{figure}
    \centering
    \includegraphics[width=\hsize]{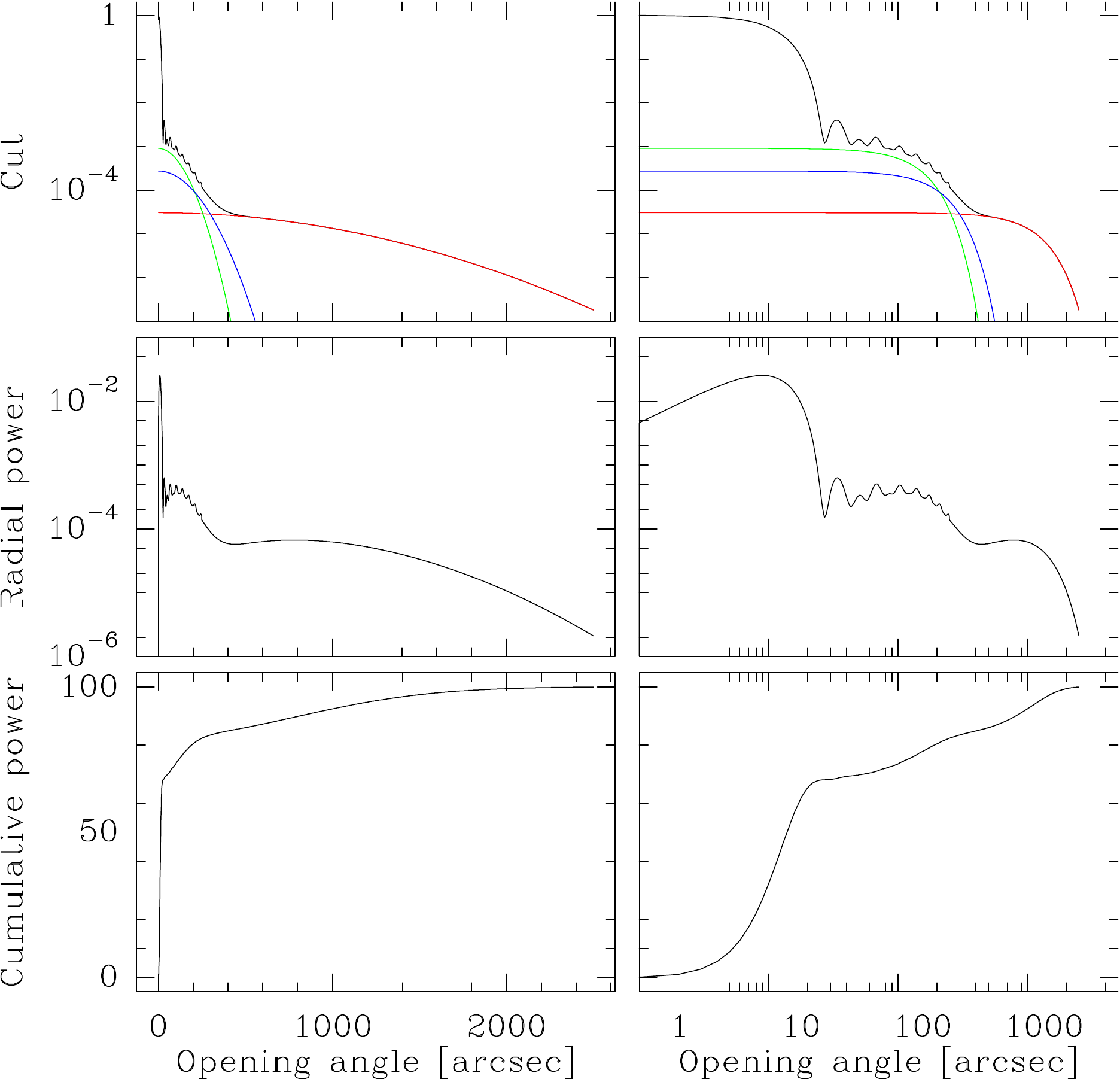}
    \caption{Characteristics of the IRAM-30m beam at the frequency
      of the \twCO{} \Jone{} line as a function of the opening angle in
      linear (left column) and logarithmic (right column) scales. Top: Cut
      of the beam profile, normalized to unity at $\theta=0''$. The black
      curve is a combination of the diffraction pattern and the three
      Gaussian error beams. The green, blue, and red curves displays the
      three error beams.  Middle: Beam power contained in circular annuli,
      normalized so that the integral is unity. Bottom: Beam power
      integrated over the solid angle sustained by the opening angle,
      normalized to 100\%.}
    \label{fig:error-beam}
  \end{figure}}

\newcommand{\FigPAWSmomentsModels}{%
  \begin{figure*}[t]
    \centering
    \includegraphics[width=\hsize{}]{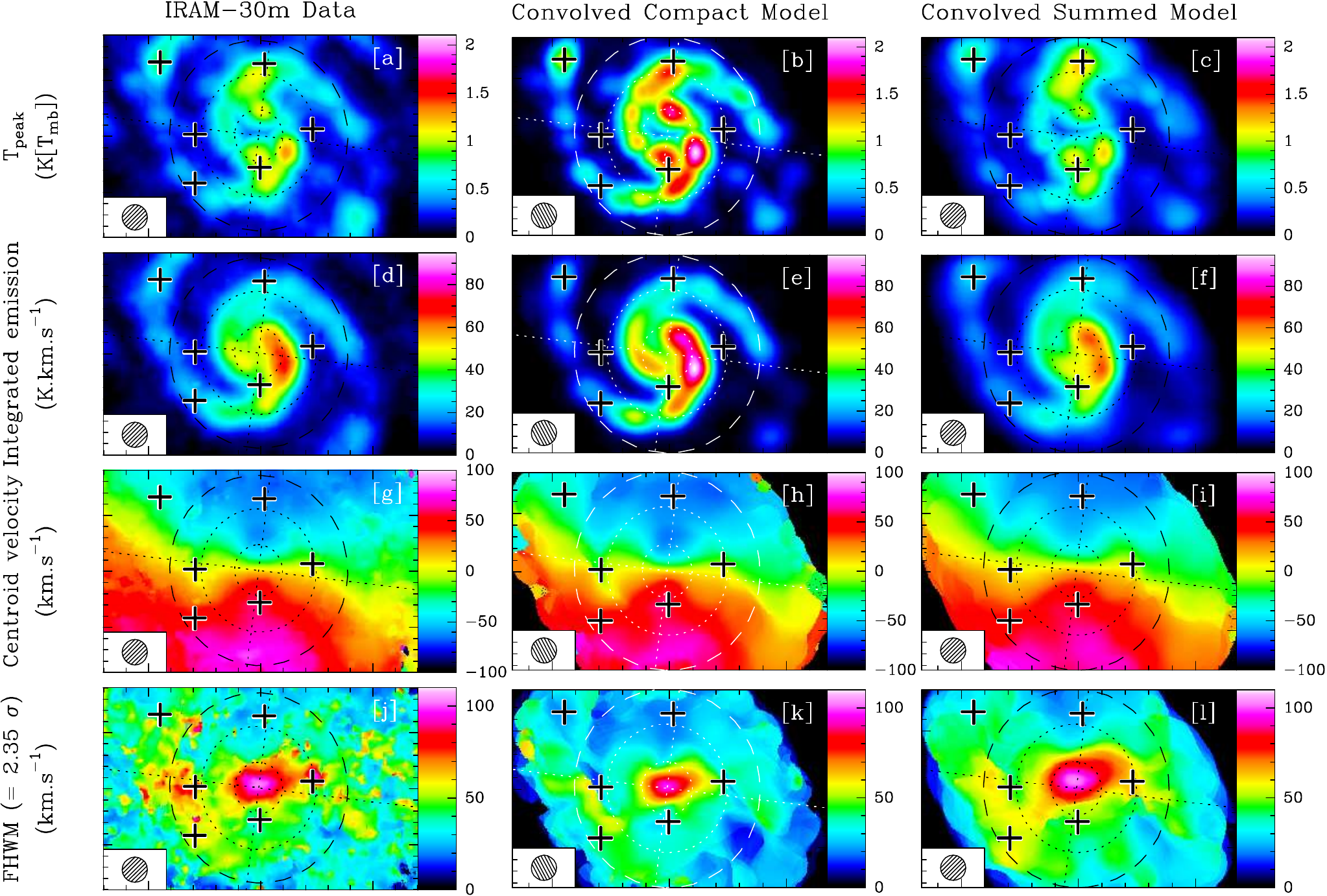}
    \caption{Comparison of the spatial distribution (from top to bottom) 
      of the peak intensity, integrated intensity, centroid velocity, and
      the line full width at half maximum (\ie{}, 2.35 times the standard
      deviation in velocity) of the \twCO{} \Jone{} emission from the
      actual observations (left column), and the two models convolved with
      the modeled 30m beam (middle and right column). The angular
      resolution is indicated by a circle in the bottom left corner of each
      panel.  The intensity scale is shown on the right-hand side of each
      panel.  The 3 images of each row share the same intensity scale to
      facilitate a meaningful visual comparison. The crosses on the images
      show the positions of the spectra displayed in
      Fig.~\ref{fig:spectra-comparison}. Other plot annotations are the
      same as in Fig.~\ref{fig:moments:paws:mixed}.}
    \label{fig:moments:models}
  \end{figure*}}

\newcommand{\FigSpectraComparison}{%
  \begin{figure*}
    \centering
    \includegraphics[width=\hsize]{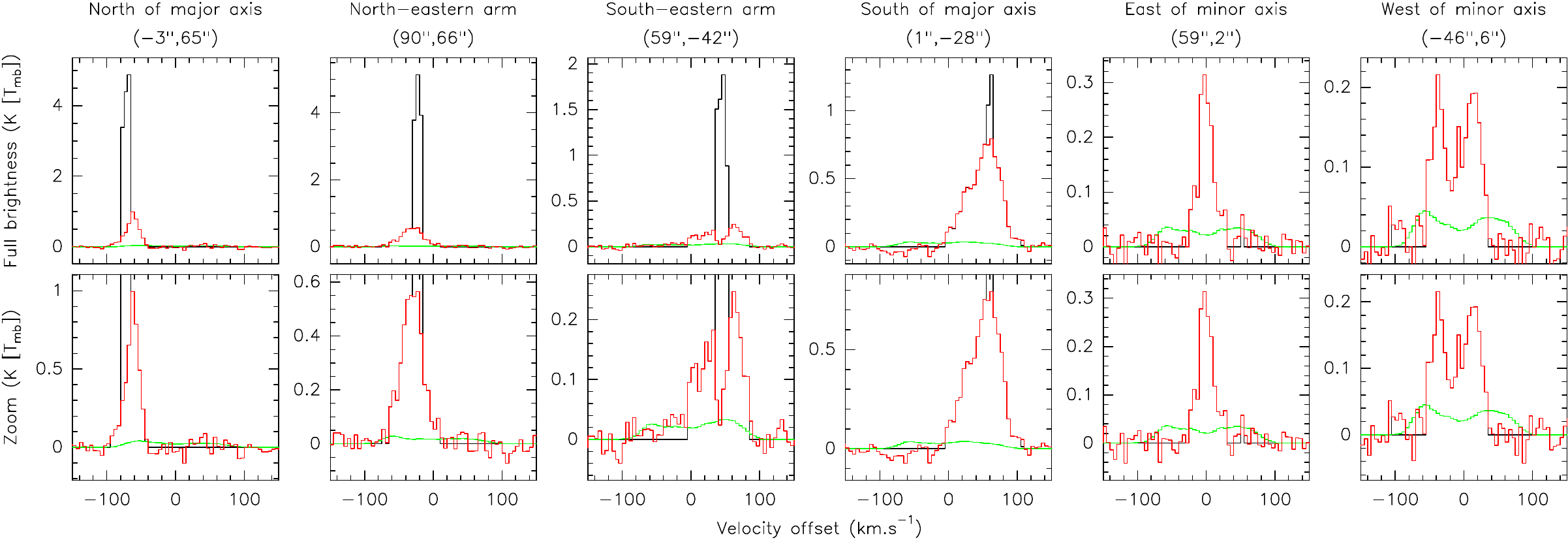}
    \caption{Comparison of the spectra of our summed model (in
      black), the extended emission as measured in the $6''$ subtracted
      cube (in red), and the error beam contribution (in green), at six
      positions where the extended emission shows different
      characteristics.  These positions are displayed as crosses on the
      images of the extended emission moments in
      Fig.~\ref{fig:moments:paws:mixed}. The bottom row shows a brightness
      zoom of the top row.}
    \label{fig:spectra-comparison}
  \end{figure*}}

\newcommand{\FigFWHMcomparison}{%
  \begin{figure}[t]
    \centering
    \includegraphics[width=\hsize{}]{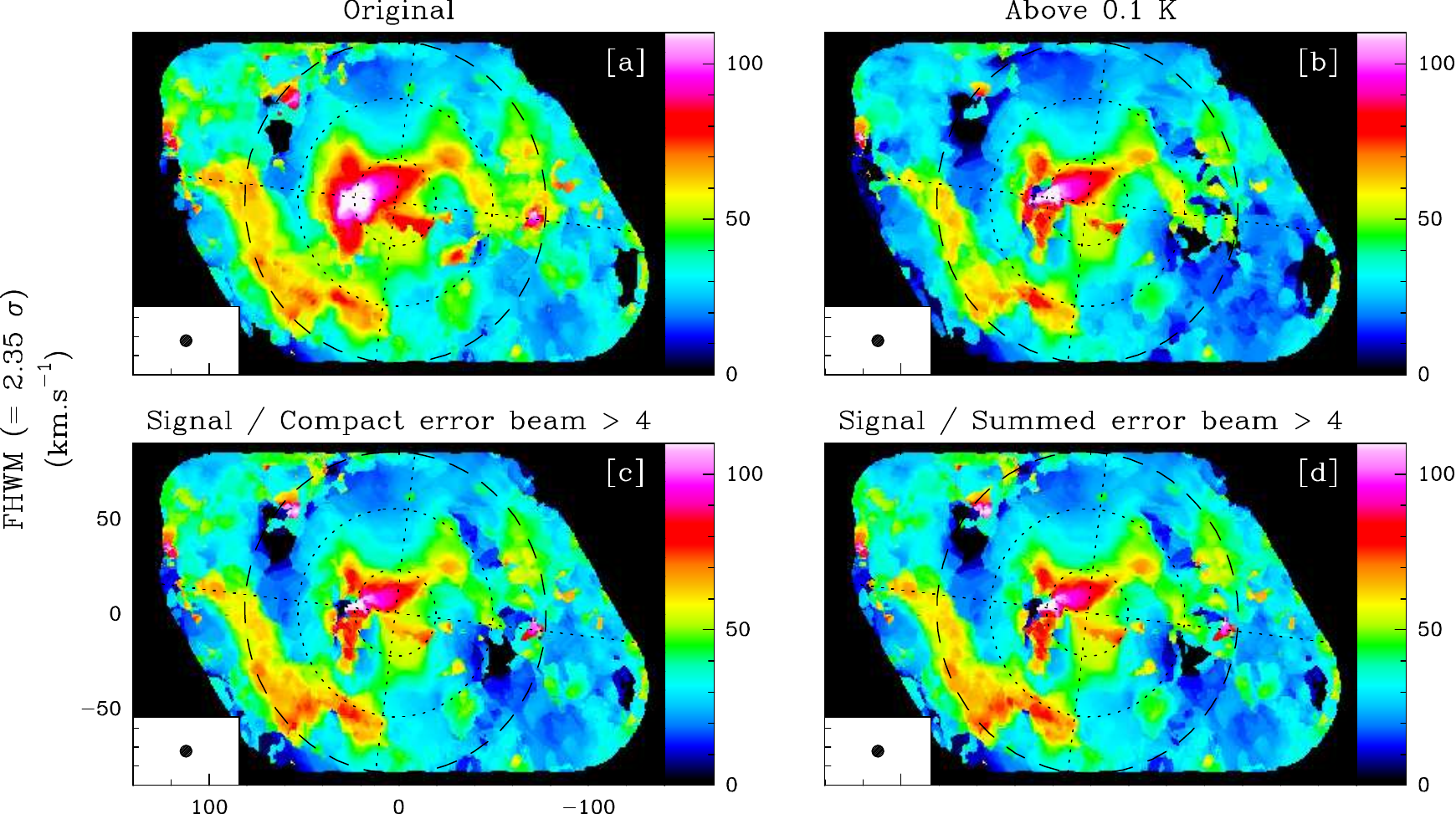}
    \caption{Comparison of the spatial distribution of
      the line full width at half maximum (\ie{}, 2.35 times the standard
      deviation in velocity) of the \twCO{} \Jone{} extended emission
      measured at $6''$. Different masking techniques were used to test the
      impact of the error beams on the computation of the line width (see
      text). The angular resolution is indicated by a circle in the bottom
      left corner of each panel. The intensity scale is shown on the
      right-hand side of each panel. The images share the same intensity
      scale to facilitate comparison. The major and minor axes are
      displayed as perpendicular dotted lines.  The dotted circles show the
      two inner corotation resonances at radii equal to $23''$ and $55''$,
      while the dashed circle shows the start of the material arms at a
      radius equal to $85''$~\citep{meidt12b}.}
    \label{fig:fwhm:comparison}
  \end{figure}}

\newcommand{\FigSDtwCOmaps}{%
  \begin{figure*}[t]
    \centering
    \includegraphics[width=\hsize{}]{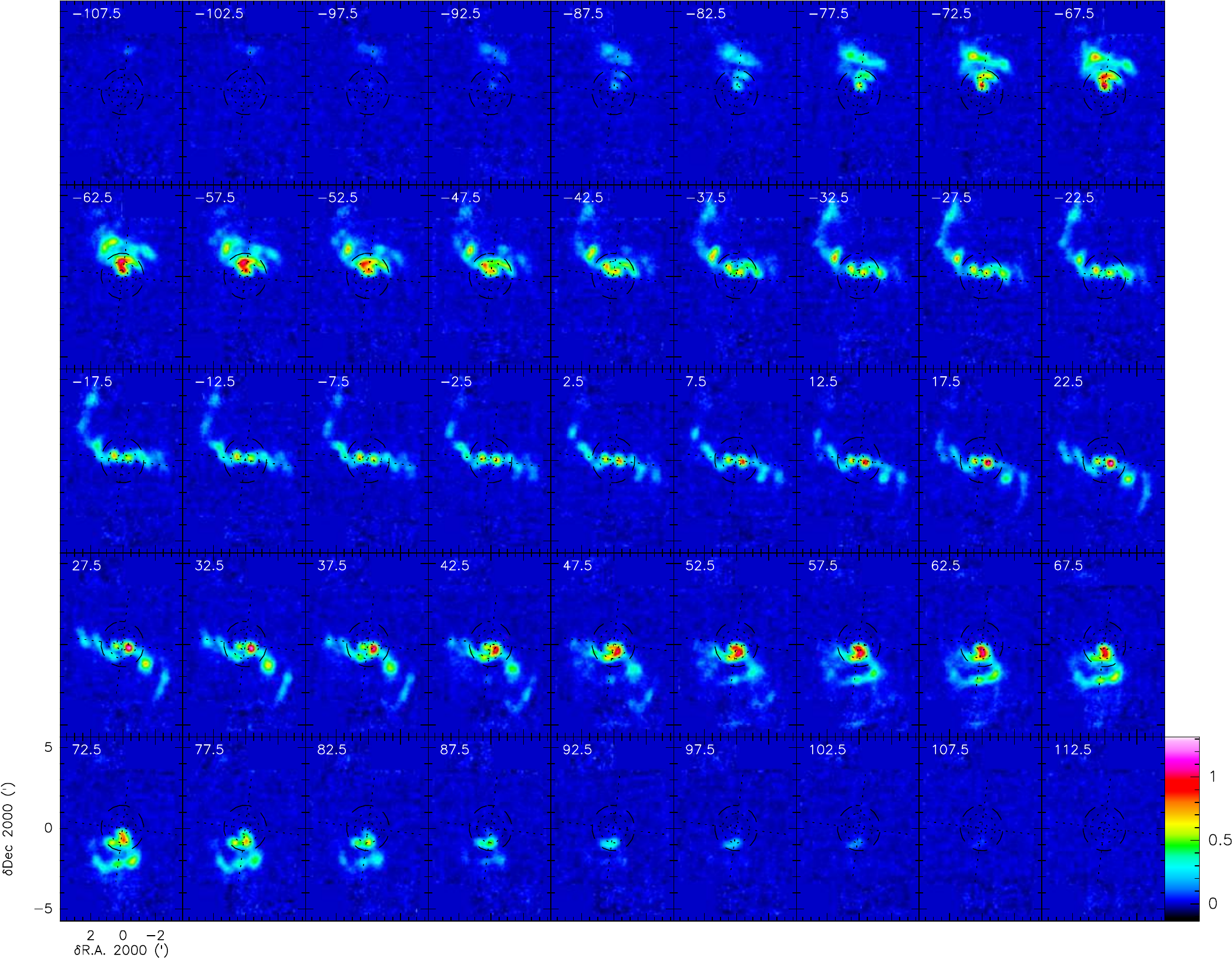}
    \caption{Channel maps of the \twCO{} \Jone{} emission obtained with the
      IRAM-30m telescope. The velocity in \kms{} of each channel is
      displayed in the top left corner of each panel. The intensity scale
      (in \Tmb{}) is shared by all the panels and it is displayed in the
      bottom right corner of the figure. The major and minor axes are
      displayed as perpendicular dotted lines. The dotted circles show the
      two inner corotation resonances at radii equal to $23''$ and $55''$,
      while the dashed circle shows the start of the material arms at a
      radius equal to $85''$~\citep{meidt12b}.}
    \label{fig:channelmaps:30m:12co10}
  \end{figure*}}

\newcommand{\FigSDthCOmaps}{%
  \begin{figure*}[t]
    \centering
    \includegraphics[width=\hsize{}]{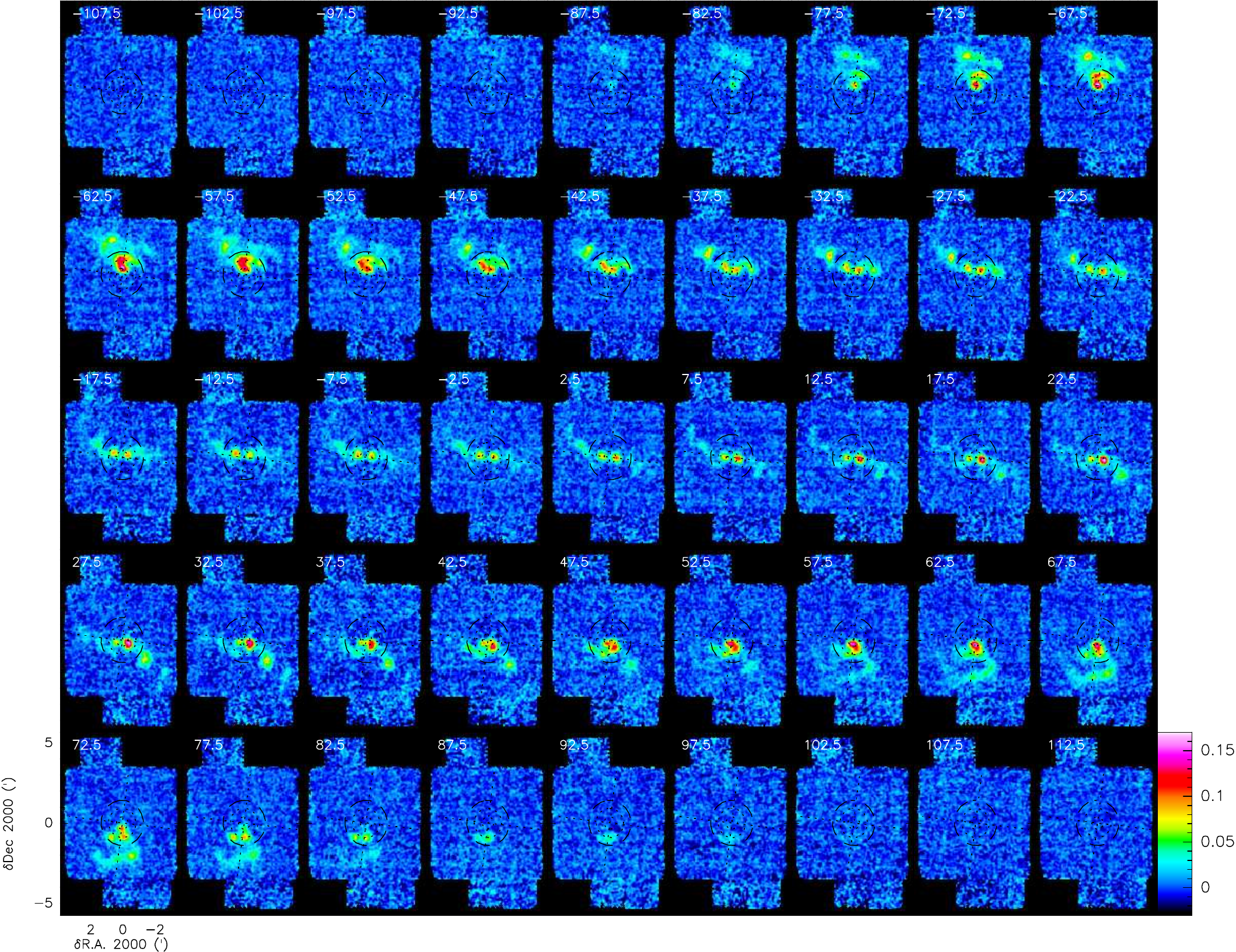}
    \caption{Channel maps of the \thCO{} \Jone{} emission obtained with the
      IRAM-30m telescope. The figure layout is the same as for
      Fig.~\ref{fig:channelmaps:30m:12co10}.}
    \label{fig:channelmaps:30m:13co10}
  \end{figure*}}

\newcommand{\FigPdBImapOne}{%
  \begin{figure*}[t]
    \centering
    \includegraphics[width=\hsize{}]{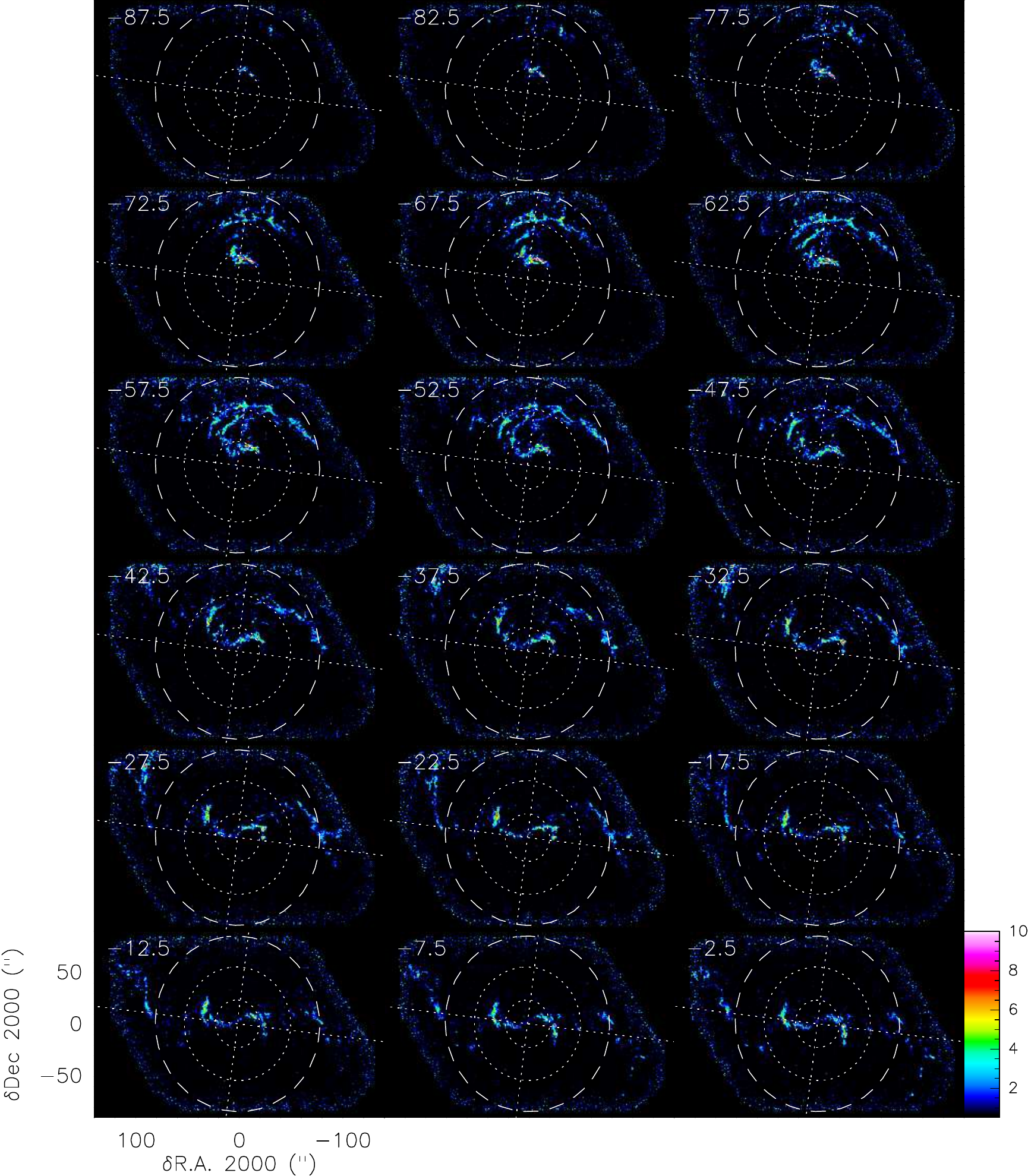}
    \caption{Channel maps of the \twCO{} \Jone{} emission obtained from the
      combination of IRAM-30m and IRAM-PdBI observations. The velocity of
      each channel in \kms{} is displayed in the top left corner of each
      panel. Only the negative velocity channels are shown here. The
      intensity scale (in \Tmb{}) is shared by all the panels and it is
      displayed in the bottom right corner of the figure. This intensity
      scale is saturated to emphasize the pixels with significant emission.
      The major and minor axes are displayed as perpendicular dotted lines.
      The dotted circles show the two inner corotation resonances at radii
      equal to $23''$ and $55''$, while the dashed circle shows the start
      of the material arms at a radius equal to $85''$~\citep{meidt12b}.}
    \label{fig:channelmaps:pdbi+30m:1}
  \end{figure*}}

\newcommand{\FigPdBImapTwo}{%
  \begin{figure*}[t]
    \centering
    \includegraphics[width=\hsize{}]{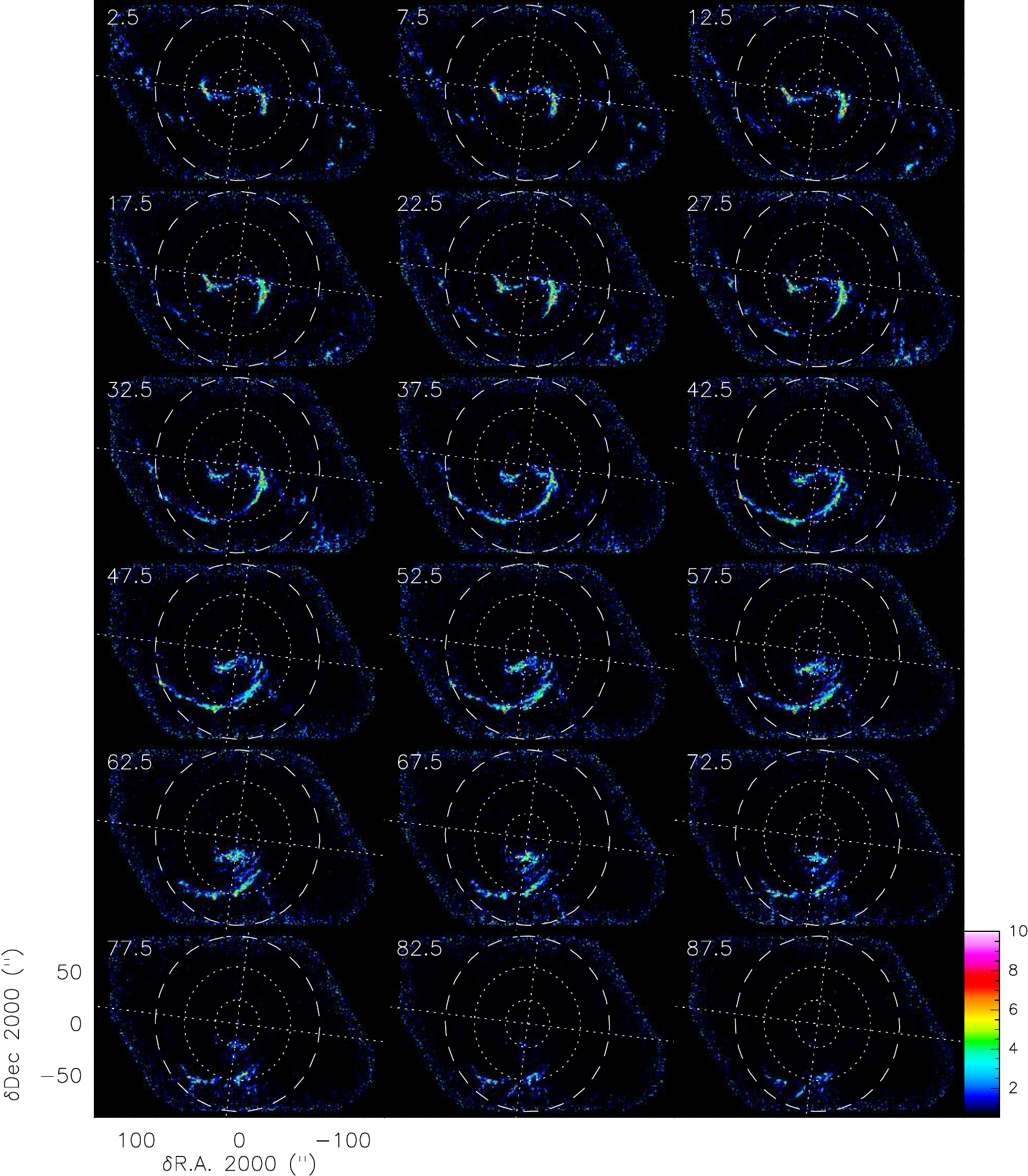}
    \caption{Same as Fig.~\ref{fig:channelmaps:pdbi+30m:1} but for the
      positive velocity channels.}
    \label{fig:channelmaps:pdbi+30m:2}
  \end{figure*}}

\newcommand{\FigExtendedCombinedMapOne}{%
  \begin{figure*}[t]
    \centering
    \includegraphics[width=\hsize{}]{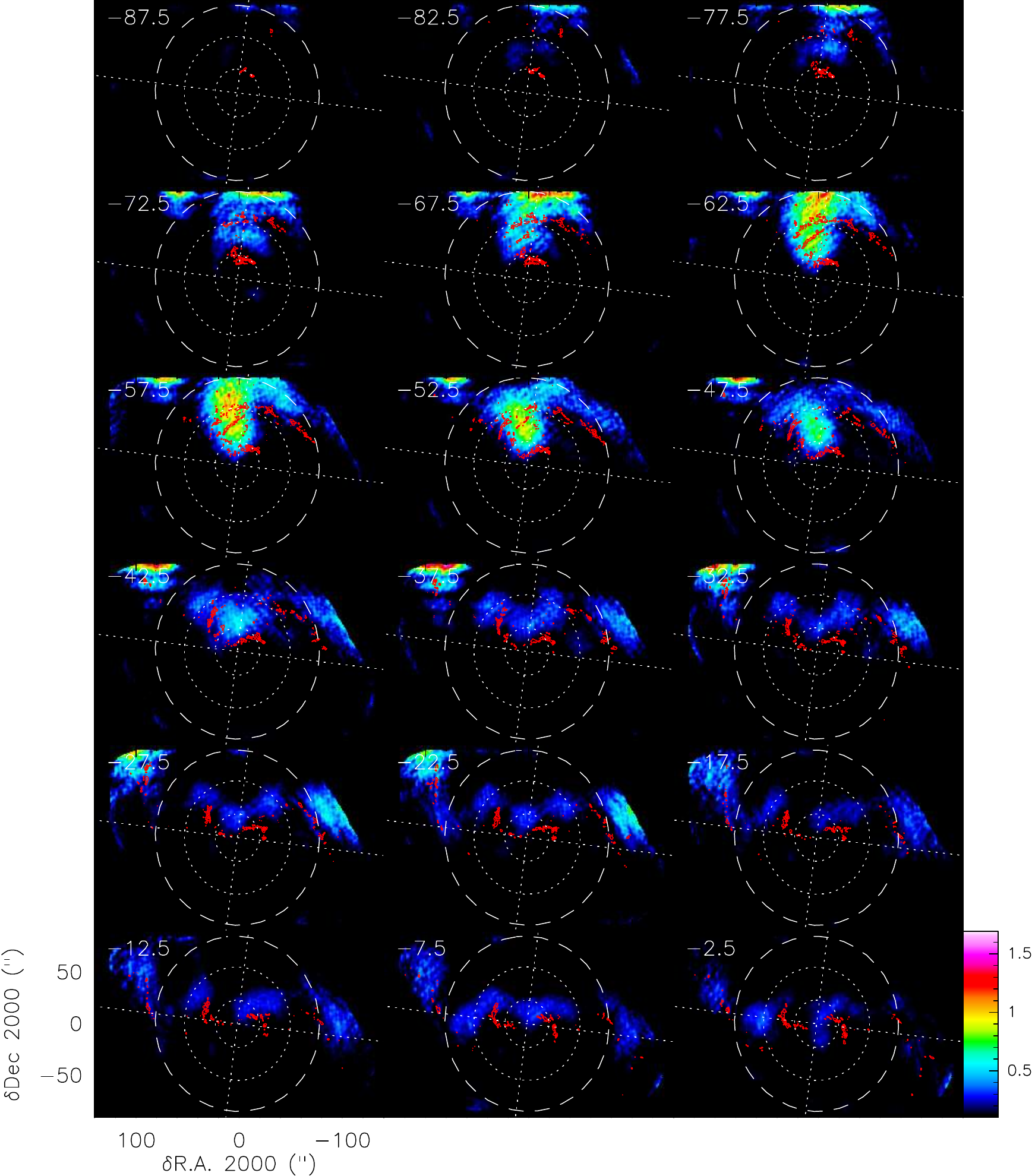}
    \caption{Contours of the channel maps of the signal-to-noise ratio of
      the combined PdBI+30m cube laid over the channel maps of the resolved
      emission. The contour levels start at a signel-to-noise ratio of 8.
      The figure layout is as in Fig.~\ref{fig:channelmaps:pdbi+30m:1}.}
    \label{fig:channelmaps:extended-vs-combined:1}
  \end{figure*}}

\newcommand{\FigExtendedCombinedMapTwo}{%
  \begin{figure*}[t]
    \centering
    \includegraphics[width=\hsize{}]{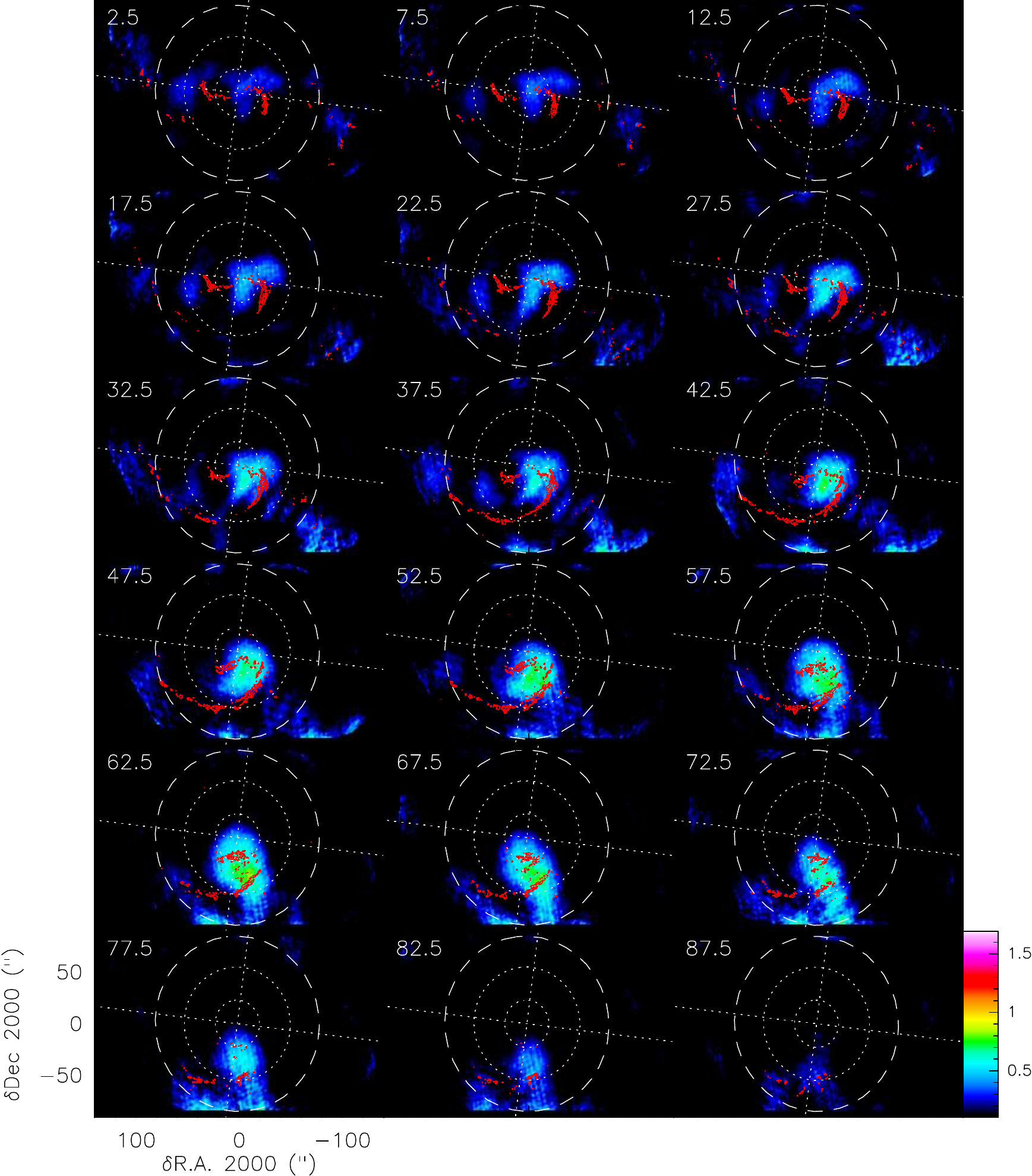}
    \caption{Same as Fig.~\ref{fig:channelmaps:extended-vs-combined:1} but
      for the positive velocity channels.}
    \label{fig:channelmaps:extended-vs-combined:2}
  \end{figure*}}

\newcommand{\FigPAWSmomentsFull}{%
  \begin{figure*}[t]
    \centering
    \includegraphics[width=\hsize{}]{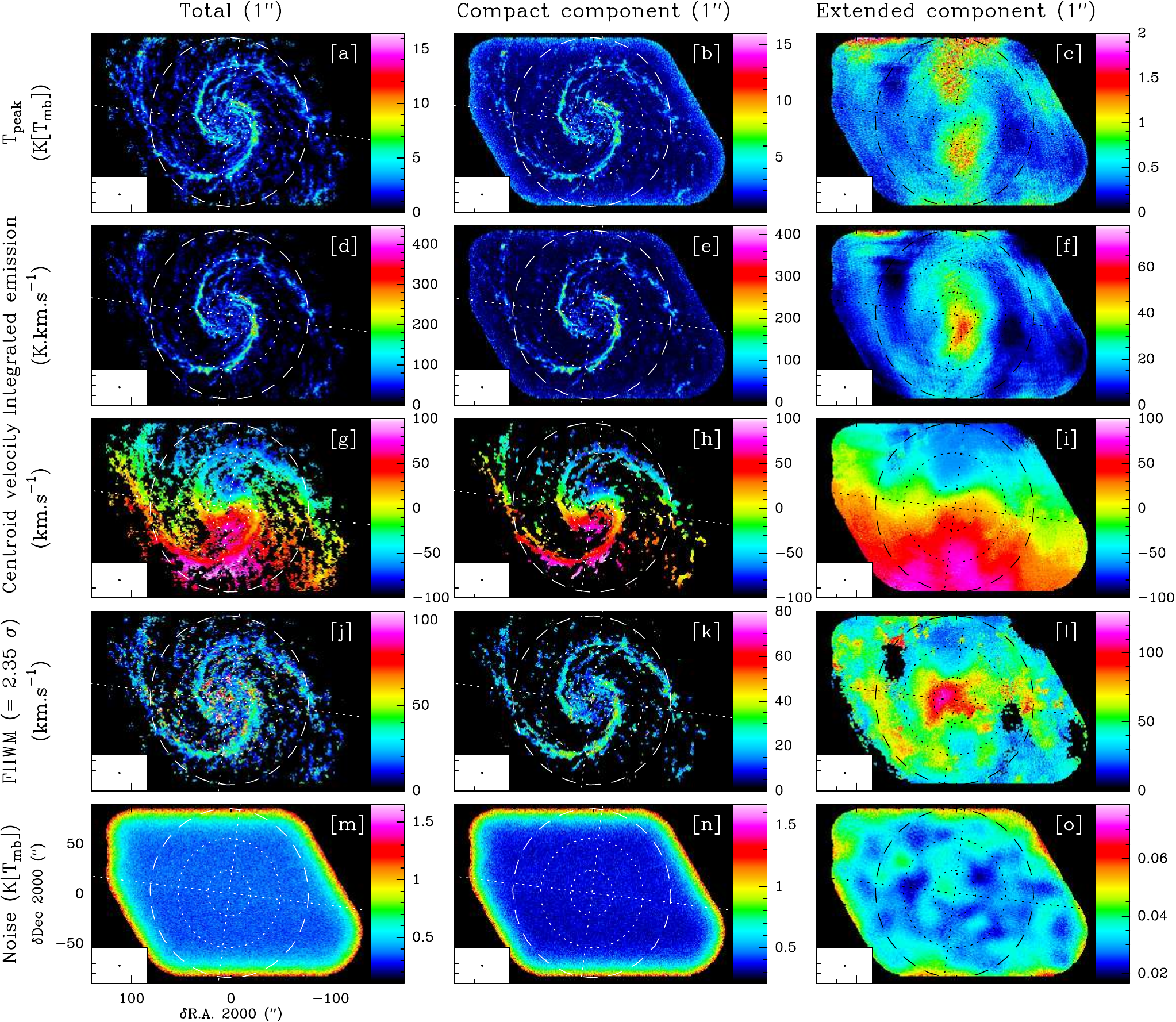}
    \caption{Comparison of the spatial distribution at 
      $\sim1''$-resolution (from top to bottom) of the peak intensity,
      integrated intensity, centroid velocity, the line full width at half
      maximum (\ie{}, 2.35 times the standard deviation in velocity) and,
      rms noise of the \twCO{} \Jone{} emission for the hybrid synthesis
      (PdBI + 30m, left column), the PdBI-only (middle column) and the
      subtraction of the PdBI-only from the hybrid synthesis cubes (right
      column). The angular resolution is shown as a circle in the bottom
      left corner of each panel. The intensity scale is shown on the right
      of each panel. The major and minor axes are displayed as
      perpendicular dotted lines. The dotted circles show the two inner
      corotation resonances at radii equal to $23''$ and $55''$, while the
      dashed circle shows the start of the material arms at a radius equal
      to $85''$~\citep{meidt12b}.}
    \label{fig:moments:paws:full}
  \end{figure*}}

\newcommand{\FigPAWSmomentsTapOne}{%
  \begin{figure*}[t]
    \centering
    \includegraphics[width=\hsize{}]{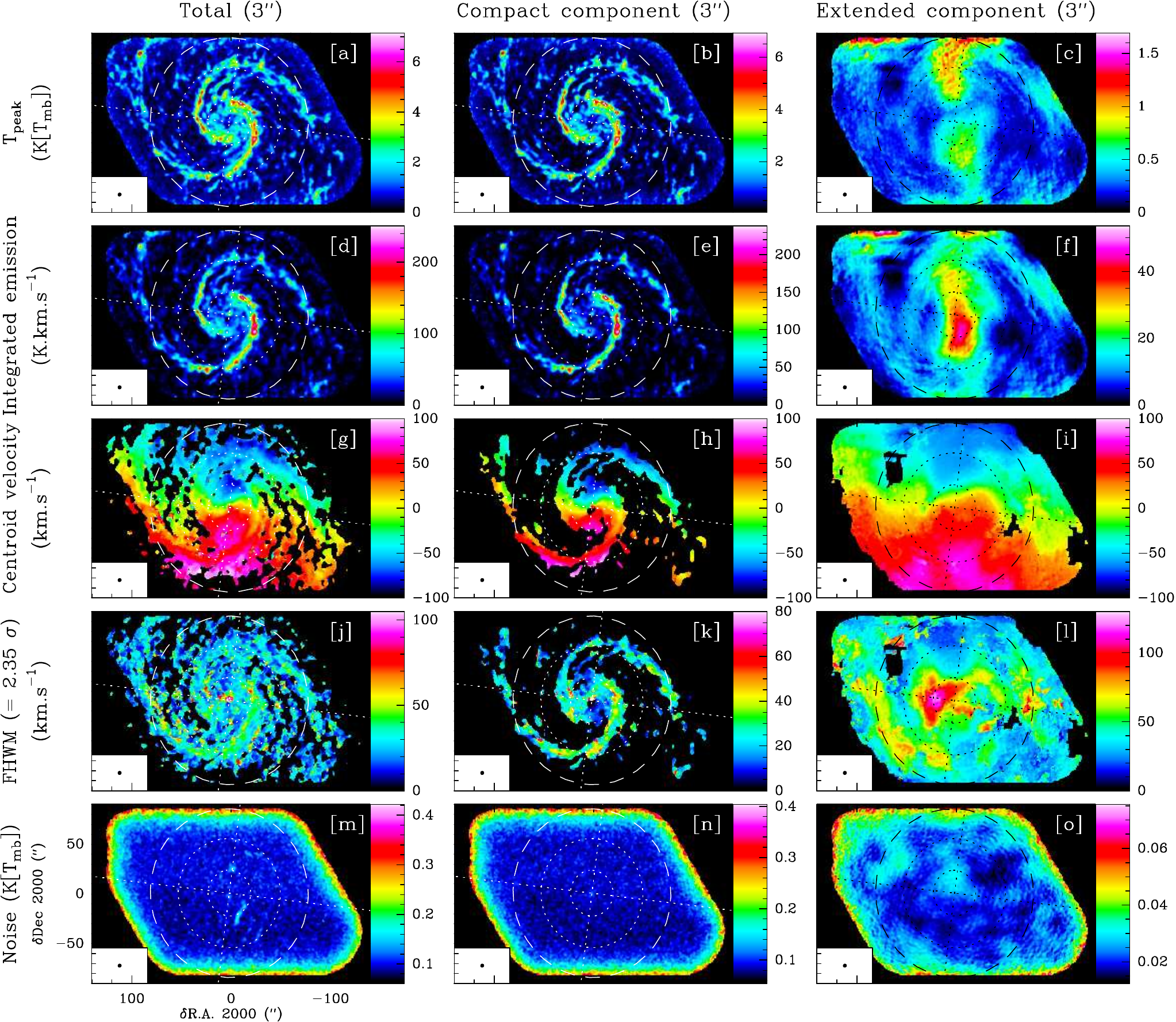}
    \caption{Same as Fig.~\ref{fig:moments:paws:full} but the comparison is
      done here at an angular resolution of $3''$.}
    \label{fig:moments:paws:tap1}
  \end{figure*}}

\newcommand{\FigPAWSmomentsTapTwo}{%
  \begin{figure*}[t]
    \centering
    \includegraphics[width=\hsize{}]{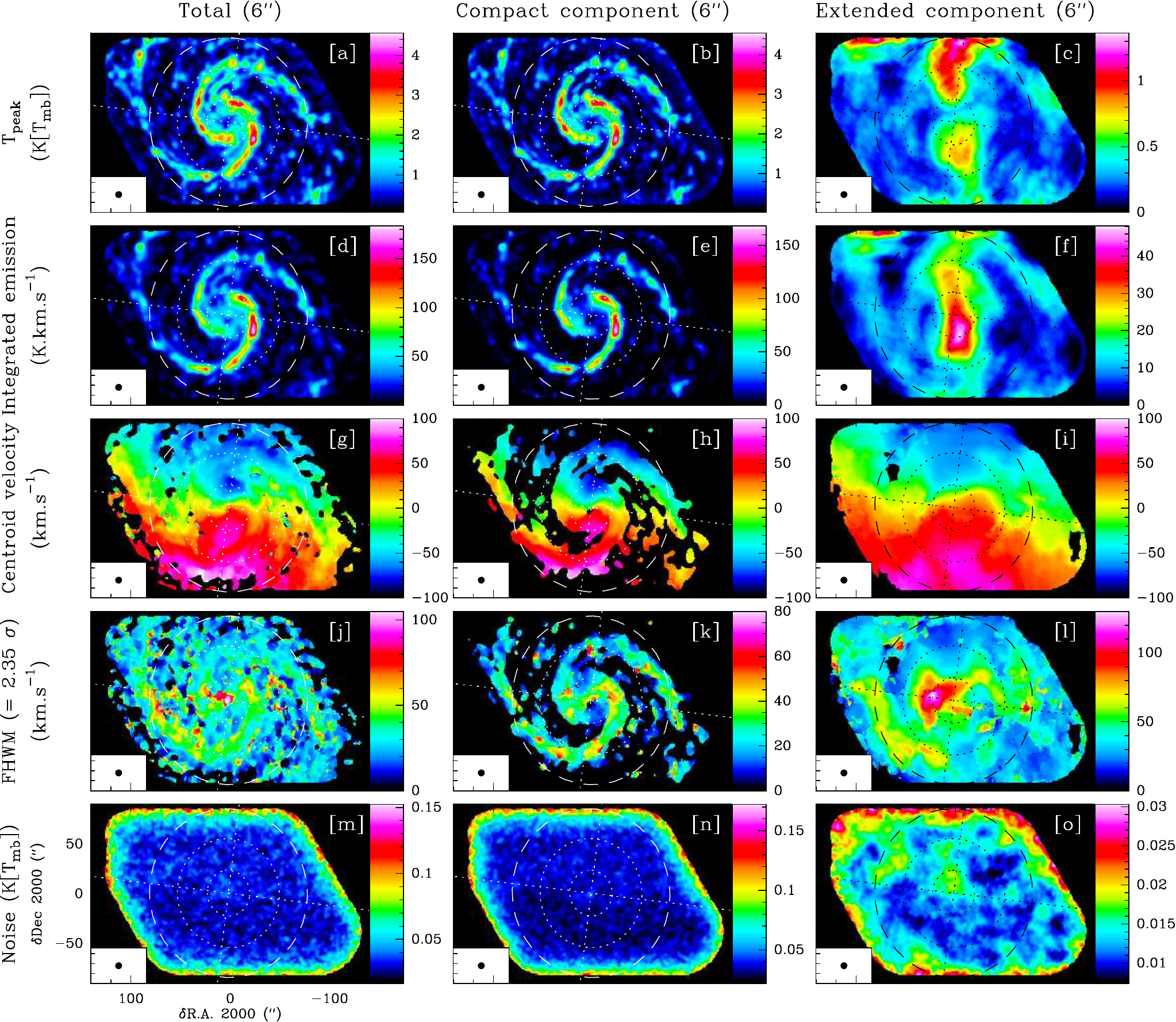}
    \caption{Same as Fig.~\ref{fig:moments:paws:full} but the comparison is
      done here at an angular resolution of $6''$.}
    \label{fig:moments:paws:tap2}
  \end{figure*}}

\newcommand{\FigSDmomentsTapTwo}{%
  \begin{figure*}[t]
    \centering
    \includegraphics[width=\hsize{}]{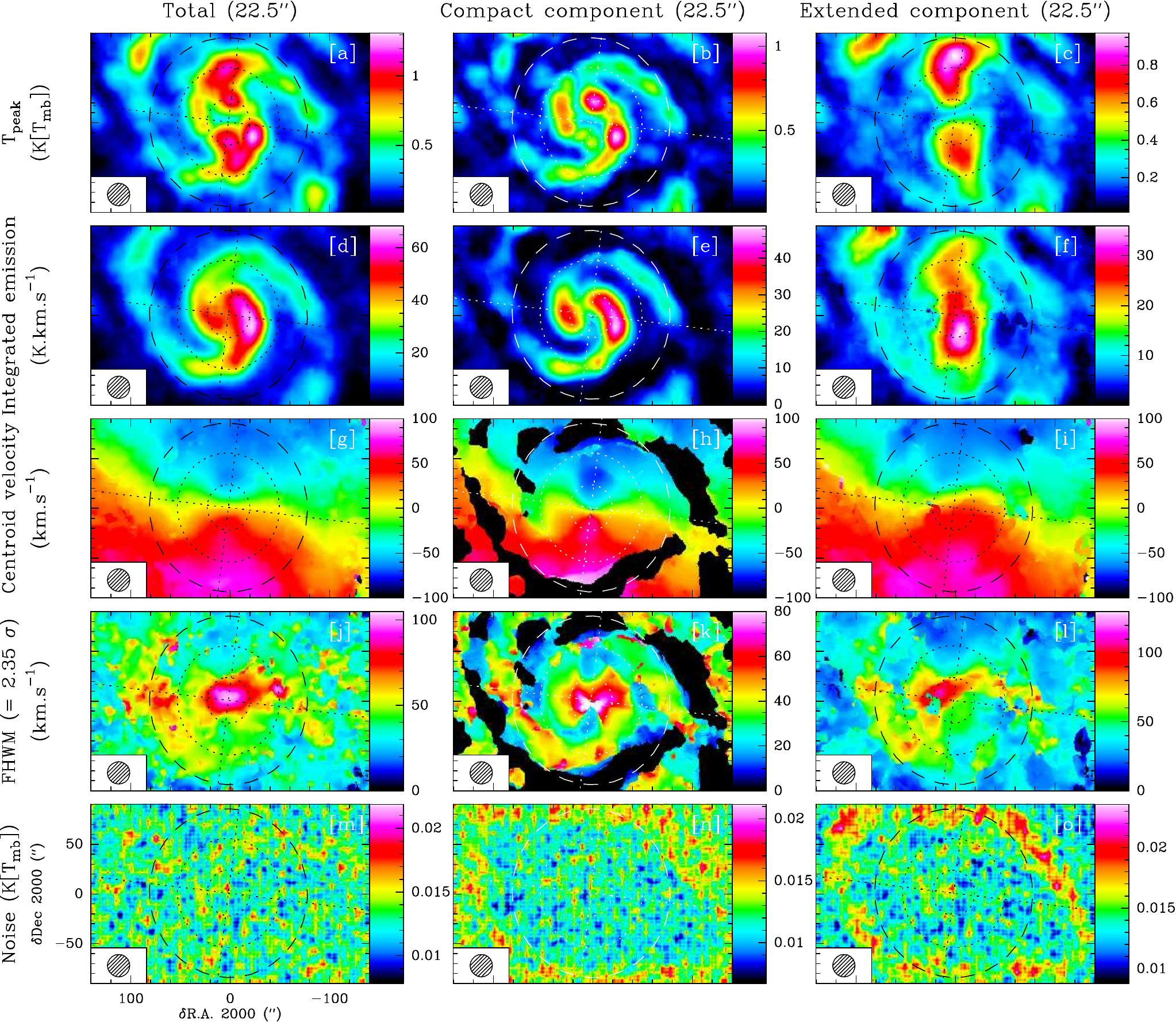}
    \caption{Comparison of the spatial distributions at 
      $22.5''$-resolution of the peak intensity (top), integrated
      intensity, centroid velocity, 2.35 times the standard deviation and
      rms noise (bottom) of the \twCO{} \Jone{} emission for IRAM-30m cube
      (left column), the $6''$ PdBI-only data cube smoothed at $22.5''$
      (right column), and the subtraction of the two previous cubes (middle
      column). The angular resolution is displayed as a circle in the
      bottom left corner of each panel. The intensity scale is shown on the
      right side of each panel. The major and minor axes are displayed as
      perpendicular dotted lines. The dotted circles show the two inner
      corotation resonances at radii equal to $23''$ and $55''$, while the
      dashed circle shows the start of the material arms at a radius equal
      to $85''$~\citep{meidt12b}.}
    \label{fig:moments:30m:tap2}
  \end{figure*}}

\begin{document}

\title{The Plateau de Bure + 30\,m Arcsecond Whirlpool Survey \\
  reveals a thick disk of diffuse molecular gas in the M51 galaxy}

\author{J\'er\^ome Pety\altaffilmark{1,2}} %
\email{pety@iram.fr}
\author{Eva Schinnerer\altaffilmark{3}} %
\author{Adam K. Leroy\altaffilmark{4}} %
\author{Annie Hughes\altaffilmark{3}} %
\author{Sharon E. Meidt\altaffilmark{3}} %
\author{Dario Colombo\altaffilmark{3}} %
\author{Gaelle Dumas\altaffilmark{1}} %
\author{Santiago Garc\'{i}a-Burillo\altaffilmark{5}} %
\author{Karl F. Schuster\altaffilmark{1}} %
\author{Carsten Kramer\altaffilmark{6}} %
\author{Clare L. Dobbs\altaffilmark{7}} %
\author{Todd A. Thompson\altaffilmark{8,9}} %

\altaffiltext{1}{Institut de Radioastronomie Millim\'etrique, 300 Rue de la Piscine, F-38406 Saint Martin d'H\`eres, France}
\altaffiltext{2}{Observatoire de Paris, 61 Avenue de l'Observatoire, F-75014 Paris, France.}
\altaffiltext{3}{Max Planck Institute for Astronomy, K\"onigstuhl 17, 69117 Heidelberg, Germany}
\altaffiltext{4}{National Radio Astronomy Observatory, 520 Edgemont Road, Charlottesville, VA 22903, USA}
\altaffiltext{5}{Observatorio Astron\'{o}mico Nacional - OAN, Observatorio de Madrid Alfonso XII, 3, 28014 - Madrid, Spain}
\altaffiltext{6}{Instituto Radioastronom\'{i}a Milim\'{e}trica, Av. Divina Pastora 7, Nucleo Central, 18012 Granada, Spain}
\altaffiltext{7}{School of Physics and Astronomy, University of Exeter, Stocker Road, Exeter EX4 4QL, UK}
\altaffiltext{8}{Department of Astronomy, The Ohio State University, 140 W. 18th Ave., Columbus, OH 43210, USA} 
\altaffiltext{9}{Center for Cosmology and AstroParticle Physics, The Ohio State University, 191 W. Woodruff Ave., Columbus, OH 43210, USA}

\shorttitle{PAWS reveals a thick disk of diffuse molecular gas in the M51 
  galaxy} %
\shortauthors{Pety et al.}  %

\begin{abstract}
  We present the data of the Plateau de Bure Arcsecond Whirlpool Survey
  (PAWS), a high spatial and spectral resolution \twCO{} \Jone{} line
  survey of the inner $\sim10\times6\kpc$ of the M51 system, and the first
  wide-field imaging of molecular gas in a star-forming spiral galaxy with
  resolution matched to the typical size of Giant Molecular Clouds
  (40\pc{}). We describe the observation, reduction, and combination of the
  Plateau de Bure Interferometer (PdBI) and IRAM-30m ``short spacing''
  data.  The final data cube attains $1.1\arcsec$-resolution over the $\sim
  270\arcsec \times 170\arcsec$ field of view, with sensitivity to all
  spatial scales from the combination of PdBI and IRAM-30m data, and
  brightness sensitivity of $0.4\K$ ($1\sigma$) in each 5\kms{}-wide
  channel map. We find a CO-luminosity of $9\times 10^{8}\Kkmspcpc$,
  corresponding to a molecular gas mass of $4\times 10^{9}\Msol$ for a
  standard CO-to-H$_2$ conversion factor.  Unexpectedly, we find that a
  large fraction, $(50\pm10)\%$, of this emission arises mostly from
  spatial scales larger than $36\arcsec \simeq 1.3\kpc$. Through a series
  of tests, we demonstrate that this extended emission does not result from
  a processing artifact.  We discuss its origin in light of the stellar
  component, the $\twCO/\thCO$ ratio, and the difference between the
  kinematics and structure of the PdBI-only and hybrid synthesis (PdBI +
  IRAM-30m) images.  The extended emission is consistent with a thick,
  diffuse disk of molecular gas with a typical scale height of
  $\sim200\pc$, substructured in unresolved filaments which fills
  $\sim0.1\%$ of the volume.
\end{abstract}


\section{Introduction}

\FigPdBIvsTHM{} %
\TabWhirlpool{} %

Along the path leading from the accretion of hot ionized gas onto galaxies
to the birth of stars, the formation and evolution of Giant Molecular
Clouds (GMCs) is the least well understood step. For example, the
dependence of their mass distributions, lifetimes, and star formation
efficiencies on galactic environment (\eg{} arm, interarm, nuclear region)
is largely unknown~\citep[for a review, see][]{mckee07}. Because the Sun's
position within the Milky Way disk makes GMC studies difficult within our
own Galaxy, observations of GMC populations in nearby face-on galaxies
offer the best way to address many of these unknowns.

Complete CO maps that resolve individual GMCs have been carried out across
the Local Group, allowing for the construction of mass functions and an
estimation of GMC lifetimes via comparison with maps at other
wavelengths~\citep[]{kawamura09}.  To date, these observations have probed
mostly low mass galaxies where \HI{} dominates the interstellar
medium~\citep[\eg{},][]{blitz07}.  The main reason is that the angular
resolution required to identify individual GMCs~\citep[typical size
\about{} 40\pc{}, \eg{},][]{solomon87} in any galaxy outside the Local
Group is extremely difficult to achieve.  Reaching such resolutions with
single dish telescopes remains impossible in all but the very closest
galaxies. This presents a major obstacle in linking our understanding of
star formation and galactic evolution. Even for M31, the closest massive
spiral galaxy to the Milky Way, the \twCO{} \Jone{} IRAM-30m map achieves a
spatial resolution of only 85\pc{} and it suffers from projection
effects~\citep{nieten06}. This is an important problem because these
massive star forming spirals dominate the mass and light budget of blue
galaxies and they host most of the star formation in the present-day
universe~\citep[\eg][]{schiminovich07}.

To remedy this situation, we used the Plateau de Bure Interferometer (PdBI)
to carry out the PdBI Arcsecond Whirlpool
Survey~\citep[PAWS,][]{schinnerer12a}. The high quality receivers and good
weather conditions allowed PAWS to map the central, molecule-bright part of
M51 at a resolution of $\sim1.1''$ $(\sim40\pc)$ while still maintaining
good brightness sensitivity (RMS $\sim 0.4\K$).  M51 is the best target for
such a program (see Table~\ref{tab:m51}). It is one of the closest $(D \sim
7.6\Mpc)$, face-on $(i \sim 21\degr)$ grand design spirals, and it has been
extensively studied at essentially all wavelengths.  In contrast to Local
Group galaxies with a resolved GMC population, the molecular gas clearly
dominates the interstellar medium inside the mapped
region~\citep[\eg{},][]{garciaburillo93,aalto99,schuster07,hitschfeld09,leroy09,koda11}.
M51 thus offers the opportunity to relate the physical properties of
molecular gas to spiral structure. We complemented the interferometric data
with a sensitive (RMS $\sim 16\mK$) map of the whole M51 system with the
IRAM-30m single-dish telescope. This allowed us to produce a hybrid
synthesis map --- a joint deconvolution of the PdBI and IRAM-30m data sets
--- that is sensitive to all spatial scales between our synthesized beam
and the PAWS field of view (see Fig.~\ref{fig:pdbi-vs-30m}).

In Section \ref{sec:obs}, we detail the observing strategy and the data
reduction. In Section \ref{sec:ext}, we show that a large portion
($\sim50\%\pm10\%$) of the emission in our hybrid maps arises from faint,
extended structures.  We provide a detailed discussion of the nature of the
gas responsible for this emission.  We summarize our conclusions in
Sect.~\ref{sec:summary}. The Appendices provide details on technical
aspects of our observations, reductions and analysis, and additional
supplementary Tables and Figures.

\section{PAWS data acquisition and reduction}
\label{sec:obs}

\FigSkyCover{} %

This section presents the observing strategy, the data reduction and the
resulting data set. Sects.~\ref{sec:obs:pdbi} and~\ref{sec:obs:30m} focus
on the PdBI and IRAM-30m data, respectively.  Sect.~\ref{sec:obs:imaging}
explains how we combined these data to produce a final set of hybrid maps
sensitive to all spatial scales.

\subsection{IRAM Plateau de Bure Interferometer data}
\label{sec:obs:pdbi}

After a discussion of the observing setup, we describe the calibration of
the interferometric data.

\subsubsection{Observations}

PdBI observations dedicated to this project were carried out with either 5
or 6 antennas in the A, B, C, and D configurations (baseline lengths from
24\m{} to 760\m{}) from August 2009 to March 2010. The two polarizations of
the single-sideband receivers were tuned at 115.090\GHz{}, \ie{}, the
\twCO{}~\Jone{} rest frequency redshifted to the LSR velocity (471.7\kms{})
of M51. Four correlator bands of 160\MHz{} per polarization were
concatenated to cover a bandwidth of $\about550\MHz$ or $\about1430\kms$ at
a spectral resolution of $1.25\MHz$ or $3.25\kms$.

We observed two 30-field mosaics, as described in
Table~\ref{tab:obs:pdbi:main} and shown in Fig.~\ref{fig:skycover}. Both
mosaics were centered such that their combination covers the inner part of
M51. The total field of view is approximately $270''\times170''$. Each
pointing was observed during $3\times15$ seconds in turn.  This allowed us
1) to observe one mosaic between two calibrations, which were taken every
22.5 minutes and 2) to minimize the dead-times due to moves from one field
position to the next, while ensuring that the integration time per
visibility (15 seconds) is short enough to avoid mixing independent $uv$
plane information in all the configurations~\citep[see, \eg{}, Appendix C.1
of][for detailed calculations]{pety10}. An inconvenient aspect of such an
observing strategy is that we obtained two data sets, observed in slightly
different conditions, implying slightly different noise properties and $uv$
coverage (\ie{} slightly different dirty/synthesized beams).

The field positions followed an hexagonal pattern, each field pointing
being separated from its nearest neighbors by the primary beam full width
at half maximum (FWHM), \Afwhm{}. Nyquist sampling requires a distance
between two consecutive pointings of $\wavelength/\dprim$ along two
orthogonal axes, where \wavelength{} is the observation wavelength and
\dprim{} is the diameter of the interferometer antennas. At PdBI, we
typically have $\Afwhm = 1.2 \wavelength/\dprim$. The hexagonal pattern
used here thus ensures Nyquist sampling along the Declination axis but a
slight undersampling along the Right Ascension axis.

The field-of-view was observed for about 169~hours of \emph{telescope} time
with 5 antennas in configuration D (19 hours) and 6 antennas in
configuration C (18 hours), B (57 hours) and A (75 hours).  Taking into
account the time for calibration and the data filtering applied, this
translated into final \emph{on--source} integration times (computed for a
6-antenna array) of useful data of 8.3~hours in D configuration, 15.2~hours
in C configuration, 43~hours in B configuration and 60~hours in A
configuration.  In each configuration, the time was approximately equally
distributed between both mosaics.

\subsubsection{Calibration}

\TabObsPdBImain{} %
\TabObsSDmain{} %
\TabNoise{} %

Standard calibration methods implemented inside the \GILDAS{}/\CLIC{}
software were used for the PdBI data.  The radio-frequency bandpass was
calibrated using observations of two bright (\about 10~Jy) quasars,
0851$+$202 and 3C279, leading to an excellent bandpass accuracy (phase rms
$\lesssim1^\circ$, amplitude rms $\lesssim 1\%$). The temporal phase and
amplitude gains were obtained from spline fits through regular measurements
of the following close-by quasars: 1418$+$546, 1308$+$326, J1332$+$473. The
flux scale was determined against the primary flux calibrator, MWC349. The
resulting fluxes of the calibration quasars are summarized in
Table~\ref{tab:fluxes}. The absolute flux accuracy is \about{} 10\%.

The data were filtered using statistical quality criteria on the pointing,
flux, amplitude and phase calibrators. The source data were flagged when
the surrounding calibrator measurements implied a phase rms larger than
$40\degr$, an amplitude loss larger than 22\%, a pointing error larger than
30\% of the primary beamwidth and/or a focus error larger than 30\% of the
wavelength. Finally, the data were also flagged when the tracking error was
larger than 10\% of the field of view. This reduces the amount of usable
data to 39\%\footnote{This number takes into account visibilities that were
  flagged because of shadowing.}, 70\%, 71\%, and 71\% of the data obtained
in the D, C, B, and A configurations, respectively.

\subsection{IRAM-30m single-dish data}
\label{sec:obs:30m}

\FigSDmoments{} %

A multiplicative interferometer filters out the low spatial frequencies,
\ie{}, spatially extended emission. We thus observed M51 with the IRAM-30m
single dish telescope on May 18-22, 2010 in order to recover the low
spatial frequency (``short- and zero-spacing'') information filtered out by
the PdBI. We describe here the observing strategy and the calibration,
baselining and gridding methods we used to obtain single-dish data whose
quality matches the interferometric data.

\subsubsection{Observations}

Table~\ref{tab:obs:30m:main} summarizes the IRAM-30m observations. We used
the EMIR receivers to map the \twCO{} \Jone{} and \thCO{} \Jone{} lines
over a $\sim 60$ square arcminute field-of-view covering the M51 system,
\ie{}, NGC\,5194 and its companion NGC\,5195. The upper sideband of the
3\mm{} separated sideband EMIR mixers (E090) was tuned at the \twCO{}
\Jone{} frequency.  The full 8 GHz bandwidth of the upper sideband was then
connected to the WILMA autocorrelator backend. This allowed us to
simultaneously measure the \twCO{} and \thCO{} lines (at 115.271 and
110.201\GHz{}, respectively).  The backend channel spacing is 2\MHz{},
which translates into a velocity channel spacing of 5.4 and 5.2\kms{} at
110 and 115\GHz{}, respectively.

We observed the galaxy in seven different patches. Four of these covered
the central $400''\times400''$ part of the galaxy in different ways. Three
additional patches extended the coverage to include the ends of the spiral
arms and the companion. Conditions during the observations varied from
``good'' summer weather ($\about 4\mm$ of precipitable water vapor) during
the first three nights to ``average'' summer weather ($\about 7\mm$ of
water vapor) over the last two nights.

We used the position-switch on-the-fly observing mode, covering each field
with back-and-forth scans along either the right ascension or declination
axes. We slewed at a speed of $\approx 8''$/sec and we dumped data to disk
every 0.5 seconds, yielding about 5.5 integrations per beam in the scanning
direction (the HPBW of the IRAM-30m telescope at the frequency of the CO
\Jone{} line is $\about 22''$).  The scan legs were separated by $8''$,
yielding Nyquist sampling transverse to the scan direction at 2.6\mm{}.
Each position in the central part was observed 34 times on average, with
observations split evenly between right ascension- and declination-oriented
scanning to suppress scan artifacts. The sky positions at the far end of
the CO spiral arms were observed 12 times so that the final effective
integration time on the extensions is somewhat shorter than on the main
field (see Fig.~\ref{fig:moments:30m}).

We observed the hot and cold loads plus the sky contribution every 12
minutes to establish the temperature scale and checked the pointing and
focus every $\sim 1$ and $\sim 4$ hours. The IRAM-30m position accuracy is
$\about 2''$.

\subsubsection{Calibration and gridding}

We reduced the IRAM-30m data using a combination of the
\GILDAS{}\footnote{See \url{http://www.iram.fr/IRAMFR/GILDAS} for more
  information about the \GILDAS{} softwares.} software suite~\citep{pety05}
and an {\tt IDL} pipeline developed for the IRAM HERACLES Large Program
\citep{leroy09}.

First, we calibrated the temperature scale of the data in {\tt GILDAS/MIRA}
based on the hot and cold loads plus sky observations~\citep{penzias73}.
The resulting flux accuracy is better than 10\%~\citep{kramer08}.  We then
subtracted the ``OFF'' spectrum from each on-source spectrum. We used {\tt
  GILDAS/CLASS} to write these calibrated, off-subtracted spectra to {\tt
  FITS} tables which we read into {\tt IDL} for further processing.

In {\tt IDL}, visual inspection indicated the presence of signal in the
$[-200,300\kms]$ velocity range around the systemic velocity of the galaxy.
About 1/16th of the 8\GHz{} bandwidth (\ie{} about 1000\kms{}) centered on
the line rest frequency were thus extracted from the calibrated spectra to
reduce the computation load of the next data reduction steps. We then fit
and subtracted a third-order baseline from each spectrum. When conducting
these fits, we use an outlier-resistant approach and exclude regions of the
spectrum that we know to contain bright emission based on previous
reduction or other observations. We experimented with higher and lower
order baselines and found a third-degree fit to yield the best results.
After fitting, we compared the rms noise about the baseline fit in
signal-free regions of each spectrum to the expected theoretical noise and
used this to reject a few pathological spectra. For the most part the data
are very well behaved and this is a minor step.

We gridded the calibrated, off-subtracted, baseline-subtracted spectra into
a data cube whose pixel size is $4''$. Doing so, we weighted each spectrum
by the inverse of the associated rms noise. We employed a gaussian
convolution (gridding) kernel with FWHM $7.1''$, $\sim 1/3$ of the primary
beam \citep[e.g., see][]{mangum07}. This gridding increases the FWHM of the
effective beam from $\sim 21.3''$ to $\sim 22.5''$ at 115\GHz{}.

After gridding, we fit a second set of third-order polynomial baselines to
each line of sight through the cube. The process of the initial fitting and
gridding is linear, so that these fits represent refinements to our initial
fits after the averaging involved in gridding. We experimented with several
more advanced processing options such as PLAITing \citep{emerson88} and
flagging of standing waves. However the data were very clean and none of
these algorithms improved the quality.

Figure~\ref{fig:moments:30m} presents the reduced, calibrated, gridded
IRAM-30m maps of the \twCO{} \Jone{} and \thCO{} \Jone{} line emission.
The figure displays the spatial distributions of noise, peak temperature,
integrated emission and centroid velocity.
Fig.~\ref{fig:channelmaps:30m:12co10} and~\ref{fig:channelmaps:30m:13co10}
(available in the electronic version only) display the channel maps of
these lines.

\subsection{Combination, imaging and deconvolution}
\label{sec:obs:imaging}

\FigDeconvolution{} %

The interferometric and single-dish data provide us with two data sets,
which sample the high and low spatial frequencies, respectively.  It is
thus possible to produce two different deconvolved results: 1) one obtained
from the interferometric data set alone, and 2) one obtained from the
combination of the interferometric and single-dish data.  While the latter
is the desired final product, as it is sensitive to all measured spatial
scales, the former is often produced because no single-dish measurements
will be available or because they have not yet been acquired. In this
paper, we will present both data cubes to emphasize the amount of flux
which is recovered in the interferometric-only data set.  Moreover, the
angular resolution of the interferometric data is not uniquely defined. It
depends on the weighting scheme chosen. For instance, it is sometimes
useful to produce data cubes at lower angular resolutions to improve the
brightness sensitivity, \ie{}, the sensitivity to extended emission. We
exploit this here in addition to producing the full resolution cube.

This section explains 1) the generic imaging and deconvolution methods used
to produce all these data cubes, 2) how these methods influence the amount
of flux recovered in the interferometric-only data, and 3) the additional
steps required to image jointly the single-dish and interferometric data.

\subsubsection{Generic methods}

Each interferometric pointing was imaged and a single dirty image was built
by linear combination of the 60 individual dirty images.  The dirty image
was then deconvolved using an adaption of the H\"ogbom \texttt{CLEAN}
algorithm. A detailed account of the \GILDAS{}/\MAPPING{} implementation of
the imaging and deconvolution processing of mosaics can be found
in~\citet{pety10}. To help the deconvolution, masks indicating the region
where to search for \texttt{CLEAN} components were defined on individual
channels from the short-spacing data cube. This cube was convolved with a
Gaussian kernel to a final angular resolution of $30''$.  The
\texttt{CLEAN} masks were then defined by all the pixels whose
signal-to-noise ratio was larger than 6. This method was designed to avoid
biasing the deconvolution by defining masks wide enough to encompass all
the detected signal from M51.

The deconvolution of each channel was stopped when a fraction of the
maximum number of clean components were found. This fraction was defined as
the ratio of the area of the current channel mask to the area of the wider
channel mask (see left panel of Fig.~\ref{fig:deconv}). The deconvolution
was assumed to have converged under three conditions.  First, the
cumulative flux as a function of the number of clean components converged
in each channel.  Second, the residual channel images looked like noise.
Both criteria indicated a satisfying convergence of the deconvolution.
Finally, we deconvolved again the data using exactly the same method except
that we doubled the maximum number of clean components.  The subtraction of
both cubes looks like noise.

The list of clean components was regularized with a Gaussian beam and the
residual image was added to obtain the final cube. As all the 30 fields of
the top and bottom mosaics were regularly observed in short cycles of 22.5
minutes, the synthesized beams do not vary inside each mosaic. However, the
two mosaics were observed at different times implying a slight difference
of the synthesized beam between both mosaics (see
Table~\ref{tab:obs:pdbi:main}).  We used the same averaged Gaussian
restoration beam for the northern and southern mosaic. This process is
valid because 1) the Gaussian fits of the synthesized beams are similar as
shown in the top right panel of Fig.~\ref{fig:deconv} and 2) the remaining
flux in the residual image is negligible or undetectable (\ie{}, below the
noise limit). The resulting data cube was then scaled from Jy/beam to
\Tmb{} temperature scale using the restoration beam size.

While the natural velocity channel spacing of the interferometer backend is
3.25\kms{} at 2.6\mm{}, we smoothed the data to a velocity resolution of
5\kms{}. This decreases the effect of correlation between adjacent
frequency channels output by the correlator. This also increases the
signal-to-noise ratio per channel (an important factor for the
deconvolution) and speeds up the processing. Signal is present between
$-110$ and $+110\kms$, relative to NGC\,5194's LSR systemic velocity of
471.7\kms{}. We thus imaged and deconvolved 120 channels producing a
velocity range of $[-297.5,+297.5\kms]$, implying that about two-thirds of
the channels are devoid of signal. On a machine with 2 octa-core processors
and 72 GB of total RAM memory, the deconvolution of the 120 channels up to
a maximum number of 320\,000 clean components typically took 38.5~hours
(human time). The deconvolution duration increases linearly with the
maximum number of clean components per channel.

\subsubsection{PdBI only}
\label{sec:pdbionly}

\FigDeconvolvedFlux{} %

To start, we imaged and deconvolved the PdBI data without the
short-spacings from the IRAM-30m observations. Achieving the convergence of
the deconvolution algorithm at a given angular resolution is
\emph{insufficient} to prove that all the flux was recovered in PdBI-only
data sets. Indeed, the absence of zero spacing implies that the total flux
of the dirty image is zero valued, and it is the deconvolution algorithm
which tries to recover the correct flux of the source. This works only when
the source is small compared to the primary beam and the signal-to-noise is
large enough. No deconvolution algorithm will succeed in recovering the
exact flux of even a point source at low signal-to-noise. Indeed, the
algorithm recovers only the flux which is above a few times the noise rms.
Adding the deconvolution residuals will not help because the dirty beam
(\ie{}, the interferometer response) has a zero valued integral, \ie{}, the
residuals always contain zero flux.

Moreover, a given interferometer needs $2^4$ as much observing time to keep
the same \emph{brightness} sensitivity when just doubling the angular
resolution (assuming similar observing conditions). Such an increase in
observing time is impractical. The brightness sensitivity thus decreases
quickly when the angular resolution improves. In the PAWS case, we
approximately doubled the observing time every time we went to the next
wider interferometer configuration, which typically doubled the angular
resolution.  This allowed us to reach a median noise of 0.4\K{} at full
resolution, \ie{} $1.16''\times 0.97''$ at a position angle (PA) of 73\,deg
(when using natural weighting of the visibilities).  While this is the best
(pre-ALMA) sensivitivity reachable for such a large mosaic, it is also much
higher than the sensitivity of 16\mK{} we reached with the IRAM-30m at an
angular resolution of $22.5''$.  This may mean that faint intensities may
be hidden in the noise of the $1''$-resolution cube.

Multi-resolution \texttt{CLEAN} algorithms (only available for single-field
observations in \GILDAS{} and thus not used for the PAWS mosaic) partly
solve this problem of \emph{brightness} signal-to-noise ratio because the
deconvolution simultaneously happens on dirty images at different
resolutions and thus different brightness noise levels.  Interferometric
brightness noise is a compromise between the synthesized angular resolution
and the time spent in the different configurations. To check what happens
with our deconvolution algorithm, we tapered the visibility weights to
increase the brightness sensitivity at the cost of losing angular
resolution, as this method is to first order similar to a Gaussian
smoothing in the image plane. We choose Gaussian tapering functions such
that we obtained synthesized resolutions of $3''$ and $6''$.
Table~\ref{tab:noise} lists the typical rms noise values for the 3
different resolutions, \ie{}, 0.4, 0.1 and 0.03\K{} at respectively $1''$,
$3''$, and $6''$.

The lower panel of Fig.~\ref{fig:flux:deconv} visualizes the flux found by
the \texttt{CLEAN} algorithm at the 3 different resolutions as a function
of velocity.  The H\"ogbom \texttt{CLEAN} algorithm finds 40\% more flux at
$3''$ than at $1''$.  On the other hand, only 4\% more flux is recovered at
$6''$ than at $3''$ even though the brightness sensitivity increases by a
factor of $\sim2.5$. Since the typical resolution of PdBI at 3\mm{} in its
most compact configuration is $6''$, it means that PdBI reaches its maximum
brightness sensitivity in this configuration.  Hence, further tapering the
data would \emph{not} recover more flux.  Recovering only a marginal
additional amount of flux when going from $3''$ to $6''$ thus implies that
we recovered at these resolution all the flux present in the
interferometric data.

\subsubsection{Hybrid synthesis (PdBI + IRAM-30m)}
\label{sec:hybrid}

The hybrid synthesis is a joint deconvolution of the PdBI and IRAM-30m data
sets. The IRAM-30m and PdBI data sets were first made consistent using the
following 4 steps. 1) We converted the IRAM-30m spectra to main beam
temperatures (\Tmb{}) using the forward and main beam efficiencies (\Feff{}
\& \Beff{}) given in Table~\ref{tab:obs:30m:main}. 2) These spectra were
reprojected on the projection center used for the interferometric data set.
3) The LSR systemic velocity was set to zero to mimic observations at the
redshifted frequency.  4) The velocity axis was resampled to 5\kms{}
channel spacing.  The last transformation introduced some correlation
between channels as the autocorrelator natural channel spacing is only
5.2\kms{} at the \twCO{} \Jone{} frequency.

Following~\citet{rodriguez08}, the \GILDAS{}/\MAPPING{} software and the
single-dish map from the IRAM-30m were used to create the short-spacing
visibilities not sampled by the Plateau de Bure interferometer. In short,
the maps were deconvolved from the IRAM-30m beam in the Fourier plane
before multiplication by the PdBI primary beam in the image plane. After a
last Fourier transform, pseudo-visibilities were sampled between 0 and
15\m{} (the diameter of the PdBI antenna). These visibilities were then
merged with the interferometric observations. The relative weight of the
single-dish versus interferometric data was computed in order to get a
combined weight density in the $uv$ plane close to Gaussian. Since the
Fourier transform of a Gaussian is a Gaussian, and the dirty beam is the
Fourier transform of the weight density, this ensures that the dirty beam
is as close as possible to a Gaussian, making this criterion optimum from
the deconvolution point-of-view. In general, the short spacing frequencies
are small compared to the largest spatial frequency measured by an
interferometer. This implies we can use the linear approximation of a
Gaussian in the vicinity of its maximum, \ie{}, one can assume the Gaussian
is constant in the range of frequencies used for the processing of the
short spacings. We thus need to match the single-dish and interferometric
densities of weights in the inner region of the $uv$ plane.

In practice, we compute the density of weights from the single-dish in a
$uv$ circle of radius $1.25\dprim$ and we match it to the averaged density
of weights from the interferometer in a $uv$ ring between 1.25 and
$2.5\dprim$. Experience shows that this gives the right order of magnitude
for the relative weight and that a large range of relative weight around
this value gives very similar final results~\citep[see \eg{} Fig.~5
of][]{rodriguez08}. In our case, this computation was independently done
for each interferometric pointing to take into account the fact that the
two mosaics used to produce the final image were observed in slightly
different conditions. The relative weight varies typically by 1\% from
pointing to pointing in each mosaic, and by $\sim 6-7\%$ between both
mosaics.

\FigMeanSpectra{} %

Contrary to interferometric-only data sets, the upper panel of
Fig.~\ref{fig:flux:deconv} shows that the deconvolved flux in a hybrid
synthesis is the same for the 3 different angular resolutions. In other
words, the deconvolved flux is independent of the brightness sensitivity
reached, as it is fully constrained by the zero spacing amplitude. Having
\emph{high} signal-to-noise zero spacing data thus ensures that the total
flux inside the deconvolved cube will be the total flux of the single-dish
data in the same field of view. Section~\ref{sec:flux-calibration} checks
this for the PAWS data set. This is linked to the fact that the dirty beam
integral is now normalized to unity. Both the dirty image and the residual
fluxes are meaningful in this case. This enables an additional check of the
convergence of the deconvolution algorithm for the hybrid synthesis
deconvolution: We checked that less than 1.2\% of the clean flux remains in
the residual cube in the $[-100,+100]\kms$ velocity range, even at $1''$
angular resolution.

Fig.~\ref{fig:channelmaps:pdbi+30m:1} and~\ref{fig:channelmaps:pdbi+30m:2}
(available in the electronic version only) display the channel maps of the
$1''$-resolution hybrid (30m+PdBI) synthesis cube at negative and positive
velocities (relative to the systemic velocity), respectively.

\section{A luminous component of extended emission}
\label{sec:ext}

This section presents the unexpected finding that about half the CO
luminosity in the hybrid map arises from a faint, extended component.
Sect.~\ref{sec:ext:existence} and~\ref{sec:ext:errorbeam} demonstrate that
this result is unlikely to reflect an artifact in the data.
Sect.~\ref{sec:ext:decomp} explores how the emission in the hybrid map
breaks apart into a bright, compact and a faint, extended component. It
also compares the structures of these distributions.
Sect.~\ref{sec:ext:thinvsthick} and~\ref{sec:ext:densevsdiffuse} consider
the nature of the CO emission components in light of 1) calculations of
vertical disk structure, and 2) the observed \twCO{}/\thCO{} ratio.
Sect.~\ref{sec:ext:discussion} discusses our interpretation of this
emission component.

\subsection{Verifying the existence of an extended component}
\label{sec:ext:existence}

This section shows that $(50\pm10)\%$ of the total flux in the PAWS field
of view is filtered out by the interferometer. After a thorough discussion
of several checks to prove that this is a true effect, we show that the
filtered emission has typical spatial scales larger than $36''$, \ie{},
$1.3\kpc\,(D/7.6\Mpc)$, where $D$ is the assumed distance of M51.

\subsubsection{Flux recovered in the PdBI-only data cubes}

Fig.~\ref{fig:mean-spectra} compares the spectra averaged over the field of
view of 1) the hybrid synthesis data set (in blue) and 2) the PdBI-only
data set (in red) for the 1, 3, and $6''$ data cubes from top to bottom.
These spectra were obtained as a mean over the PAWS field of view
contracted by about 1 primary beamwidth to decrease the influence of
increasing noise at the map edges. The differences between the hybrid
synthesis and PdBI-only spectra are displayed in white.

The main result is that only half of the total flux is recovered in the
PdBI-only data set. Indeed, only 37\% of the total flux is recovered at
$1''$ but this is attributed to the relatively ``low'' brightness
sensitivity reached at this resolution (see Section~\ref{sec:pdbionly}).
Approximately 50\% of the total flux is recovered both at $3''$ and $6''$,
while the brightness sensitivity differs by a factor of 2.5 between them.
We stress that although the deconvolution recovers 44\% more flux at $6''$
than at $1''$, the difference amounts to only 13\% of the total flux
present in the PAWS field of view.

\subsubsection{Coordinate registration}

We checked the overall registration of the IRAM-30m moment-0 images against
the PdBI-only data. We convolved these ``reference'' maps to the resolution
of the IRAM-30m data. We then repeatedly shifted the images relative to one
another and recorded the cross-correlation between the IRAM-30m data and
the other images. The overall registration of the IRAM-30m with the PdBI
data appears to agree within $\sim 1''$. In addition, we obtained the same
agreement with the BIMA SoNG~\citep{helfer03} and CARMA~\citep{koda11}
data. This good agreement is expected because radio-observatories check the
coordinate registration directly against the radio-quasars used to define
the equatorial coordinate frame.

\FigFluxComp{} %
\TabComp{} %

\subsubsection{Flux calibration}
\label{sec:flux-calibration}

As a test of our calibration strategy, we compared the flux within the
hybrid synthesis data cube with recent CO surveys of M51 by BIMA
\citep{helfer03} and CARMA \citep{koda11}. Although conceptually
straightforward, this comparison must be done with care since each dataset
has different angular resolution, channel width, noise characteristics and
field of view. To obtain the most meaningful comparison, we smoothed the
BIMA, CARMA and hybrid synthesis data cubes to the same angular resolution
as the IRAM-30m data $(22\farcs5)$ and interpolated them onto a common
$(l,m,v)$ grid with square pixels of $4''$ size, and a velocity channel
spacing of 5\kms.  The grid uses a global sinusoid (GLS) projection, and is
centered at RA 13:29:54.09, Dec $+$47:11:38.0 (J2000).

To define the region for the flux comparison, we constructed a mask of
significant emission within the IRAM-30m data cube using the dilated mask
technique described in Appendix~\ref{sec:masking}. In summary, this mask is
obtained by the contiguous extension down to an intensity level of 2 times
the rms noise $\sigma$ of all the pixels whose intensity is above
$5\sigma$. The IRAM-30m $\sigma$'s were estimated for each sightline using
the median absolute deviation. The resulting mask was applied to all four
data cubes.  The total integrated flux within a spatial region
corresponding to the PAWS field of view contracted by $\sim1$ primary beam
width was then computed for each data set.

The results are listed in Table~\ref{tab:comp}.  The flux of the combined
hybrid synthesis cube agrees with the flux of the IRAM-30m to within 1\%,
which is expected since the algorithm that we have used to combine the
interferometer and single-dish data is designed to conserve flux. The
integrated fluxes estimated from the PdBI+30m, CARMA+45m and BIMA+12m
surveys agree to within 10\%, which is the typical accuracy of the absolute
flux calibration of radio observatories at millimeter wavelengths.

Figure~\ref{fig:flux:comp} shows the global spectrum of the four surveys
from the comparison region (bottom panel) and the relative flux difference
per velocity channel between PAWS and the three other data sets (top
panel). In general, the flux differences for individual velocity channels
are consistent with the integrated measurements, \ie{}, there is $\sim$10\%
(5\%) less flux in each channel of the BIMA+12m (CARMA+45m) data cube
compared to PdBI+30m.

These tests confirm the IRAM-30m flux calibration accuracy and its correct
transmission to the hybrid synthesis data set. As a test of the flux
accuracy of the PdBI-only data set, we compared the efficiencies resulting
from the interferometric flux calibration to the well-known PdBI antenna
efficiencies as measured through regular holographies. These agree within
$\sim 10\%$. The key point here is that the flux calibration is relatively
straightforward at 3\mm{}, because the antennas have an excellent
efficiency and the atmosphere was mostly transparent when the data was
acquired (as confirmed by the achieved system temperatures listed in
Table~\ref{tab:obs:pdbi:detail}).

\subsubsection{Amplitude of the brightness Fourier transform as a function 
  of the $uv$ distance}
\label{sec:amplitude}

\FigAmpVsRadiusWide{} %
\FigAmpVsRadiusNarrow{} %

It is important to test how well the hybrid synthesis and the PdBI-only
data cubes agree with each other at high spatial frequencies. To do this,
Fig.~\ref{fig:amp:rad:100m} compares the azimuthal average of the amplitude
of the Fourier transform of the PdBI-only and the hybrid synthesis data
cubes at $6''$ resolution as a function of the $uv$ distance for every
third velocity channel between $[-102.5,+107.5]\kms$.  In such a
representation, the amplitude at the zero $uv$ radius is equal to the total
flux present in the channel map.  We thus normalized all the curves inside
each panel so that their zero-spacing values represent the fraction of the
total flux available in the hybrid synthesis cube at this velocity channel.
For reference, 1) the power spectrum would be computed as the square of
these curves, 2) the dashed vertical lines indicate the minimum radius
directly measured by the interferometer, \ie{}, $\rmin \simeq 15.1\m$.
Moreover, noise and signal behave differently after undergoing a Fourier
transformation.  Noise (\eg{}, signal-free channels at the edges of the
velocity range in the figure) will give a shape consistent with the Fourier
transform of the Gaussian restoration beam~\citep[see Chapter 17
of][]{bracewell00}. On the other hand, signal channels display the product
of Fourier transform of the true source brightness and the Fourier
transform of the Gaussian restoration beam as a function of the $uv$
spacing. Hence, channels with significant line signal fall more quickly
than the Gaussian restoration beam.

The good agreement between the amplitudes from the hybrid and PdBI-only
data at radii larger than \rmin{} provides confidence in the deconvolution
results of the compact sources (whose angular extent is smaller than
$\wavelength/\rmin \sim 35''$) in the PdBI-only data set even though the
short-spacings are missing. Below \rmin{}, the two curves diverge because
the deconvolution can only ``extrapolate'' up to the zero $uv$ radius for
the PdBI-only data set, while it ``interpolates'' for the hybrid synthesis
data set. As the amplitude at zero spacing is proportional to the total
flux, these plots clearly indicate that PdBI-only data miss a large
fraction of the flux even for the $6''$ resolution cube, whose
deconvolution is immune to low signal-to-noise effects.

Moreover, the change of slope visible in Fig.~\ref{fig:amp:rad:100m} for
the hybrid synthesis amplitude between \about{} 5 and 15\m{} is \emph{not}
a processing artifact. Indeed, Fig.~\ref{fig:amp:rad:30m} displays a zoom
of the azimuthal average of the amplitude of the Fourier Transform of the
IRAM-30m and the hybrid synthesis data cubes at $22.5''$ and $6''$
resolution (green and pink curves, respectively). These curves cannot be
directly compared because they result from two different measurement
equations (the single-dish and the interferometric ones). The solution is
well-known: We just have to apply to the IRAM-30m curve the standard
processing steps needed to produce short-spacing visibilities (see
Sect.~\ref{sec:hybrid}). Indeed, after deconvolution of the single-dish
brightness by the $22.5''$ single-dish beam and multiplication with the
$6''$ interferometric primary beam, the amplitudes of the Fourier transform
of both data sets agree well.  Since a Gaussian of $22.5''$ FWHM has a flux
equal to 77, 45, and 21\% of its maximum at 5, 10, and 15\m{}, the observed
change of slope of the deconvolved 30m data cannot be attributed to
imprecision in the deconvolution of the IRAM-30m beam.  This change of
slope in flux is responsible for the failure of the deconvolution to
extrapolate the correct total flux at the zero $uv$ radius for the
PdBI-only data set.

\subsubsection{Impact of the IRAM-30m error beam}
\label{sec:ext:errorbeam}

The beam efficiency of the IRAM-30m telescope at the frequency of the
\twCO{}~\Jone{} line is predicted to be 0.78, based on the interpolation of
the measured efficiencies at 86, 145, 210, 260, and 340\GHz{} using the
Ruze formula\footnote{For details, see
  \url{http://www.iram.es/IRAMES/mainWiki/Iram30mEfficiencies}}.  Moreover,
the forward efficiency is 0.95 at the same frequency.  This implies that
about 22\% of the measured flux in a given direction results from beam
pick-up from solid angles outside the main beam. Of this amount, 5\%
originates from the rear lobes, which mainly collects diffuse emission from
the telescope backside. Hence, about 17\% of the measured flux is picked up
by the rings of the telescope diffraction pattern and the error beams due
to the limited surface accuracy of the telescope. In
Appendix~\ref{sec:errorbeam}, we assess in-depth how much these effects
contribute to the extended emission. Here, we summarize the main results.

We estimate that the peak brightness due to the error beam contribution is
at most 55\mK{}, \ie{}, about 3.5 times the median noise level of the
IRAM-30m observations. The median value of the error beam contribution is
19\mK{}, while pixels brighter than 5 times the noise level have a median
of 140\mK{} in the extended emission measured at $6''$ (see
Sect.~\ref{sec:noise+signal}). The emission associated with the error beams
has a very different signature from that of extended emission both in space
and velocity. In particular, the putative error beam contribution
translates into much wider lines than actually observed.  Moreover, the
typical angular scales of the error beams are large, implying that the flux
is scattered at low brightness level over wide regions of the sky. Finally,
we estimate that at most 10\% of the total flux in the PAWS field of view
can be due to error beam scattering.

Taken together, these arguments lead us to the conclusion that the IRAM-30m
error beam cannot explain the presence and properties of the extended
emission.

\subsubsection{About half the CO luminosity arises from an extended 
  emission component}

\FigAmpVsRadiusDiff{} %

We now consider why $(50\pm10)\%$ of the total flux is missing in the
PdBI-only data set. The direct interpretation is that this missing flux has
been filtered out because the PdBI does not measure visibilities at
spacings shorter than $\sim15\m$, \ie{} the well-known short-spacing
problem.  We now wish to quantify the spatial scales at which the emission
filtered by the PdBI is structured. To do this, Fig.~\ref{fig:amp:rad:diff}
compares as a function of the $uv$ distance 1) the amplitude of the
\emph{difference} of the Fourier transforms between the hybrid synthesis
and the PdBI-only data cubes (red plain curve), with 2) the noise of the
hybrid synthesis Fourier transform (pink dashed curve), computed by
averaging the signal-free channels. For all the velocity channels showing
some signal, the amplitude of PdBI-filtered emission is (much) larger than
the noise level only up to a radius of $\sim 15\m$. This implies that a
very large fraction of the missing flux is structured only at spatial
scales larger than $\sim \wavelength/\rmin = 36''$. However, the difference
and noise curves stay close to each others up to $35\m$ for a few velocity
channels (\eg{}, $v=-42.5$ or $+32.5\kms$). This indicates there is
probably a small fraction of the missing flux which is structured at
spatial scales between $15''$ and $36''$. In the following, we will thus
state that the missing flux is structured mostly at spatial scales larger
than $\sim \wavelength/15\m = 36''$ or $1.3\kpc\,(D/7.6\Mpc)$, where $D$ is
the distance of M51.  Conversely, this also means that the flux recovered
in the PdBI-only cubes is structured mostly at spatial scales smaller than
$36''$.

\subsection{Structure of compact and extended emissions}
\label{sec:ext:decomp}

In order to investigate the nature of the emission filtered by the PdBI
interferometer, we need to image it. To do this, we subtracted at each
angular resolution (1, 3 and $6''$) the PdBI-only from the hybrid synthesis
data cubes to determine the properties of the emission filtered by the
PdBI. We were careful to image both data sets on the same spatial and
spectral grid. We also used the same weighting scheme, deconvolution
method, stopping criterion and restoration beam.

In the following, we will refer to the 3 cubes as hybrid synthesis
(PdBI+30m), PdBI-only, and subtracted (\ie{}, hybrid synthesis minus
PdBI-only) cubes. The PdBI-only and subtracted cubes are equivalent to a
decomposition of the hybrid synthesis signal into two kinds of source
morphology: Compact (angular scales $\la 36''$) and extended (angular
scales $\ga 36''$) sources. This decomposition is a convenient instrumental
side effect. As such, it is arbitrary and only a multi-scale analysis would
deliver the full spatial distribution of the emission.  However,
Fig.~\ref{fig:amp:rad:100m} and~\ref{fig:amp:rad:30m} show that the Fourier
transform amplitude changes its slope between \about{} 5 and 15\m{}. Thus,
the intrinsic distribution of spatial scales of M51 is such that our
decomposition makes sense.

For completeness, we carried out this decomposition for the 1, 3 and $6''$
cubes. While this gives a taste of multi-scale approaches, two of these
cubes stand out. The PdBI-only $1''$ data cube is the one where the
``compact'' sources are best resolved while the subtracted $6''$ data cube
is the one which gives the most accurate description of the extended
component.  Finally, we smoothed the PdBI-only $6''$ cube to the resolution
of the IRAM-30m data cube $(22.5'')$ and we subtracted them from the 30m
data.  This allows us to obtain an idea of the relative contribution of
both emission components at the typical resolution of single-dish
observations as discussed in Sect.~\ref{sec:diffuse:m51}.

In the rest of this section, we first describe the statistical
distributions of noise and signal for the different data cubes, as this
allows us to discuss how beam dilutions and signal-to-noise ratios evolve
with angular scale.  We then comment on the 2-dimensional spatial
distributions of the line moments. We also present their azimuthal
averages, which we will later use to analyse the vertical structure of both
components.  Finally, we show the kinematics of the two components along
the M51 major axis.

\subsubsection{Noise and signal distributions}
\label{sec:noise+signal}

\FigNoiFilling{} %
\FigSigFilling{} %

In order to quantify the $(l,m,v)$-volume filling factors and the beam
dilution properties of the compact and extended emission, we now describe
the distribution of the noise and brightness values.

Figure~\ref{fig:filling:noi} displays the cumulative histograms of the rms
noise for the 4 different resolutions (1, 3, 6, and $22.5''$), and for the
3 kind of cubes: Hybrid synthesis (top), PdBI-only (middle) and subtracted
(bottom).  The left column displays the raw histograms while the right
column presents the histograms normalized by their median value, as this
eases the comparison of shapes. The histograms show that the noise
distributions are well centered on the median value, implying relatively
uniform noise properties.  The IRAM-30m single-dish data display an
increase of the histogram population at high noise values because the noise
distributions are homogeneous over the inner $400''\times400''$ field of
view but they increase on the northern and southern patch.  A similar
effect is seen for the interferometric data, which is directly linked to a
noise increase at the edges of the field because of the correction of
primary beam attenuation.  Moreover, in the single-dish data, the \twCO{}
median noise level (16\mK{}) is about twice as high as the \thCO{} one
(7.5\mK{}) because the \twCO{} \Jone{} line is closer to an oxygen
atmospheric line.

The brightness noise levels of the hybrid synthesis data cubes are mostly
set by the interferometric radiometric noise because the increase of
integration time from the IRAM-30m to the PdBI does not match the gain in
angular resolution (see Sect.~\ref{sec:pdbionly}). Hence, the median rms
noise (see Table~\ref{tab:noise}) and the noise cumulative histograms (see
Fig.~\ref{fig:filling:noi}) are almost identical for the PdBI-only and
hybrid synthesis data cubes. On the other hand, the brightness noise levels
for the subtracted data cubes are mostly set by the single-dish data.
Indeed, we subtracted two data sets whose noise is partially correlated:
The noise coming from the PdBI visibilities is common.  Subtracting both
data cubes result in noise properties close to the IRAM-30m noise
properties. In particular, the median rms noise in the subtracted data cube
is more than one order of magnitude smaller than the median rms noise of
the hybrid synthesis cube at $1''$ resolution.

Figure~\ref{fig:filling:sig} shows the cumulative histograms of the cube
brightness using a similar layout as Fig.~\ref{fig:filling:noi}. In the
right column, the histograms are normalized by the maximum brightness.  The
histograms were computed using the full PAWS field of view but only the
parts above the $5\sigma$ brightness noise level are displayed. The hybrid
synthesis and PdBI-only histograms are similar both qualitatively and
quantitatively. The histograms are displaced towards higher brightness
values when the angular resolution increases, implying that all the
structures above the $5\sigma$ noise levels experience large beam dilution
effects.  For example, the maximum brightness increases by more than one
order of magnitude from 1.3 to 16\K{}, when increasing the resolution from
22.5 to $1''$. On the other hand, the histograms of the subtracted cubes
are identical within the noise constraints (even for the normalized
histogram of the $1''$ cube as the value of the maximum brightness is
relatively uncertain). Indeed, the maximum brightness of the extended
component evolves only from $0.97\pm0.07$ to $1.36\pm0.07\K$ from $22.5''$
(the IRAM-30m resolution) to $6''$ (the PAWS resolution for which the
extended emission is best defined). This implies that beam dilution is
negligible for the subtracted emission. This is consistent with this
emission being structured mostly at spatial scales larger than $36''$.

The $(l,m,v)$-volume filling factors and the median brightness values were
computed on all the cube pixels with detected signal, \ie{} their
brightness is larger than 5 times their noise. Using this definition, the
brightness of the compact component (measured on the $1''$ PdBI-only cube)
ranges from 2.0 to 16.0\K{}, with a median value of 2.5\K{} and it fills
less than 2\% of the $(l,m,v)$-volume. The brightness of the extended
component (measured on the $6''$ subtracted cube) ranges from 0.07 to
1.36\K{}, with a median value of 0.14\K{} and it fills $\sim30\%$ of the
$(l,m,v)$-volume.

The comparison of the noise and signal distributions of the cubes at
different angular resolutions makes it clear why the deconvolution of the
PdBI-only data recovers more flux at $3''$ than $1''$ but almost the same
flux at $6''$ and $3''$. Indeed, the subtracted emission has a median
brightness of 0.14\K{}, which is $\sim0.35$ times the hybrid synthesis
brightness noise level at $1''$ but $\sim1.3$ and 4 times the hybrid
synthesis noise level at $3''$ and $6''$, respectively. In other words, any
emission present in the PdBI-only data is only recovered when its
signal-to-noise is large enough (typically $\ge5$).

\subsubsection{Spatial distribution of the line moments}

\FigPAWSmomentsMixed{} %

Figure~\ref{fig:moments:paws:mixed} summarizes the properties of the
decomposition of the total emission into the compact and the extended
emission. The spatial distributions of the peak temperature, line
integrated emission (the 0th order moment), centroid velocity (the 1st
order moment), line full width at half maximum (computed as 2.35 times the
line 2nd order moment), and the noise are presented from top to bottom.
The total and compact emissions are presented at the best possible PAWS
resolution ($1''$), while the extended emission is displayed at the
resolution where it is best measured, \ie{}, $6''$.
Figures~\ref{fig:moments:paws:full},~\ref{fig:moments:paws:tap1},~\ref{fig:moments:paws:tap2},
and~\ref{fig:moments:30m:tap2} (available in the electronic version only)
show the same decomposition at fixed angular resolutions from 1, 3, 6, and
$22.5''$, respectively. This demonstrates how the different moments of each
component of the emission vary with spatial resolution.

The deconvolved intensity distribution is corrected for primary beam
attenuation, which makes the noise level spatially inhomogeneous. In
particular, the noise strongly increases near the edges of the field of
view (see, \eg{}, panels [m] to [o] of Fig.~\ref{fig:moments:paws:mixed}).
To limit this effect, the deconvolved mosaic is truncated at its edge,
giving an almost parallelogram field of view of $\sim10.5$ square
arcminutes. A comparison with the IRAM-30m data
(Fig.~\ref{fig:moments:30m:tap2}[a-f]) shows that signal is present near or
slightly beyond the edges of the PAWS field of view. This implies that the
signal at the edge of the PAWS field of view is probably less well
deconvolved from the contribution of emission outside the PAWS field of
view. This effect is more important at the northern edge and to a lesser
extent at the southern edge than at the western and eastern edges.

Two kinds of artifact appear in the peak temperature and integrated
intensity maps of the $1''$ subtracted cube, which shows the extended
component. First, a moire effect due to the undersampling of the field
pointings in the mosaic appears as a slight modulation of the intensity at
a typical spatial scale of $\sim18''$ (see
Fig.~\ref{fig:moments:paws:full}[f]). This is due to power aliasing in the
$uv$ plane~\citep{pety10}. Second, the chicken-pox aspect at a spatial
scale close to the synthesized resolution is a known artifact of the
deconvolution method that we used (see
Fig.~\ref{fig:moments:paws:full}[c]). The overall spatial repartition of
the extended component is nevertheless correct as evidenced by the
comparison with the spatial distributions of the moments at $3''$ and
$6''$, where the impact of these artifacts becomes negligible.

The subtracted cube reveals extended emission whose peak temperature
distribution is barely detected in the $1''$ hybrid synthesis cube, as the
signal-to-noise ratio of this emission ranges between 0.2 and 3.5 (see
Fig.~\ref{fig:moments:paws:mixed}[a-c]). This is why a relatively complex
dedicated masking technique was devised to compute meaningful 1st and 2nd
order moments for the $1''$ hybrid synthesis cube (see
Section~\ref{sec:masking} for a detailed description). The peak temperature
(Fig.~\ref{fig:moments:paws:mixed}[c]) and integrated emission
(Fig.~\ref{fig:moments:paws:mixed}[f]) maps are maximum along the major
axis of the galaxy. This is expected because emission in a given velocity
channel extends over a large 2D area near the major axis, while is mostly
extended in one spatial dimension along the minor axis. Hence, the
interferometer will recover ``extended'' emission along the minor axis much
better than along the major axis. This projection effect thus minimizes
emission along the minor axis in the subtracted cube.  However, there is
more than a major axis trend in the subtracted cube.  Indeed,
Figure~\ref{fig:channelmaps:extended-vs-combined:1}
and~\ref{fig:channelmaps:extended-vs-combined:2} (available in the
electronic version only) show an overlay of the signal-to-noise contours of
the $1''$ PdBI-only data cube onto the signal of the $6''$ subtracted cube
for a set of channels at negative and positive velocities. These figures
suggest that the extended emission fills the central $55''$, bounded by the
inner edge of the spiral arms, and then falls on the convex side of the
arms at larger radii (out to $\sim85''$).

While the peak temperature map exhibits a symmetric spatial distribution
relative to the galaxy center, the integrated emission peaks in the
southern part. Extended emission is completely absent in the
$1'\times1'$-areas located roughly southwest $(-60'',-20'')$ and northeast
$(+60'',+40'')$ of the center. The maps of centroid velocity indicate
differences between the kinematics of the compact and extended emission.
This is best seen when following the 0\kms{}-isovelocity line, \ie{},
NGC\,5194's systemic velocity, on Fig.~\ref{fig:moments:paws:tap2}[h and i]
or on Fig.~\ref{fig:moments:30m:tap2}[h and i]. The linewidth of the
extended component (Fig.~\ref{fig:moments:paws:mixed}[l]) is largest inside
a central circle of $\sim 35''$ radius. Linewidths are on average much
larger for the extended than for the compact emission. The clearest
exception is at the galaxy center, \ie{}, at radii smaller than $2.5''$,
where the compact emission has high peak temperature, large linewidth, and
large integrated emission. This is reminiscent of the properties of
molecular gas in the inner 180\pc{} of our Galaxy~\citep{morris96}.
Alternatively, \citet{kohno96} and \citet{matsushita04} interpret these as
gas being entrained by the AGN radio jet.

We verified that the large linewidth of the extended component is not
caused by the contribution from the error beams. Details are provided in
the Appendix~\ref{sec:errorbeam:results}.

\subsubsection{Azimuthal averages}
\label{sec:azimuthal-averages}

\FigPAWSaverages{} %
\FigPVmajor{} %

Figure~\ref{fig:aver:paws} shows the azimuthal average (and the associated
dispersion) around the kinematic center of the deprojected images of the
peak temperature, the integrated emission, the rotational velocity, the
modulus of the centroid velocity gradient, and the line full width at half
maximum computed for the $1''$ PdBI-only (blue curves), the $1''$
subtracted (red curves), and the $22.5''$ IRAM-30m (green curves) cubes.

For the compact emission (PdBI-only cube), the inner $5''$ clearly display
high peak temperatures and large FWHMs, implying large integrated line
emissions.  The molecular ring dominates from $\sim 10$ to $40''$, where
the integrated emission and the peak temperatures are larger than in the
disk (radii larger than $\sim 40''$). The peak temperature seems to
increase from the ring to the outer disk, while the integrated emission
stays mostly constant outside $r\sim80''$.  On the other hand, the velocity
FWHM decreases slightly from $\sim 25\kms$ at a radius of $\sim 30''$ to
$20\kms$ at radii larger than $\sim 100''$.

The extended emission (subtracted cube) has a typical peak temperature of
0.75\K{} in the central region and 0.5\K{} in the disk. The FWHM decreases
by a factor of 2 from $\sim 100\kms$ at $0''$ to $\sim 45\kms$ at $50''$
and it then varies between 40 and 50\kms{} in the outer disk.  Both
properties result in a regular decrease of the integrated emission from
$\sim 45\Kkms$ at the center to $\sim 15\Kkms$ at $50''$. It then varies
between 10 and 15\Kkms{}.

The compact emission has on average a peak temperature twice as large as
the extended emission. In contrast, the extended emission has a velocity
FWHM at least twice as large as the compact emission (except near the
center). Both effects almost compensate to yield similar integrated line
emissions for both components.

The dispersion of the peak temperature and of the integrated emission is
larger for the compact emission than for the extended one. This is a
consequence of the fact that the compact emission is structured at all
scales down to or below the angular resolution while the filtered flux is
structured mostly at scales larger than $36''$. The dispersion of the FWHM
measurement is similar in both the compact and extended emissions. Indeed,
its azimuthal average is computed only where there is enough signal to
define it (\ie{}, on a small fraction of $360\degr$), while the peak
temperature and integrated emission are averaged over $360\degr$ (at least
up to a radius of $85''$).

The rotation velocity of each component was measured by fitting tilted
rings with fixed systemic velocity to the line-of-sight velocity field
using the \texttt{GIPSY} task \texttt{ROTCUR}. In both cases, we assume the
kinematic center listed in Table~\ref{tab:m51}, and we adopt a constant
position angle $(173^\circ{})$ and inclination $(21^\circ)$, as estimated
from the more radially extended \HI{} emission mapped at lower resolution
by the THINGS project~\citep[see][for more details]{colombo12b}. The middle
black curve, labeled ``Model'', is a 3-parameter fit of the measured
rotation curve.  The inner part of this fit (inside $100''$) compares well
with what we would expect if the stars (traced at 3.6\mum{}) dominate the
baryonic mass~\citep{meidt12b}.

The rotational velocity of the extended emission (red curve) is increasing
almost monotonically with radius, while it oscillates twice for the compact
emission (blue curve) as an effect of the streaming motions and corotation
resonances (and not the bulge).  Moreover, the centroid velocity of the
extended emission is typically closer to NGC\,5194's systemic velocity by
50\kms{} at radii smaller than $35''$ where an inner stellar bar dominates
the dynamics. At larger radii, it overlaps the rotational velocity curve of
the compact emission. The modulus of the centroid velocity gradient is
around 6 and $10\kms/''$ for the extended and compact components,
respectively.  Hence, the kinematics of the extended emission vary much
more smoothly on the plane of the sky than the kinematics of the compact
emission, which is very much affected by streaming motions (This is further
discussed in Sect.~\ref{sec:scale-height:application}).

Any intrinsic behavior beyond a radius of $85''$ must be interpreted with
caution as the azimuthal averages reach the edges of the field of view in
its smallest dimension. Two effects happen: 1) The noise increases sharply
at the mosaic edges (see the bottom left panel of
Fig.~\ref{fig:moments:paws:mixed}), and 2) the azimuthal averages miss the
outside interarm regions, which occupy a larger and larger fraction of the
area as the distance from the center increases. However, the comparison of
the averages of the extended and compact emission at each radius is
meaningful as the averages are made on the same ellipse portions.

\subsubsection{Kinematics along the major axis}

Figure~\ref{fig:pv:major} compares the position-velocity diagrams along the
major axis of the compact emission (PdBI-only, green contours) and the
extended (grey image) emission. The red curve is the measured rotation
curve. It matches well the overall velocity variation along the major axis.
The middle blue curve is the 3-parameter fit of the measured rotation
curve, implying an overall inclination of the galaxy on the plane of sky of
$21\pm3\degr$ (see Sect.~\ref{sec:azimuthal-averages}). The two other blue
curves show the same velocity model for inclinations of $15\degr$ and
$27\degr$, in order to give an indication of the effect of inclination on
the kinematics.

This diagram confirms that the linewidth is much larger for the extended
than for the compact emission (with the possible exception of the molecular
gas at the center of the galaxy). The extended emission has a parallelogram
shape with gas emitting at forbidden velocity in the $[-35'',+35'']$ radius
range.  This is a typical signature of nuclear bar
kinematics~\citep[\eg{},][]{binney91,garciaburillo95,garciaburillo99}. The
distribution of the extended emission in PV-space might in addition or
otherwise indicate that it lags the compact emission.  The parallelogram
shape is not symmetric, and emission is absent in the region near $(+15'',
-25\kms)$.  At and inside this position (and its mirror, near $-15''$),
emission in the extended component that falls below the rotational
velocities exhibited by the compact emission (red curve) might arise with a
genuine lag.

\subsection{Interpretation 1: Two CO disks -- Thin and thick}
\label{sec:ext:thinvsthick}

After an intermediate summary of the main observational properties of the
two CO emission components, we will translate them into physical properties
of the gas traced by the \twCO{} \Jone{} emission. To do this, we will
first summarize the expressions for the gas scale height, mid-plane
pressure and gas density as a function of the gas and stellar surface
densities and vertical velocity dispersions~\citep{koyama09}. We will then
discuss how to apply these expressions to M51. In particular, we will show
how the contribution of the streaming motions to the CO linewidth can be
estimated to derive an accurate vertical gas velocity dispersion.

\subsubsection{Intermediate summary and consistency checks}

The emission filtered out by the PdBI interferometer accounts for about
half of the total flux imaged in the hybrid (PdBI+30m) synthesis.  The
subtraction of the PdBI-only from the hybrid synthesis cubes shows \twCO{}
\Jone{} emission mostly structured at angular scales larger than $36''$,
\ie{}, $\sim1.3\kpc$. Its brightness temperature ranges from 0.1 to 1.4\K{}
with a median value of 0.14\K{}. It covers about 30\% of the PAWS field of
view.  Its spatial distribution surrounds the bright spiral arms.  While
the integrated line emission peaks at 48\Kkms{} near the major axis in the
south-western galaxy quadrant, the peak brightness along the major axis
ranges from 0.7 to 1.4\K{} in the northeast and 0.5 to 0.8\K{} in the
south-western quadrant. This emission is thus faint and extended. In
contrast, the emission observed by the interferometer is compact and
bright. Indeed, its brightness temperature ranges from 2 to 16\K{} with a
median value of 2.5\K{} and it covers less than 2\% of the PAWS field of
view.  Rotational velocities estimated from the centroid velocity map of
the extended component are closer by $50-100\kms$ to the systemic velocity
than the ones of the compact component inside a circle of $35''$ radius.
Outside this radius, the rotational velocities of both components have the
same order of magnitude.  But, the rotational velocity curve of the compact
component oscillates around the modeled velocities, while the rotational
velocity curve of the extended component smoothly increases and mostly lies
below the modeled velocities. The line FWHM of the extended component is
twice as large as that of the compact component.

\FigHIstackedSpectra{} %
\TabHIstackedFit{} %

We made two additional consistency checks. First, \texttt{CPROPS}
decomposes the $1''$ hybrid synthesis cube into two different components:
1) ``Clouds'' which account for 55\% of the total flux, and 2)
``Intercloud'' gas which ``surrounds'' the GMCs and accounts for the
remaining flux~\citep{colombo12a}. These numbers remain stable when the
analysis is done using the $6''$ hybrid synthesis cube.  This decomposition
result reflects the fact that we have a faint extended component in
addition to the bright compact \twCO{} emission. Second, we checked whether
existing \HI{} observations are consistent with the possibility of having a
narrow and broad linewidth component. We started from the THINGS \HI{} cube
imaged at $11.9''\times10.0''$ resolution with natural
weighting~\citep{walter08}, as it maximizes the signal-to-noise ratio.  We
applied the shuffle method of \citet{schruba11}, \ie{}, we shifted each
spectrum of the \HI{} cube to a common velocity scale by removing the
systematic velocity field structure as measured from the centroid velocity.
The spectra were then averaged over the PAWS field of view.  Only a dual
Gaussian can accurately fit the \HI{} spectrum.
Figure~\ref{fig:stacked:hi} displays the shuffled, averaged spectrum, its
Gaussian decomposition and the residuals.  Table~\ref{tab:stacked:fit:hi}
presents the quantitative results of the dual Gaussian fit. The total flux
is divided approximately equally in both components while the FWHM is
approximately twice as large in one of the components. This is consistent
with a separation into two emission components with very different
linewidths.

\subsubsection{Expressions for the gas scale heights,
  mid-plane pressures, and gas densities}

Using the vertical momentum and Poisson equations averaged over the
horizontal plane of the galaxy, \citet{koyama09} obtained a second-order
differential equation for the averaged vertical density profile. Solving
it, they showed the averaged gas density (\rhogas{}) and pressure (\Pgas{})
are approximately given by Gaussian profiles of the height $z$, \ie{},
\begin{equation}
  \rhogas(z) = \rho_0\,\exp\paren{-\frac{z^2}{2\Height{}^2}}
  \quad \mbox{and} \quad
  \Pgas(z) = P_0\,\exp\paren{-\frac{z^2}{2\Height{}^2}}.
\end{equation}
In these equations, the gas scale height \Height{} is given by
\begin{equation}
  \Height{} = 
  \frac{\sigvz}{\sqrt{4\pi\Grav\,\rhostar}} \frac{1}{\A + \sqrt{\A^2+1}},
\end{equation}
where \sigvz{} is the thermal plus turbulent velocity dispersion
perpendicular to the galactic disk, \rhostar{} is the stellar density,
\Grav{} is the gravitational constant, and \A{} is a dimensionless factor
that measures the relative densities of the gaseous and stellar disks. It
can be expressed as
\begin{equation}
  \A = \sqrt{\frac{\Grav \, \siggas^2}{2 \rhostar \, \sigvz^2}},
\end{equation}
where \siggas{} is the gas surface density. Once the gas vertical scale
height is known, the gas mid-plane density and pressure can easily be
derived with
\begin{equation}
  \rho_0 = \frac{\siggas}{\sqrt{2\pi}\,\Height{}} 
  \quad \mbox{and} \quad
  P_0 = \sigvz^2\,\rhogas_0.
\end{equation}
These expressions for the gas scale height, mid-plane pressure, and
mid-plane density take into account 1) gravity forces that both the stars
and gas exert, and 2) turbulent and thermal hydrodynamic pressures.
However, they still are lower limits as they neglect any contribution from
the magnetic field.

Finally, for an isothermal, self-gravitating stellar disk, the stellar
surface density, the stellar volume density, the stellar vertical scale
height, \Height{\star}, and the stellar vertical velocity dispersion,
\sigvstar{}, are linked via
\begin{equation}
  \label{eq:isothermal-disk}
  \rhostar = \frac{\sigstar}{2\Height{\star}},
  \quad \mbox{and} \quad
  \Height{\star} = \frac{\sigvstar^2}{\pi\Grav\,\sigstar}.
\end{equation}
The A factor can then be rewritten as
\begin{equation}
  \label{eq:toomre}
  A \sim \frac{\Qstar}{\Qgas},
\end{equation}
where \Qstar{} and \Qgas{} are the Toomre's gravitational stability
parameters for the stellar and gas disks.

\subsubsection{Application to M51}
\label{sec:scale-height:application}

Figure~\ref{fig:scale-height} displays how the previous expressions are
applied to the case of M51 as a function of galactocentric radius for the
total gas at $22.5''$ resolution (green curves), the extended component at
$6''$ resolution (red curves) and the compact component at $1''$ resolution
(blue curves). Radial zones where the results should be interpreted with
caution are highlighted in grey: 1) at radii larger than $85''$, the
azimuthal averages start to reach the edges of the observed field of view;
and 2) from 0 to $\sim45''$, the assumption that the stars are distributed
in a disk breaks down, so that the stellar scale height and volume density
are not well constrained (see below).

\FigScaleHeight{} %
\FigFWHMvsCVI{} %

The gas mass surface densities for each component are computed from the
azimuthal averages of the \twCO{} \Jone{} integrated emission using the
Galactic value of the \Xco{} factor and taking into account the presence of
helium. The stellar surface density is derived from the 3.6\mum{}
emission~\citep{meidt12a}. The stellar velocity is computed following
\citet{bottema93} and \citet{boissier03} who showed that, for a flat
exponential disk, it falls off from the central dispersion $\sigma_0$
according to
\begin{equation}
  \sigvstar = \sigma_0\,\exp\paren{-\frac{r}{2\,H_B}},
\end{equation}
where $r$ is the galaxy radius and $H_B$ is the disk scale length of the
$B$-band. \citet{mcelroy95} estimates the central stellar velocity
dispersion in M51 to be $\sigma_0 = 113\kms$ and \citet{trewhella00}
estimates that $H_B=2.82\kpc$. This gives us an estimate for the stellar
scale height and volume density, according to Eq.~\ref{eq:isothermal-disk}.
We emphasize that this assumes an isothermal, self-gravitating stellar
disk. This assumption clearly breaks down in the center of M51, where the
bulge and nuclear bar dominate, \ie{}, at radii less than $\sim25''$ where
the 3.6\mum{} surface brightness profile exhibits steepening, and near the
location of the bar corotation radius as estimated from gravitational
torques~\citep[][]{meidt12b}. In fact, the application of
Eq.~\ref{eq:isothermal-disk} results in a stellar scale height that
decreases with radius from $\sim45''$ to the galaxy center.  We therefore
opted to adopt a constant, lower limit to the scale height inside this zone
by extrapolating the value given by Eq.~\ref{eq:isothermal-disk} at a
radius of $45''$ inward.

The vertical velocity dispersion of the gas can be estimated as the line
2nd order moment for a face-on galaxy. The inclination of M51 onto the line
of sight is small but non-zero, implying that the line 2nd order moment is
only a first order approximation.  Indeed, the systematic motions averaged
inside the beam of the observations contribute to the line 2nd order
moment. This is more problematic in the case of M51 because the streaming
motions are known to be large for this galaxy.  Appendix~\ref{sec:vz:disp}
shows that the vertical velocity dispersion can be estimated as
\begin{equation}
  \sigvz^2 \sim \aver{\paren{\vobs-\vcent}^2} - \bracket{\cvi{} \, \frac{\beamwidth}{2.35}}^2,
\end{equation}
where \aver{} symbolizes the brightness-weighted average over the line
profile, \vobs{} is the velocity projected along the line of sight,
\vcent{} the line centroid velocity, \cvi{} the modulus of the gradient of
the centroid velocity, and \beamwidth{} the resolution beamwidth of the
observations. In this equation, the first term is the square of the second
moment and the second term estimates the contribution of unresolved
systematic motions.

Figure~\ref{fig:fwhm:cvi} diplays the joint distributions of these two
quantities for the IRAM-30m cube and the $1''$ PdBI-only and subtracted
cubes. At the resolution of the IRAM-30m data, the unresolved systematic
motions contribute significantly to the value of the second moment. This
behavior is clearly split in the decomposition between compact and extended
components at $1''$. The unresolved systematic motions are negligible for
the extended component. For the compact component, they contribute to less
than 34\% of the linewidth for 50\% of the data. Hence, the streaming
motions are seen in the compact component but they are not seen in the
extended component. We stress that the subtracted cube measures the
extended component at an angular resolution of $1''$ because it results
from the subtraction of two cubes whose resolution is $1''$, namely the
hybrid synthesis and the PdBI-only cubes. Beam smearing of unresolved
systematic motions must therefore be considered only at angular scales
lower than $1''$. As a corollary, if the large linewidths are due to beam
smearing of unresolved streaming motions, then the linewidths should
decrease when increasing the imaging angular resolution. This effect is
observed in both the hybrid synthesis and PdBI-only cubes (panels [j] and
[k] of Fig.~\ref{fig:moments:paws:full} to~\ref{fig:moments:paws:tap2}),
while the second moment of the subtracted cube (panels [l] of the same
figures) stays basically constant when increasing the angular resolution
from $22.5''$ to $1''$. We thus deduced that the large linewidths of the
extended component are not caused by unresolved streaming motions.

As the unresolved systematic motion can be larger than the second moment
for the compact component, we only used the second moments to compute the
vertical velocity dispersion, implying that the scale heights and mid-plane
pressures derived are slightly more robust for the extended component than
for the compact component and the IRAM-30m data.

\subsubsection{Results}
\label{sec:ext:thinvsthick:results}

Using these inputs, we computed the stellar vertical scale height, and
volume density. We also computed the gas-to-stellar density ratio (\A{}
factor), the gas mid-plane pressures, densities, and scale heights for all
the molecular gas (using the IRAM-30m data), the compact component (using
the PdBI-only data), and the extended component (using the subtracted
cube). We neglect the contribution from atomic gas traced by \HI{}
emission, as this gas represents only between 2.5 and 30\% of the molecular
gas mass for radii from 0 to $130''$~\citep[see Fig.~42
of][]{leroy08,schuster07}.

Our estimates agree very well with expectations. For example, the A factor
is less than 1 for the extended component, while it is equal to $\sim2.5$
for the compact component for radii larger than 0.5\kpc{}.  Using
Eq.~\ref{eq:toomre}, this implies that $Q_\emr{compact} \sim 0.3
Q_\emr{extended}$, where $Q_\emr{compact}$ and $Q_\emr{extended}$ are the
Toomre factors for the compact and extended components, respectively. Said
otherwise, the compact component is more likely to form stars than the
extended component. The gas mid-plane pressure is about the same for both
components and for the total gas. This probably reflects pressure
equilibrium. From the mid-plane pressure, we estimate a molecular fraction
$\Sigma_{\HH}/\Sigma_{\HI}$ close to what is observed: Using the empirical
formula which relates the molecular fraction $\Sigma_{\HH}/\Sigma_{\HI}$ to
the mid-plane pressure~\citep[]{blitz06}
\begin{equation}
  \Sigma_{\HH}/\Sigma_{\HI} = \paren{\frac{P_0}{4.3\times10^4\Kpccm}}^{0.92},
\end{equation}
we find a molecular fraction of $\sim26$ and $\sim4$ at radii equal to 1.0
and 2.5\kpc{}, respectively, in good agreement with the results
of~\citet{leroy08}. Based on the high molecular fraction, we thus expect a
tight correlation between the star formation rate and the molecular gas
surface density, but probably little relation to the atomic gas surface
density, as found by~\citet{kennicutt07}.

The scale height of the extended component is almost constant, its value
varying slighty betwen 190 and 250\pc{} for radii larger than 0.5\kpc{}.
The scale height of the compact component varies more. It is $\sim40\pc$
between $\sim1.5$ and 3.5\kpc{}. It varies between $\sim20$ and 40\pc{}
beyond $\sim3.5\kpc$ and it decreases to $\sim10\pc$ at $\sim0.5\kpc$. In
other words, the scale height of the compact emission is 5-6 times smaller
than the scale height of the extended component from $\sim1.5$ to
$3.5\kpc$. The compact component is between 5 and 20 times (typically 10
times) denser than the extended component. At a radius of 2.5\kpc{}, the
extended and compact components have an average volume density of $\sim1$
and 10\,\HH\,\pccm{}, respectively. For reference, the distribution of the
molecular gas density in the Solar Neighborhood is~\citep{ferriere01,cox05}
\begin{equation}
  \frac{n}{\HH\,\pccm} = 0.29\,\exp\bracket{-(z/81\emr{pc})^2},
\end{equation}
\ie{}, a scale height of $57\pc (= 81\pc/\sqrt{2})$ and a mid-plane gas
density of 0.29\,\HH\,\pccm{}. The volume density is several orders of
magnitude smaller than the expected densities of molecular gas. This is due
to the fact that the molecular gas fills a small fraction of the galactic
volume.

\FigThCOvsTwCO{} %

\subsection{Interpretation 2: A mixture of dense and diffuse gas}
\label{sec:ext:densevsdiffuse}

The averaged volume density of the compact component is typically one order
of magnitude larger than the one of the extended component, pointing toward
different kinds of molecular gas. In this section, we recall that 1) bright
\twCO{} \Jone{} emission traces diffuse as well as dense gas, and 2) the
value of the $T(\twCO)/T(\thCO)$ ratio may be used to discriminate between
dense and diffuse gas. We will then check this ratio for M51.

\subsubsection{Bright \twCO{} \Jone{} emission also traces diffuse gas}

Bright \twCO{} \Jone{} emission is generally associated with dense cold
(typically $n\sim10^4-10^5\pccm$ and $T\sim10-20\K$) molecular gas, where
all hydrogen is molecular and all carbon is locked in CO.  However,
\citet{pety08} and \citet{liszt09} found surprisingly bright \twCO{}
\Jone{} lines (up to $\sim10\K$) in the nearby environment of Galactic
diffuse lines of sight ($\Av \sim1$), where the hydrogen is partly atomic
and partly molecular, and where the carbon occurs mostly in ionized form,
\ie{}, \Cp{}.  \citet{liszt10} explain such large \twCO{} brightnesses in
diffuse warm gas (typically $n\sim100-500\pccm$ and $T\sim50-100\K$) by the
fact that the gas is subthermally excited gas.  Indeed, large velocity
gradient radiative transfer methods~\citep{goldreich74,scoville74} show
that 1) $\WCO/\NCO$ is large because of weak CO excitation in warm gas
(50-100\K{}), and 2) $\WCO \propto \NCO$ until the opacity is so large that
the transition approaches thermalization. Hence, relatively bright \twCO{}
lines may at least trace either diffuse warm or dense cold molecular gas.

This is surprising because it is often argued that CO cannot survive
outside dense molecular gas since chemical models predict that several
magnitude of visual extinction are required so that CO survives
photo-dissociation. However, many absorption measures in the UV and
millimeter domains show that CO is present in gas whose hydrogen column
density is as low as
$10^{12}\pscm$~\citep{liszt08,sheffer08,sonnentrucker07}. The key point
here is that the CO chemistry in diffuse gas is still far from being
understood.  \citet{sonnentrucker07} found that a plot of the $\log(\NCO)$
as a function of the $\log(\NHH)$ can only be correctly fitted by two
power-law relationships, with a break at $(\NCO = 1.3 \times 10^{14}\pccm,
\NHH = 2.5 \times 10^{20}\pccm)$, corresponding to a change in the
production route for CO.  The production routes of CO are well understood
only in the regime of higher density gas. In diffuse gas, \citet{liszt00}
and \citet{liszt07} showed that if the amount of \HCOp{} observed in the
diffuse gas is fixed as a model parameter, it is easy to get large amount
of CO in UV illuminated gas through electron recombination of \HCOp{}
\begin{equation}
  \HCOp + \emr{e^-} \rightarrow \emr{CO} + \emr{H}.
\end{equation}
\citet{visser09} later confirmed this result. The next (still unsolved)
question is how large quantities of \HCOp{} form in the diffuse gas.

\subsubsection{The value of the $T(\twCO)/T(\thCO)$ ratio may 
  discriminate between dense and diffuse gas}

As bright CO \Jone{} lines may be associated with diffuse or dense gas,
this line alone cannot be used to differentiate both scenarios.
\citet{liszt10} argue that the $T(\twCO)/T(\thCO)$ ratio can discriminate
diffuse and dense gas. Indeed, from wide ($\sim 0.5\degr^2$) \twCO{} and
\thCO{} \Jone{} maps of the Galactic plane, \citet{polk88} measured an
average ratio of line intensities $\R=T(\twCO)/T(\thCO)$ of $6.7\pm0.7$.
Matching the beam areas at the two frequencies introduces an upward
correction factor of 1.1, \ie{}, $\R = 7.5\pm0.8$. This value is higher
than the typical factor measured for the core of Giant Molecular Clouds,
\ie{}, $\sim 3-5$~\citep[see \eg][]{frerking82}. Using the observational
fact that $T(\twCO)/T(\thCO) \sim 10-20$ in diffuse molecular
gas~\citep[see \eg][]{knapp88}, \citet{polk88} deduced that this diffuse
gas component significantly contributes to the large-scale \twCO{} emission
of the Galaxy. More recently, \citet{goldsmith08} deduced from \twCO{} and
\thCO{} wide-field mapping of the Taurus giant molecular cloud that about
half the mass of the gas traced by the \twCO{} \Jone{} emission comes from
diffuse gas.

The radiative and chemical properties of CO explain the large values of
this ratio in diffuse molecular gas. From measurements of CO absorption
lines against extra-galactic continuum background sources, \citet{liszt98}
showed that in local diffuse clouds, the ratio of \twCO{} and \thCO{}
column densities is in the range $ 15 \le N(\twCO)/N(\thCO) \le 54$, with
this ratio declining with increasing $N(\twCO)$. Hence, this ratio usually
differs strongly from the local interstellar ratio of the C elemental
abundances, typically $\twC/\thC = 60$. The value of the
$N(\twCO)/N(\thCO)$ ratio is due to the competition between the
fractionation of CO with \Cp{} and selective photo-dissociation of the two
CO isotopologues~\citep{liszt07}. Fractionation of CO with \Cp{}, \ie{},
\begin{equation}
  \twCO + \thCp \leftrightarrow \thCO + \twCp,
\end{equation}
enriches CO in \thCO{}. But selective photodissociation (\ie{}, the fact
that \twCO{} better self-shields from UV illumination than \thCO{}) more
than counters this effect in diffuse gas.

\subsubsection{In M51}
\label{sec:diffuse:m51}

Figure~\ref{fig:moments:30m} indicates that the morphology of the \twCO{}
and \thCO{} moments are strikingly similar where both tracers are detected.
To quantify this, Fig.~\ref{fig:13co-vs-12co} shows bi-dimensional
histograms of the integrated intensity (top row) and brightness (bottom
row) of the \thCO{} \Jone{} versus \twCO{} \Jone{} emission. Both sets of
histograms display similar linear relationships whose width is typically
related to the noise levels of both tracers. The left column displays the
histogram computed for the full field of view, while the three other
columns show how this histogram decomposes as a function of radius.  For
reference, we drew the same three lines on all histograms.  The slope of
these lines were chosen to follow the ridges of the brightness histograms
in the different radial ranges.

While the typical value for the $\twCO/\thCO$ ratios is about 8, there is a
slight but significant increase of this ratio from the inside to the
outside of the galaxy: This ratio is typically 7 (blue lines and
histograms) for radii below $35''$, 8.3 (green lines and histograms) for
radii between 35 and $150''$ and 11 (red lines and histograms) for radii
beyond $150''$. For comparison, this ratio increases from 4.6 to 10.0
across the Milky-way molecular ring, \ie{} from 0.5\Rsun{} to
\Rsun{}~\citep{liszt84}. This was an early indication that molecular gas
near the Solar Circle has a high proportion of diffuse material.

\FigMassDistribution{} %

At the angular resolution of the IRAM-30m ($22.5''$), the
$T(\twCO)/T(\thCO)$ increases from 6 to 11 with a typical value of \about{}
8 in M51. Assuming a Galactic [\twC]/[\thC] elemental ratio, the comparison
with our Galaxy points towards the interpretation that a significant
fraction of the CO emitting gas is diffuse. Is this visible in the emission
of \Cp{}? \citet{nikola01} published the first map of the 158\mum{}
[\ion{C}{ii}] emission at $55''$ resolution for M51 obtained with the
Kuiper Airborne Observatory. The interpretation of this emission is
difficult because it originates mostly from the warm ionized medium, though
the cold neutral medium contributes significantly to the [\ion{C}{ii}]
emission. In the cold neutral medium, the density solution is degenerate,
\ie{}, the medium could have either a low ($\sim100-300\pccm$) or a high
($10^3-10^6\pccm$) density. They conclude that ``A large fraction of the
overall [\ion{C}{ii}] emission in M51 can originate in an underlying
extended medium.'' A $12''$-resolution [\ion{C}{ii}] maps was observed as
part of a Herschel/PACS guaranted time project (PI: C.~Wilson, Parkin
\etal{}, in prep.). The comparison of this map with the $22.5''$
decomposition between compact and extended sources (see
Fig.~\ref{fig:moments:30m:tap2}) will probably shed light on the origin of
the [\ion{C}{ii}].

\subsection{Discussion}
\label{sec:ext:discussion}

Here we discuss the structure of the gas that emits the extended component
--- in particular the fact that it could be substructured like drops in fog
--- and the amount of gas that is extra-planar. We then summarize results
about extra-planar gas traced in \HI{} and CO. We finish with a discussion
of the possible origin of extra-planar CO emission.

\subsubsection{Structure of the extended component}

Giant molecular clouds are often thought to be composed of small clumps
that are unresolved, \ie{} a set of point sources. Hence, the
interferometer should recover all the flux of this set of point sources and
short-spacings should not be needed for extra-galactic observations.  This
argument is incorrect for two reasons. The first reason is that the limited
sensitivity of the interferometers limits the power of deconvolution to
recover the flux of point sources at low signal-to-noise ratio (See
Sect.~\ref{sec:pdbionly} and~\ref{sec:hybrid}).

The second reason why the PdBI-only data may not recover the full flux of
the source is more fundamental. We first assume that the source is a set of
unresolved components filling a volume that projects onto a plane-of-sky
area larger than the interferometer primary beam. If the unresolved
components are typically separated by an angular distance larger than the
synthesized beamwidth, the interferometer will indeed recover all the flux
because the GMC is observed as a set of separated point sources.  In
contrast, the interferometer will filter out most of the source flux when
the typical angular separation between the unresolved components is smaller
than the synthesized beamwidth because the source emission appears like a
flat source. Well known examples of this effect are 1) fog which appears
flat while made of water drops, and 2) a XIXth century pointilliste
painting which appears flat when unresolved by the eye. For this reason, it
is not possible to know before the observation of large complex sources
like nearby galaxies whether the short-spacings will be needed, and
multi-resolution \texttt{CLEAN} algorithms cannot help to solve this
ambiguity.

To explore this further, we speculate that the CO extended component is
made of diffuse gas as found in the envelopes of giant molecular clouds.
This gas would have a typical mid-plane density $\la 1000\pccm$, and a
temperature $\la200\K$. In such conditions, the \twCO{} \Jone{} is
subthermally excited. Moreover, Sect.~\ref{sec:ext:thinvsthick:results}
indicates that the average volume density is 1\pccm{}, implying a typical
volume filling factor of 0.1\%.  The structure of this gas is probably
filamentary, as observed in our Galaxy.

\subsubsection{Vertical mass distribution and extra-planar gas}

Figure~\ref{fig:mass:dist} displays the vertical distribution of the
molecular gas volume density, its decomposition into two components of
different scale heights, and the percentage of mass above a given galactic
height \abs{z}. In order to sample a large fraction of the sensible
parameter space, we computed 4 different cases: Two fractions of flux in
the extended component (either 30 or 50\%{}) and two scale height ratios (5
and 10) between the compact and extended components.  We see that the
fraction of flux in both components has a significant impact on the gas
distribution close to the galatic mid-plane.  However, the percentage of
the total mass above a given height $z$ is only marginally affected. In
contrast, doubling the ratio of the scale height doubles the galaxy height
above which a given mass is located. For instance, if we assume that the
smallest scale height is 40\pc{} and a scale height ratio of 5, only 2\% of
the total mass is above 400\pc{} while this proportion increases to 20\%
when the scale height ratio is 10. We thus estimate that between 2 and 20\%
of the molecular gas is extra-planar, \ie{} it lies at a galactic height at
least 10 times larger than the scale height of the dense molecular gas.

\subsubsection{\HI{} thick disk}

Lagging, thick \HI{} layers have long been detected in external galaxies
\citep[see, \eg{}, ][in NGC 253, NGC 4559, NGC 891 and NGC
6946]{boomsma05,barbieri05,oosterloo07,boomsma08}. The \HI{} gas in these
galaxy halos amounts to 3-30\% of the total \HI{} mass. For edge-on
galaxies, the \HI{} emission extends up to 12-22\kpc{} from the galactic
mid-planes.

\citet{miller09} studied the prototypical face-on spiral galaxy M83. They
found a spatially extended component rotating in the same sense but
40-50\kms{} more slowly in projection, with a line-of-sight velocity
dispersion of $10-15\kms$. The spatially extended structures are coincident
with the optical spiral arms. They interpreted this component as a
vertically extended disk rotating in the same sense but about 100\kms{}
more slowly than the kinematically cold, thin disk. It contains 5.5\% of
the total \HI{} mass within the stellar disk of the galaxy.

\subsubsection{A CO thick disk in the edge-on NGC 891 galaxy}

Among the previous galaxies, NGC 891 is particularly interesting. Its
characteristics are very similar to the Milky Way and
\citet{garciaburillo92} found in this system the first evidence for
extra-planar CO emission in an edge-on galaxy.  They detected molecular
emission 1-1.4\kpc{} above the disk, wide-spread along the major axis. The
associated Gaussian scale height is $\geq 600\pc$, \ie{} typically 2 to 3
times larger than that deduced here for M51. This CO scale height is
confirmed by a new IRAM/HERA map, which now images most of the thick disk
of the galaxy (Garcia-Burillo et al., in prep.), and it is consistent with
the scale height derived from the PAH emission~\citep{rand08}.

\citet{garciaburillo92} estimate this halo emission represents less than
20\% of the molecular mass in the disk.  However, they also state that 40
to 60\% of the thin disk emission must be in a low density component
(diffuse gas) to explain the typical value (8.5) of the ratio of the
\twCO{} \Jone{} to \thCO{} \Jone{} brightnesses.  Hence, the halo molecular
emission could be the tip of an independent component mixed with the dense
molecular thin disk. In this case, the extended component would amount for
a much larger fraction of the entire molecular gas. Indeed, in a follow-up
study \citet{sofue93} mapped the molecular gas at a distance of 3.5\kpc{}
from NGC 891's galaxy center, \ie{}, at the ring northern side. Using a
dual Gaussian fit through the emission profile extracted perpendicular to
the galaxy plane, they estimate 1) that the extended component has a
typical scale height of 2.1\kpc{} and 2) that it could account for 50\% of
the entire molecular gas.

\subsubsection{A diffuse CO thick disk in our Galaxy?}

Mapping 3 strips at constant longitude within a latitude range of
$[-3.5\degr,+3.5\degr]$, \citet{dame94} detected a thick molecular disk in
the inner Galaxy about 3 times as wide as that of the dense central CO
layer, and comparable in width to the thin \HI{} layer, \ie{} the height
profile can be fitted with a Gaussian FWHM of $\sim230\pc$. This data also
suggests that the high latitude gas lies mainly above the spiral arms.  The
mass of this gas would be 15\% of the total if it belongs to a distinct
molecular component.

Combining \Ebv{} reddening, \HI{} absorption and \twCO{} emission
measurements along many diffuse lines of sight in the Milky Way,
\citet{liszt10} recently found that the CO luminosity per \HH{} molecule
for diffuse gas is standard, \ie{} $\Xco = 2.0 \times 10^{20}\pscmpKkms$.
This standard value was introduced by~\citet{polk88} to take into account
the contribution of the diffuse gas to the \Xco{} factor. \citet{liszt10}
and \citet{liszt12} then deduced that the diffuse gas contribution to the
total CO luminosity seen looking down on the Milky Way is 0.47\Kkms{} to be
compared to 0.75\Kkms{}, the result from Galactic surveys. Hence, the
contribution of diffuse gas to the total CO luminosity in our Galaxy would
be between 39\% $(=100\times0.47/(0.47+0.75))$ and 63\%
$(=100\times0.47/0.75)$, depending on the fraction of the low brightness
diffuse component that was already detected by the Galactic surveys.

The typical sensitivity of wide-field CO surveys is 1\Kkms{}. In addition,
they have been mostly observed before the advent of wide bandwidth
receivers/spectrometers, which were installed in current facilities from
$\sim2005$ onwards. The total velocity bandwidth of the well-known
widefield Galactic CO surveys is typically only $\sim300\kms$. Finally, the
extension of the diffuse component we detect in M51 is large, several
hundred of parsecs.  Hence, such a component in our Galaxy could have been
confused with low level baselines in wide-field CO surveys.

\subsubsection{Possible origin of the extra-planar CO emission}

The following three scenarios are typically evoked to explain gas outside
the disk \citep[e.g.][]{putman12}: expelled gas from the disk via galactic
fountains or chimneys, accretion from the inter-galactic medium and/or
tidal debris from galaxy interactions. Except for the first scenario, it is
expected that the involved component reaches very large (more than a few
kpc) scale heights.

Therefore we focus in the following discussion on observational and
theoretical evidence for a potential flow of dense cold material from the
disk to the halo caused by massive star formation. The most extreme case to
be considered might be molecular outflows driven by stellar winds or
nuclear AGN, as seen, e.g., in CO in the nearby starburst galaxy M82
\citep{walter02}, and recently by Herschel in several nearby (U)LIRGs
\citep{sturm11}. However, typically these phenomena are restricted to the
centers of galaxies.

Evidence for dense material outside the thin disk comes from observations
of distinct optical extinction features in edge-on galaxies that trace
extra-planar dust. \cite{howk05} summarizes the findings on extra-planar
dust in nearby galaxies: extra-planar dust is found out to scale heights of
$z \le 2\kpc$, correlates well with regions of massive star formation in
the disk and the total amount of star formation, and appears to reside in
cold dense gas with densities of $n(\H) > 25\pccm$.  In this context the
detection of abundant extra-planar PAH emission is interesting.
\cite{rand11} find PAH emission out to $\sim0.5\,-\,1\kpc$ above the disk
plane in two edge-on spiral galaxies with detected extra-planar dust and
detect emission from the mid-IR $17\mum$ \HH{} line out to distances of
2\kpc{} in their targets. They speculate that massive star formation in the
disk is the cause for the extra-planar cold interstellar medium detected.

The resolved CO emission in M51 is preferentially found along the convex
side of the spiral arms where massive star formation is ongoing and has an
inferred scale height of $z \sim 200\pc$. These two findings are very
similar to results of the studies of extra-planar dust and PAH emission.
Thus we speculate that indeed star formation via galactic fountains or
chimneys has transported some of the molecular material away from the disk.

Several simulations of disk galaxies with star formation also suggest that
this explanation might be valid. The simulations have shown that cold ($T
\lesssim 200\K$), dense gas can reach heights of $100-200\pc$ above the
plane, but only when stellar feedback is
included~\citep{wada08,koyama09,dobbs11,acreman12,hill12}.  \citet{dobbs11}
find the scale height of the cold gas is $\sim 50-100\pc$, depending on the
level of feedback in the simulation.  From the top panel of Figure 9
of~\citet{dobbs11}, it is seen that gas at heights of a few 100 parsecs is
around $10^{-25}-10^{-24}\unit{g}\,\pccm$ or $20-200\,\HH\,\pccm)$. Gas at
these densities is not typically molecular. However, the top of Figure 14
of~\citet{dobbs08} shows that gas which has already been dense and
molecular can retain a high molecular fraction down to densities of
$\sim2\,\HH\,\pccm$ before the \HH{} is photodissociated.  Thus the
simulations indicate that a possible explanation of an extended diffuse
component is that stellar feedback pushes gas out to large distances above
the plane, but this gas remains molecular. The likelihood of this depends
on the gravitational contribution from the stars/gas in the vertical
direction, the local chemistry of \HH{} formation and destruction, the
effects of the stellar feedback and the surface density of the gas.

\section{Summary and conclusions}
\label{sec:summary}

We described in detail the calibration and construction of the PAWS \twCO{}
\Jone{} imaging of the central $\sim10\times6\kpc$ ($\sim270''\times170''$)
in M51, using observations from both the PdBI and 30m telescope.  The
achieved spatial resolution of 40\pc{} ($1.1''$) is close to the typical
size of galactic GMCs, and at least 10 times smaller in area than previous
interferometric maps obtained at OVRO, BIMA and
CARMA~\citep[]{aalto99,helfer03,koda11}.  The median brightness sensitivity
of 0.4\K{} in 5\kms{} channel spacing corresponds to $8.6\Msol\,\pc^{-2}$.
The total flux in the PAWS field of view and in a velocity range of
$\pm120\kms$ around the LSR systemic velocity is 64\% of the total M51 flux
of $1.4\times10^{9}\Kkmspcpc$, \ie{}, a molecular gas mass of
$6.2\times10^{9}\Msol$ (helium included).  The mean CO integrated intensity
and molecular mass surface density inside the PAWS field of view are
$18\Kkms$ and $77\Msol\pc^{-2}$, respectively.
  
The interferometer recovers only $(50\pm10)\%$ of the total flux.  The
remaining flux is mostly distributed on scales larger than 1.3\kpc{}
($36''$). Hence, the flux is about equally distributed into a spatially
extended and a spatially compact component. Using the hybrid synthesis
(PdBI+30m) and the PdBI-only deconvolved results, we established that the
extended component has the following properties: 1) It has a median
brightness temperature of 0.14\K{}, about 18 times fainter than the compact
component. 2) It covers about 30\% of the PAWS field of view, about 15
times as much as the compact component. 3) Its plane-of-sky kinematics
evolves smoothly, while plane-of-sky kinematics of the compact component
are more affected by streaming motions. 4) Its linewidth is typically twice
as large as that of the compact component.  5) Outside the region dominated
by the galaxy bulge and inner nuclear bar, its typical scale height is
$\sim200\pc$, or five times the compact component scale height.  6) Its
typical gas mid-plane pressure is $\sim(2-4)\times10^5\Kpccm$, in
approximate pressure equilibrium with the compact component. 7) Its typical
average mid-plane gas volume density is $\sim1\HH\,\pccm$, 10 times less
dense than the compact component.

We estimated between 2 and 20\% of the total molecular mass to be
extra-planar, \ie{} at galactic heights larger than 400\pc{}. We emphasized
that, while the emission of the extended component is mostly structured at
spatial scale larger than $36''$, it is probably made of unresolved
filamentary structures, which could typically fill 0.1\%{} of the volume of
the extended component.  The $T(\twCO)/T(\thCO)$ ratio at $23.6''$
resolution ranges from $\sim7$ to $\sim11$ when going from the inner to the
outer part of the galaxy, in agreement with a mixture of dense and diffuse
gas evolving from completely molecular in the inner galaxy to half atomic
in the outer galaxy.

Atomic thick disks are very common. Evidence for a molecular thick disk
exists in at least the edge-on NGC\,891 galaxy. A thick molecular disk
about 3 times as wide as the dense central CO layer was detected in the
inner part of our Galaxy. We thus interpret the extended component of M51
as a diffuse CO thick disk. The underlying physical interpretation of the
\twCO{} \Jone{} emission is different. If the gas is dense, it fills a
small fraction of the interstellar volume; it is confined by ram or
turbulent pressure (if not gravitationally bound); and it is on the verge
of forming stars. If the gas is diffuse, it is a warmer, low pressure
medium filling a large fraction of the interstellar volume; it contributes
more the mid-IR or PAH emission; and it is probably not gravitionally bound
or about to form stars.

\acknowledgements{}%

We thank the IRAM staff for their support during the observations with the
Plateau de Bure interferometer and the 30m telescope.  DC and AH
acknowledge funding from the Deutsche Forschungsgemeinschaft (DFG) via
grant SCHI 536/5-1 and SCHI 536/7-1 as part of the priority program SPP
1573 'ISM-SPP: Physics of the Interstellar Medium'.  CLD acknowledges
funding from the European Research Council for the FP7 ERC starting grant
project LOCALSTAR.  TAT acknowledges support from NASA grant \#NNX10AD01G.
During this work, J.~Pety was partially funded by the grant
ANR-09-BLAN-0231-01 from the French {\it Agence Nationale de la Recherche}
as part of the SCHISM project (\url{http://schism.ens.fr/}).  ES, AH and DC
thank NRAO for their support and hospitality during their visits in
Charlottesville.  ES thanks the Aspen Center for Physics and the NSF Grant
\#1066293 for hospitality during the development and writing of this paper.

We thank J.~Koda for providing the CARMA+45m data. J.~Pety thanks
F.~Combes, D.~Downes, S.~Guilloteau, P.~Gratier, H.~Liszt, V.~Pi\'etu, and
J.~Uson for illuminating comments about the interpretation of the data.

{\it Facilities:} \facility{IRAM:Interferometer}, \facility{IRAM:30m}.


\begin{appendix}

\section{CO luminosity and molecular gas masses}
\label{sec:co:luminosity}

\subsection{Computations}

The CO luminosity, \Lco{}, is estimated from the main beam temperature,
\Tmb{}, as
\begin{equation}
  \Lco = \cbrace{\sum_{l,m,v}\,\Tmb(l,m,v)}\,\D{lmv},
\end{equation}
where $(l,m,v)$ are the indices of the position, position, velocity data
cube and \D{lmv} is the volume of one pixel of this cube computed as
\begin{equation}
  \D{lmv} = \D{v}\,\bracket{\frac{\D{l}}{1''}\,\frac{\D{m}}{1''}\,\paren{37\pc\frac{D}{7.6\Mpc}}^2}.
\end{equation}
The uncertainty on this luminosity is computed as
\begin{equation}
  \uncert{\Lco} = \sqrt{\sum_{l,m,v}\,\independent(l,m,v)\,\uncert{\Tmb^2(l,m,v)}}\,\D{lmv},
\end{equation}
where \uncert{\Tmb(l,m,v)} is the noise level for the pixel $(l,m,v)$ and
$\independent(l,m,v)$ is the factor which reflects the correlations between
pixels. If we assume that the noise level is independent of the velocity
channel and that the velocity channels are uncorrelated, we obtain
\begin{equation}
  \uncert{\Lco} = \sqrt{\sum_{l,m}\,n_{v}(l,m)\,\independent(l,m)\,\uncert{\Tmb^2(l,m)}}\,\D{lmv},
\end{equation}
where $n_{v}(l,m)$ is the number of velocity channels for the position
(l,m). The correlation between spatial pixels can be approximated as the
inverse of the number of pixels in the resolution area, \ie{},
\begin{equation}
  \independent(l,m) = \frac{8\log(2)\D{l}\D{m}}{2\pi\Theta\theta},
\end{equation}
where $\Theta$ and $\theta$ are respectively the FWHM major and minor axes
of the Gaussian beam. It is independent of the pixel position $(l,m)$.

The mass of the associated molecular gas, \Mhh{}, is then
\begin{equation}
  \Mhh \pm \uncert{\Mhh} = \Xco \, \paren{\Lco\pm\uncert{\Lco}},
\end{equation}
where \Xco{} is the CO-to-\HH{} conversion factor. We will use the standard
Galactic conversion factor
\begin{eqnarray}
  \Xco &=& 2.0 \times 10^{20}\HH\pscmpKkms \\
       &=& 4.35 \Msol\pc^{-2} / (\Kkms).
\end{eqnarray}
The last value includes a factor of 1.36 by mass for the presence of
helium.

\subsection{Applications to M51}

The direct sum of the pixel brightness over the field of view observed with
the IRAM-30m map and between the $[-200,+300]\kms$ velocity range indicates
that the total CO luminosity of M51 (including the companion) is $1.442
\times 10^{9} \pm 4\times 10^{5}\Kkmspcpc$. The same computation in the
$[-120,+120]\kms$ velocity range (where mainly M51a emits) gives $1.441
\times 10^{9} \pm 3\times 10^{5}\Kkmspcpc$. Finally, the direct sum of the
pixel brightness over the field of view where the line integrated emission
is measured with a signal to noise ratio larger than 3 gives $1.428 \times
10^{9} \pm 2\times 10^{5}\Kkmspcpc$. It is clear that the uncertainty on
the result decreases when the summed volume of the cube is reduced because
most of the original volume is devoid of signal. But reducing the volume
introduces some kind of bias. However, the total luminosity does not vary
significantly in both three results. We thus assume that most of the signal
is contained within the smallest volume probed here.  This in particular
implies that the luminosity associated with the companion but outside the
$[-120,+120]\kms$ velocity range is negligible.  The uncertainties are
small compared to the absolute flux uncertainty \citep[$\la
10\%$,][]{kramer08} and the distance uncertainty (13\%). We quote them once
above to show that the radiometric noise is negligible when estimating the
CO luminosity.  The surface where emission is detected is 38.3 square
arcminutes, \ie{}, $1.9\times10^8\pc^2$. The associated total mass, mean
brightness, and mass surface density are respectively
$6.2\times10^{9}\Msol$, $7.6\Kkms$, and $33\Msol\pc^{-2}$. These values are
summarized in Table~\ref{tab:m51}.

The direct sum of the pixel brightness over the PAWS field of view and
between the $[-120,+120]\kms$ velocity range indicates that the final
hybrid synthesis data cube contains a total CO luminosity of $9.1\times
10^{8}\Kkmspcpc$. This corresponds to a molecular gas mass of $4.0\times
10^{9}\Msol$. The total surface covered was 10.5 square arcminutes, \ie{}
$5.1\times10^{7}\pc^2$. The mean brightness and mean surface density are
respectively $18\Kkms$ and $77\Msol\pc^{-2}$.

\section{Masking techniques}
\label{sec:masking}

\TabTotalFlux{} %
\FigTotalFlux{} %

In the previous section, we have seen that the line 0th order moment is
much noisier when the velocity range used for integration includes a large
number of channels devoid of signal. This effect is amplified when
computing the line 1st and 2nd order moments. It is thus desirable to limit
the velocity range of integration to channels where signal is detected. As
the systematic motions are large for a galaxy, using a single velocity
range for all the sky positions unavoidably implies moving the velocity
range of integration from position to position. Defining such moving
velocity ranges is never straightforward. In the case of the PAWS hybrid
synthesis data set, it is even more complex because a large fraction of the
flux lies at brightnesses barely detected (1 to $3\sigma$).

Moreover, masking out noisy channels always implies the risk to bias the
result. We therefore explored several alternative masking methods. In order
to quantify the benefit over cost of different techniques, we compare the
noise and signal aspects of 0th order moments images
(Fig.~\ref{fig:total:flux}) and the total luminosity found inside the kept
velocity channels (Table~\ref{tab:total:flux}).  Direct sum over a given
velocity range produces an integrated intensity map that is dominated by
noise (see Fig.~\ref{fig:total:flux}[a]) because CO emission is typically
only detected in a few velocity channels along each line-of-sight. But it
produces a robust estimate for the total CO luminosity.

The first alternative we tried is a simple {\it sigma-clipping} method,
whereby pixels containing emission with low significance were excluded. We
tested various brightness thresholds between 1 and 5$\sigma$, where
$\sigma$ is the standard deviation of the noise fluctuations estimated
using $\sim25$ emission-free channels for each line-of-sight. The
integrated intensity images were constructed by summing all unmasked (i.e.
`good') pixels across the observed LSR velocity range, i.e.
$[-297.5,+297.5]$\kms. As expected, lower thresholds produce maps that are
dominated by noise peaks, especially in the interarm region. Higher
thresholds produce maps with a cleaner appearance, but also do a poorer job
of recovering the total luminosity (see Table~\ref{tab:total:flux}). For
thresholds above $\sim3\sigma$, it is evident that genuine low surface
brightness emission in the interarm region is omitted. A map constructed
using a $3\sigma$ threshold is shown as an example in
Fig.~\ref{fig:total:flux}[b].

For the second method, which we call the {\it dilated mask} method, we
defined islands of significant emission in the hybrid synthesis data cube
by selecting peaks above a threshold of $t\sigma$ across two contiguous
velocity channels. This preliminary mask was expanded to include all
connected pixels with emission greater than $p\sigma$. Several combinations
of $t$ and $p$ values were tested, using $t \in [3,7]$ and $p \in [1.2,3]$.
The final mask was then applied to the original data cube, and the
integrated intensity image was constructed by summing unmasked pixels
within the LSR velocity range $[-297.5,+297.5]$\kms. We consider the map
obtained for $(t,p) = (5,1.2)$, shown in Fig.~\ref{fig:total:flux}[c], as
the best map produced using this method. Higher $t$ values tended to omit
genuine low surface brightness emission, whereas lower $t$ values include a
larger number of spurious noise peaks, especially in the interarm region
and at the edges of the survey field. Setting $p=1.2$ produced optimal
results in the sense that faint emission around the edge of bright
structures is retained within the mask, but the algorithm does not
misidentify large patches of noise as genuine emission. The total CO
luminosity within the map shown in Fig.~\ref{fig:total:flux}[c] is only
$7.6 \times10^{8}$~\Kkmspcpc, however, indicating that a significant
fraction of the CO emission within the survey field is not recovered in
this image.

For the third method, which we call the {\it smooth-and-mask} method
\citep[e.g.][]{helfer03}, we generated a mask by convolving the original
hybrid synthesis cube to a coarser spatial resolution using a Gaussian
smoothing kernel with FWHM $\theta$. The RMS noise for each sightline
within the smoothed cube was estimated from $\sim25$ emission-free
channels, then pixels in the smoothed cube with emission below a
significance threshold $m\sigma$ were blanked. After transferring this mask
back to the original data cube, the integrated intensity image was
constructed by summing unmasked pixels across the LSR velocity range
$[-297.5,+297.5]$\kms. As for the previous methods, we experimented with
different combinations of the $\theta$ and $m$ parameters. We consider the
map obtained using a 3\farcs6\ kernel and a $5\sigma$ threshold to be the
best produced using this method. The total CO luminosity within this map,
which is shown in Fig.~\ref{fig:total:flux}[d], is
$6.9\times10^{8}$~\Kkmspcpc.

For the fourth method, we integrated the PdBI+30m data cube over a narrow
velocity range (which we refer to as the `integration window'), centered on
the radial velocity at the peak of the \HI{} line profile for each
line-of-sight. We call this method the `\HI{} prior method'. The \HI{}
velocity template was constructed using the \HI{} data cube from THINGS
\citep{walter08}, which covers the entire disk of M51 at $\sim11\arcsec$
resolution. Integration windows with velocity widths between 10 and
100\kms{} were tested. The rationale behind this approach is that CO
emission in nearby galaxies is mostly associated with high brightness \HI{}
emission \citep[e.g.][]{schruba11,engargiola03}.  Comparison with maps in
the other panels of Fig.~\ref{fig:total:flux} suggests that CO emission in
the nuclear region of M51 is not well-recovered by this method. One
advantage of this approach, however, is that it yields an upper limit on
the CO integrated intensity for pixels without detectable emission, and may
therefore be more suitable for some types of quantitative analysis. The map
for a 50~\kms{} integration window is shown in
Fig.~\ref{fig:total:flux}[e]. The total luminosity in this map is
$8.4\times10^{8}$~\Kkmspcpc.

Finally, we used the strengths of two of these methods to build our best
mask. We started by building two 2D masks based on the second
(\emph{dilated mask}) and fourth (\emph{\HI{} prior}) approaches. In each
case, we constructed a 3D dilated mask from the hybrid synthesis datacube.
However, this technique also catches noise patches at ``wrong'' velocities.
To remove these, we computed the associated centroid velocity map, and we
then produced a 2D mask where the CO velocity centroids differed by less
than 30\kms{} from a map of the \HI{} velocity field. Using this 2D mask,
we could build a 3D filtered mask.

The difference between these two 3D filtered masks comes from the different
thresholds used to produce initial 3D dilated masks. The first dilated mask
used $(t,p) = (4,1)$, which allows us to find isolated faint point sources.
However, these parameters also catches isolated noise at ``forbidden''
velocities, which will considerably bias the line-of-sight CO centroid
velocities. The comparison of the CO and HI centroid velocities may thus
kill perfectly valid lines of sight. This happens in particular in the M51
central part. To avoid this effects, the second dilated mask used $(t,p) =
(10,1.5)$. The high signal-to-noise ratio (10) starting point ensures that
no outlier velocities will be caught in the 3D mask but it will miss
isolated low brightness signal. The two 3d filtered masks thus identify
complementary kind of signal. We finally make the union of these 3D
filtered masks to obtain the final mask. The resulting 0-moment map, which
has a total luminosity of $8.5\times10^{8}\Kkmspcpc$, is shown in
Fig.~\ref{fig:total:flux}[f]. This map contains 93\% of the total
luminosity computed from direct summation. The comparison of these maps
with the ones of the other methods shows that it is the most successful at
recovering all genuine emission within the data cube. This mask was thus
used to compute the 1st and 2nd order moments of the hybrid synthesis cube
at $1''$ resolution.

\section{Modeling the impact of the IRAM-30m error beam}
\label{sec:errorbeam}

In this appendix, we assess how much the error beams of the IRAM-30m
contribute to the extended emission. To do this, we first convolved a model
of the source emission in the PAWS field of view 1) with the ideal 30m beam
and 2) with a model of the true 30m beam, including the diffraction pattern
and the error beams.  We then subtracted the convolved maps to estimate the
contribution of the diffraction pattern and error beams to the extended
emission. We first describe the 30m beam and the source models used, before
discussing the results.

\subsection{Beam model}

\TabErrorBeam{} %
\FigErrorBeam{} %

After measuring the 30m beam pattern at 3.4, 2, 1.3 and 0.8\mm{} from total
power scans across the Moon around full Moon and new Moon, \citet{greve98}
modelled the beam using antenna tolerance theory.  They deduced that the
beam consists of the diffracted beam, two underlying error beams, which can
be explained by the panel dimensions, and a beam deformation mostly due to
large-scale transient residual deformations of the telescope structure. We
scaled their results to the frequency of the \twCO{} \Jone{} line to model
the error beams we use.  The diffraction pattern was computed for an
antenna illuminated by a Gaussian beam of 12.5\,dB edge taper and a ratio
of the secondary-to-primary diameter (blockage factor) of 0.067, using the
prescription of~\citet{goldsmith98}.  Figure~\ref{fig:error-beam} shows the
properties of the resulting beam and Table~\ref{tab:error:beam} lists the
typical scales, amplitudes, and contributions to the total power of the
different beam components.

The diffraction pattern and two of the three error beams are linked to the
structure of the primary mirror of the 30m, which have not changed since
the study of \citet{greve98}. Hence the typical angular scales of the
different error beams stay constant over time. However, the 30m primary
surface accuracy improved in 2000 after several holography campaigns.
Moreover, the thermal balance of the telescope primary surface was also
improved around this time. Both effects explain why the beam efficiencies
of the 30m are significantly higher today than the values provided
by~\citet{greve98}: \eg{}, 0.78 instead of 0.72 at the \twCO{} \Jone{}
frequency. We used slightly different values of the 30m beam efficiency in
our analysis for historical reason. As no newer detailed measurements of
the error beams are available, we used 0.72 for consistency with the work
by \citet{greve98} to model the beam here, while we applied 0.75 to the
data.  As the current best estimate of the beam efficiency is 0.78, both
values present only variations of a few percent and do not alter our
conclusions. In particular, our estimates of the error beam contribution
can be seen as a firm upper limit.

\subsection{Source model}
\label{sec:errobeam:source}

\FigPAWSmomentsModels{}%

The best proxy for the source model is the result of this project, \ie{}
the distribution of the emission measured in the PAWS field of view. We
made two different source models. First, we want to check whether the error
beam contribution of the compact sources alone can account for the extended
emission. We started with the $1''$ PdBI-only data cube, and we set to zero
all pixels whose brightness was below 3 times the noise level in order to
avoid spurious sources. We multiplied all pixel brightnesses by a factor of
about 2, needed to recover the total flux in the PAWS field of view. This
assumes that all extended emission is spurious and that any flux associated
with it should indeed belong to compact sources. We will refer to this cube
as the compact model.

In contrast, we assume for the second model that the extended emission is
genuine. To model the extended component, we start with the $6''$
subtracted cube, in which we set to zero all pixels whose brightness was
below 3 times the noise level. We then added all the brightness of the
$1''$ PdBI-only data cube whose brightness is above 3 times the noise
level. The flux contained in this model is then 3\% larger than the flux
measured in the PAWS field of view. We will refer to this cube as the
summed model.

\subsection{Modeling results}
\label{sec:errorbeam:results}

\FigSpectraComparison{}%

We convolved each model with the true 30m beam model. We normalized the
result by the main beam efficiency to get to the main beam temperature
scale. Fig.~\ref{fig:moments:models} compares the moments of both modeled
30m emission with the measured emission. All the moments were computed
using the same position-position-velocity mask (the one of the data) to
ensure a meaningful comparison. The compact model produces much higher peak
temperatures and integrated emission in the arms, while it understimates
the flux in the inter-arm region. The summed model provides a much better
match to the observations.

We subtracted from the previous cubes the associated brightness model
convolved with the ideal 30m beam. This yields the contribution of the
error beams to the signal measured with the 30m. In particular, it allows
us to estimate the error beam contribution to the extended emission, as we
showed in Section~\ref{sec:amplitude} that the flux of the extended
emission is mostly structured at scales larger than $36''$, \ie{} at scales
larger than the ideal 30m beam FWHM. For both source models, the flux
scattered into the error beams is less than 20\% the total flux in the PAWS
field of view but the peak brightness due to the error beam contribution is
only 55\mK{}, \ie{}, about 3.5 times the median noise level of the 30m
observations.  The typical angular scales of the error beams are large,
implying that the flux is scattered at low brightness level over wide
regions of the sky. The median value of the error beam contribution is
19\mK{}, while pixels brighter than 5 times the noise level have a median
of 140\mK{} in the extended emission measured at $6''$ (see
Sect.~\ref{sec:noise+signal}).  Moreover, flux above 0.1\K{} in the
extended emission cube represents 79\% of the flux in this cube (we
subtracted the flux contributed by the error beams in the same mask to
compute this value). We deduce that the error beams contribute less than
20\% of the flux in the extended emission, \ie{}, less than 10\% of the
total flux in the PAWS field of view. We speculate that the baselining of
the 30m spectra removed a large fraction of the flux associated with the
error beams.

\FigFWHMcomparison{}%

Fig.~\ref{fig:spectra-comparison} compares the spectra of our summed model,
the extended emission, and the contribution from the error beams at six
positions, which sample different ratios of compact to extended brightness
as well as different characteristics of the extended emission. These
spectra illustrate that the emission associated with the error beams has a
very different signature in space and velocity from the extended emission
distribution. In particular, the spectra resulting from the error beams
have much wider linewidths than the observed spectra of the extended
emission. We finally investigate the impact of the error beams on the
moments of the extended emission. We recomputed the moments of the $6''$
extended emission cube in two different ways. First, we only include the
pixels brighter than two times the peak brightness (\ie{}, 0.1\K{}) due to
the error beams.  In the second test, we include only the pixels brighter
than four times the contribution of the error beam for each source model.
The three resulting moment maps are similar to the original one.  This is
also true for the second moment or linewidth displayed in
Fig.~\ref{fig:fwhm:comparison}. We thus conclude that the large linewidths
of the extended emission are genuine.

\section{Estimating the vertical velocity dispersion}
\label{sec:vz:disp}

The velocity of a parcel of gas in the galaxy frame is
\begin{equation}
  \vgal = (\vec{\vp},\vec{\vz}),
\end{equation}
where (\vec{\vp},\vec{\vz}) respectively are the in-plane and perpendicular
to the galactic plane velocity components. In the observing frame, the
Doppler effect allows us to infer an observed velocity, \vobs{}, which is
the projection of \vgal{} along the line of sight. If \up{} is the
projection of \vec{\vp} on the galaxy major axis and \i{} is the
inclination of the galaxy plane onto the line of sight, we get
\begin{equation}
  \vobs = \up \, \sin \i + \vz \, \cos \i.
\end{equation}
In the optically thin limit, a line is an histogram of all the \vobs{}
values along each line of sight. For a \emph{resolved} perfectly face-on
galaxy,
\begin{equation}
  \vobs = \vz.
\end{equation}
The centroid velocity would thus be constant and the linewidth would give
the dispersion of the velocity distribution along the vertical axis under
the condition that the velocity has a symmetric distribution of velocity
along any line of sight. For a \emph{resolved} edge-on galaxy,
\begin{equation}
  \vobs = \up.
\end{equation}
If the velocity is only rotational and the density is much higher in the
spiral arm, the centroid velocity is then biased towards the velocities in
the spiral arm and the velocity dispersion gives an idea of the velocity
content along the line of sight. This is why we still have some information
about the rotation curve of the galaxy.

For any other inclination, the centroid velocity, \vcent{}, is
\begin{equation}
  \vcent = \aver{\vobs} = \aver{\up} \, \sin \i,
\end{equation}
as long as the galaxy disk is not warped and the distribution of the
velocity perpendicular to the galactic plane is symmetric, \ie{}
$\aver{\vz} = 0$. Now, the velocity dispersion is computed as
\begin{equation}
  \aver{\paren{\vobs-\vcent}^2} = 
  \aver{\bracket{\vz \, \cos \i + \paren{\up - \aver{\up}} \, \sin \i }^2}.
\end{equation}
Assuming that there are no correlations between vertical and horizontal
motion, we obtain
\begin{equation}
  \aver{\paren{\vobs-\vcent}^2} = 
  \aver{\vz^2} \, \cos^2 \i + \aver{ \paren{\up - \aver{\up} }^2} \, \sin^2 \i.
\end{equation}
If we also assume that the velocity field is only made of a systematic
motion (\usys{}) parallel to the galaxy plane plus an isotropic 3D
turbulent motion of typical dispersion (\vturb{}), then
\begin{equation}
  \aver{\vz^2} = \vturb^2
  \quad \mbox{and}
\end{equation}
\begin{equation}
  \aver{ \paren{\up - \aver{\up}}^2} = \vturb^2 + \aver{\paren{\usys - \aver{\usys}}^2}.
\end{equation}
Note that $\aver{\usys} = \aver{\up}$ and the turbulent component yield
only one time \vturb{} because \up{} is an unidimensional velocity
component. Hence,
\begin{equation}
  \aver{\paren{\vobs-\vcent}^2} = 
  \vturb^2 + \aver{\bracket{\paren{\usys - \aver{\usys}} \,\sin \i}^2},
\end{equation}
where \aver{\bracket{\paren{\usys - \aver{\usys} \,\sin \i}}^2} is the
contribution of the systematic in-plane motions inside the beam to the line
2nd order moment. We can estimate this contribution as
\begin{equation}
  \aver{\bracket{\paren{\usys - \aver{\usys}} \,\sin \i}^2} \sim
  \bracket{\cvi{} \, \frac{\beamwidth}{\sqrt{8\ln2}}}^2,
\end{equation}
where \beamwidth{} is the resolution beamwidth of the observations and
\cvi{} is the modulus of the centroid velocity 2D-gradient. In our case,
these last two quantities can be measured. This thus yields an estimation
of the galaxy vertical velocity dispersion as the square root of
\begin{equation}
  \aver{\vz^2} = \vturb^2 \sim 
  \aver{\paren{\vobs-\vcent}^2} - \bracket{\cvi{} \, \frac{\beamwidth}{2.35}}^2. 
\end{equation}

\section{Additional material}

This section displays additional material that completes the description of
the observations and the presentation of the data cubes. All these tables
and figures are only made available online.

Table~\ref{tab:obs:30m:detail} presents the weather conditions during the
IRAM-30m observing run. Table~\ref{tab:obs:pdbi:detail} exhaustively lists
the interferometric sessions observed during the IRAM large program PAWS.
Table~\ref{tab:fluxes} gives the evolution of the calibrator fluxes as a
function of time, a key intermediate in the absolute amplitude calibration
of the data.

Figures~\ref{fig:channelmaps:30m:12co10}
and~\ref{fig:channelmaps:30m:13co10} present the channel maps of the
IRAM-30m single-dish data for the \twCO{} and \thCO{} \Jone{} lines.
Figures~\ref{fig:channelmaps:pdbi+30m:1}
and~\ref{fig:channelmaps:pdbi+30m:1} show the channel maps of the hybrid
synthesis data cube at $1''$ angular resolution, the main data product of
the PAWS large program.
Figures~\ref{fig:channelmaps:extended-vs-combined:1}
and~\ref{fig:channelmaps:extended-vs-combined:2} overlay the channel maps
of the bright compact emission as contours over the channel maps of the
faint extended emission displayed in color.

Finally,
Figures~\ref{fig:moments:paws:full},~\ref{fig:moments:paws:tap1},~\ref{fig:moments:paws:tap2},
and~\ref{fig:moments:30m:tap2} summarize the properties of the
decomposition of the hybrid synthesis emission into the PdBI-only and the
subtracted emissions, respectively at 1, 3, 6, and $22.5''$.

\TabObsSDdetail{} %
\TabObsPdBIdetail{} %
\TabFlux{} %

\clearpage{} %
\newpage{} %

\FigSDtwCOmaps{} %
\FigSDthCOmaps{} %
\FigPdBImapOne{} %
\FigPdBImapTwo{} %
\FigExtendedCombinedMapOne{} %
\FigExtendedCombinedMapTwo{} %

\FigPAWSmomentsFull{} %
\FigPAWSmomentsTapOne{} %
\FigPAWSmomentsTapTwo{} %
\FigSDmomentsTapTwo{} %

\end{appendix}


\bibliographystyle{apj} %
\bibliography{pety-paws} %

\end{document}